\documentclass[a4paper,11pt]{article}
\pdfoutput=1 % if your are submitting a pdflatex (i.e. if you have
\usepackage{jheppub} % for details on the use of the package, please
\usepackage{amsmath,graphicx,color,xcolor,epsfig}
\usepackage{epstopdf}
\usepackage{hyperref}

\DeclareMathOperator{\Tr}{Tr}

\title{
Mass spectrum of the hidden-charm pentaquarks in the compact diquark model
}

\author[a,1]{Ahmed Ali,\note{Corresponding author.}}
\author[b]{Ishtiaq Ahmed,}
\author[c]{M. Jamil Aslam,}
\author[d]{Alexander Ya. Parkhomenko,}
\author[b]{and Abdur Rehman}

\affiliation[a]{Deutsches Elektronen-Synchrotron DESY, D-22607 Hamburg, Germany}
\affiliation[b]{National Centre for Physics, Quaid-i-Azam University Campus, 
                Islamabad 45320, Pakistan}
\affiliation[c]{Physics Department, Quaid-i-Azam University, Islamabad 45320, Pakistan}
\affiliation[d]{P.\,G.~Demidov Yaroslavl State University,  
                Sovietskaya 14, 150003 Yaroslavl, Russia}

\emailAdd{ahmed.ali@desy.de}                                                        
\emailAdd{ishtiaqmusab@gmail.com}
\emailAdd{muhammadjamil.aslam@gmail.com}
\emailAdd{parkh@uniyar.ac.ru}
\emailAdd{Abdur.Rehman@fuw.edu.pl}

\abstract{
The LHCb collaboration have recently updated their analysis of the resonant $J/\psi\, p$ mass 
spectrum in the decay $\Lambda_b^0 \to J/\psi\, p\, K^-$, making use of their combined Run~1 
and Run~2 data. In the updated analysis, three narrow states, $P_c (4312)^+$, $P_c (4440)^+$, 
and $P_c (4457)^+$, are observed. 
The spin-parity assignments of these states are not yet known. We interpret 
these narrow resonances as compact hidden-charm diquark-diquark-antiquark pentaquarks. 
Using an effective Hamiltonian, based on constituent quarks and diquarks, we calculate 
the pentaquark mass spectrum for the complete $SU (3)_F$ lowest $S$- and $P$-wave multiplets, 
taking into account dominant spin-spin, spin-orbit, orbital and tensor interactions. The resulting 
spectrum is very rich and we work out the quark flavor compositions, masses, and $J^P$ quantum numbers 
of the pentaquarks. However, heavy quark symmetry restricts the observable states in $\Lambda_b$-baryon,
as well as in the decays of the other weakly-decaying $b$-baryons, $\Xi_b$ and $\Omega_b$. 
In addition, some of the pentaquark states are estimated to lie below the $J/\psi\, p$ threshold 
in $\Lambda_b$-decays (and corresponding thresholds in $\Xi_b$- and $\Omega_b$-decays). They decay 
via $c \bar c$ annihilation into light hadrons or a dilepton pair, and are expected to be narrower 
than the $P_c$-states observed. We anticipate their discovery, as well as of the other pentaquark 
states present in the spectrum at the LHC, and in the long-term future at a Tera-$Z$ factory. 
}

\begin{document}

\preprint{DESY 19-052}

\maketitle
\flushbottom

\section{Introduction}
\label{sec:introduction}

Recently, the LHCb collaboration has presented an updated account of the resonant $J/\psi\, p$ 
mass spectrum in the decay $\Lambda_b^0 \to J/\psi\, p\, K^-$, based on the combined Run~1 and 
Run~2 data, adding up to 9~fb$^{-1}$~\cite{Aaij:2019vzc}. In this analysis, which supersedes 
their earlier findings from 2015~\cite{Aaij:2015tga}, nominal fits of the data have been performed 
with an incoherent sum of Breit-Wigner amplitudes, which have resulted in the observation of three 
peaks, whose masses, decay widths (with 95\%~C.L. upper limits), and the ratio~${\cal R}$, defined as 
\begin{equation}
{\cal R} \equiv \frac{{\cal B} (\Lambda_b \to P_c^+\, K^-) \, 
                      {\cal B} (P_c^+ \to J/\psi\, p)}
                     {{\cal B} (\Lambda_b \to J/\psi\, p\, K^-)} ,
\label{eq:R-def}
\end{equation}  
are given in table~\ref{tab:LHCb-data-2019}.
The state $P_c (4450)^+$ in the 2015 data~\cite{Aaij:2015tga}, is now replaced by two narrow states, 
$P_c (4440)^+$ and $P_c (4457)^+$. In addition, a third narrow peak, $P_c (4312)^+$, having the 
mass $M = (4311.9 \pm 0.7^{+6.8}_{-0.6})$~MeV, is also observed. The spin-parity,~$J^P$, assignments 
of the three narrow states, which are crucial to decipher the underlying dynamics of the pentaquark 
states, are not yet determined. The broad peak $P_c (4380)^+$ from the earlier data~\cite{Aaij:2015tga} 
is neither confirmed nor refuted, as the current LHCb analysis is not sensitive to broad resonances. 
Hence, it is entirely conceivable that more $P_c^+$-like structures are present in the $J/\psi\, p$ 
invariant-mass spectrum, anticipated in the compact pentaquark interpretation, similar to the excited 
$\Lambda^*$-baryon spectrum in the $K\, p$ channel, expected in the quark model and confirmed 
in data~\cite{Tanabashi:2018oca}. For the discussion of the 2015 LHCb data on pentaquarks and other 
multiquark hadrons and references to the earlier works see the reviews~\cite{Tanabashi:2018oca,% 
Ali:2017jda,Esposito:2016noz,Chen:2016qju,Guo:2017jvc,Olsen:2017bmm,Lebed:2016hpi,Hosaka:2016pey,% 
Karliner:2017qhf,Yuan:2018inv,Liu:2019zoy,Brambilla:2019esw} and~\cite{Ali:2019roi}. 

The new pentaquarks reported by LHCb~\cite{Aaij:2019vzc} have triggered a surge in theoretical papers 
interpreting the three narrow resonances as loosely-bound hadronic-molecule states in various 
incarnations~\cite{Chen:2019bip,Chen:2019asm,Guo:2019fdo,Liu:2019tjn,He:2019ify,Xiao:2019aya,% 
Huang:2019jlf,Shimizu:2019ptd,Xiao:2019mvs,Zhang:2019xtu}. A more up to date list of references 
can be seen in the proceedings of a recent Workshop on Exotic Hadrons~\cite{Skwarnicki:2019}.
Indeed, in the mass region of interest, a number of kinematical thresholds are present, such~as 
$\Sigma_c^+ \bar D^0$ ($E_{\rm thr} = 4317.73 \pm 0.41$~MeV), 
$\Sigma_c^{*+} \bar D^0$ ($E_{\rm thr} = 4382.3 \pm 2.4$~MeV), 
$\Lambda_c^{*+} \bar D^0$ ($E_{\rm thr} = 4457.09 \pm 0.35$~MeV), and
$\Sigma_c^+ \bar D^{*0}$ ($E_{\rm thr} = 4459.9 \pm 0.9$~MeV)~\cite{Tanabashi:2018oca}. 
The masses of the observed resonances in the hadron molecule interpretation are essentially 
the sums of their respective hadronic constituents, which is striking. Thus, not only the three 
observed narrow pentaquarks can be accommodated, having the spin-parity and isospin ($J^P = 1/2^-$, 
$I = 1/2$) and ($J^P = 3/2^-$, $I = 1/2$), but several other states, also with a negative parity 
due to their assumed $S$-wave character, are predicted. These emphatic claims are at best 
tentative and await a confirmation of the spin-parity assignments of the three observed states. 
The nature of the $P_c (4312)^+$ as a candidate for the $\Sigma_c^+ \bar D$ molecular state was 
also analyzed based on the $S$-matrix principles~\cite{Fernandez-Ramirez:2019koa} and it was 
found that the attraction in this system is not enough for a bound state. 

We emphasize that the decays $\Lambda_b \to \Sigma_c^{(*)} + X$ are anticipated to be suppressed 
by Heavy Quark Symmetry (HQS), due to the mismatch of the spectator diquark spin (a conserved quantity 
in the HQS limit) in $\Lambda_b$- and $\Sigma_c^{(*)}$-baryons. This suppression is well-known,
but is often ignored in the literature on the hadron molecular approach for the new $P_c$-states, 
though there are some rare exceptions, see, for example,~\cite{Weng:2019ynv}. There is an ample 
evidence of this suppression in the PDG tables, in which the non-leptonic transitions 
$\Lambda_b \to \Lambda_c^{(*)} + X$ dominate and the $\Lambda_b \to \Sigma_c^{(*)} + X$ transitions 
are rather sparse~\cite{Tanabashi:2018oca}. In the exact HQS limit, such decays are forbidden, but as 
the HQS is not an exact symmetry and is broken by power ($1/m_b$) and QCD ($\mathcal{O} (\alpha_s (m_b))$) 
corrections, they are allowed but have reduced decay rates. It has important implications 
for the analysis of the $P_c$-states in the hadron molecule approach as well.
In particular, it implies that the branching ratios ${\cal B} (\Lambda_b \to P_c^+\, K^-)$, 
which in the hadron-molecule interpretations of the $P_c$-states are supposed to be induced 
by the intermediate $(\Sigma_c\, \bar D^{(*)})\, K^-$ states, followed by rescattering, are 
subject to the HQS-implied suppression. Since the ratio~${\cal R}$ in~(\ref{eq:R-def}), which 
is the product ${\cal B} (\Lambda_b \to P_c^+\, K^-) \, \times {\cal B} (P_c^+ \to J/\psi\, p)$ 
in units of ${\cal B} (\Lambda_b \to J/\psi\, p\, K^-)$, is well-measured for all three states 
by LHCb (see table~\ref{tab:LHCb-data-2019}), the branching ratios ${\cal B} (P_c^+ \to J/\psi\, p)$ 
are not expected to be small in the hadron-molecule interpretation. It is relevant to point out 
that model-dependent 90\% C.L. upper limits on ${\cal B} (P_c^+ \to J/\psi\, p)$ of~4.6\%, 2.3\%, 
and~3.8\% for $P_c^+ (4312)$, $P_c^+ (4440)$, and $P_c^+ (4457)$, respectively, have been posted 
by the GlueX collaboration~\cite{Ali:2019lzf}, assuming they have $J^P = 3/2^-$ (i.\,e., $S$-wave 
$J/\psi\, p$ states). While this is currently not robust enough an argument against the $P_c$'s 
being hadron molecules, due to inherent assumptions about the theoretical estimates of the 
photoproduction cross sections, but improved measurements will soon test this interpretation 
in the $P_c$-photoproduction experiments~\cite{Ali:2019lzf,Blin:2016dlf}. 
The role of the photoproduction process $\gamma\, p \to J/\psi\, p$ in constraining the models 
of the new $P_c$-states is noted in~\cite{Cao:2019kst,Wang:2019krd}.
Another process to measure the $P_c$ production is the antiproton-deuterium collisions,
in which case the charm quark and antiquark entering the charmonium state~$J/\psi$ or~$\eta_c$ 
are the result of the proton-antiproton annihilation~\cite{Voloshin:2019wxx}. Thus, the internal 
consistency of the hadron molecule approach for the new $P_c$ states remains to be checked.

In the preceding Letter~\cite{Ali:2019npk}, two of us have argued that there also exists  
a {\it prima facie} case for the three narrow $P_c^+$-resonances to be considered as candidates 
for compact pentaquarks. The basic idea of this approach is that highly correlated colored diquarks 
play a key role in the physics of multiquark states~\cite{Maiani:2004vq,Lipkin:1987sk,Jaffe:2003sg}, 
and they are at work in the underlying dynamics of the $P_c^+$-states. A brief account of the new 
pentaquarks was presented in~\cite{Ali:2019npk} in the effective Hamiltonian framework based on 
constituent quarks and diquarks, using isospin and heavy quark symmetry. In this paper, we extend 
the compact diquark template to cover the complete $SU (3)_F$-multiplets of compact hidden-charm 
pentaquarks in the lowest $S$- and $P$-wave states. Some of them can be searched for in the decays 
of the $\Lambda_b$-baryon, but some others can only be reached in the decays of the other 
weakly-decaying $b$-baryons,~$\Xi_b$ and~$\Omega_b$.

\begin{table}[tb]
\begin{center}
\begin{tabular}{|ccccc|} 
\hline
    State      &         Mass [MeV]             &         Width [MeV]           & (95\% CL) &      $\mathcal{R}\, [\%]$       \\ \hline 
$P_c (4312)^+$ & $4311.9 \pm 0.7^{+6.8}_{-0.6}$ &   $9.8 \pm 2.7^{+3.7}_{-4.5}$ &  ($< 27$) & $0.30 \pm 0.07^{+0.34}_{-0.09}$ \\[1mm] 
$P_c (4440)^+$ & $4440.3 \pm 1.3^{+4.1}_{-4.7}$ & $20.6 \pm 4.9^{+8.7}_{-10.1}$ &  ($< 49$) & $1.11 \pm 0.33^{+0.22}_{-0.10}$ \\[1mm] 
$P_c (4457)^+$ & $4457.3 \pm 0.6^{+4.1}_{-1.7}$ &   $6.4 \pm 2.0^{+5.7}_{-1.9}$ &  ($< 20$) & $0.53 \pm 0.16^{+0.15}_{-0.13}$ \\[1mm] \hline 
\end{tabular}
\end{center} 
\caption{
Masses, decay widths (with 95\%~C.L. upper limits), and the ratio~$\mathcal{R}$, 
of the three narrow $J/\psi\, p$ resonances observed by the LHCb collaboration 
in the $\Lambda_b \to J/\psi\, p\, K^-$ decay~\cite{Aaij:2019vzc}.}
\label{tab:LHCb-data-2019} 
\end{table}

The pentaquark dynamics depends upon how the five constituents, i.\,e., the 4~quarks and an antiquark, 
are structured. Since quarks transform as a triplet~$\tt 3$ of the color $SU (3)$-group, the diquarks 
resulting from the direct product $\tt 3 \otimes 3 = \bar 3 \oplus 6$, are thus either a color 
anti-triplet~$\tt \bar 3$ or a color sextet~$\tt 6$. Of these only the color~$\tt \bar 3$ configuration 
is kept, as suggested by perturbative arguments. Another justification is that the color sextet diquarks 
are heavier than the corresponding color anti-triplet ones, and can be integrated out for spectroscopic 
considerations. This remains to be quantified, which can be eventually done in lattice QCD. 
Both spin-0 and spin-1 diquarks are, however, allowed. 
In the case of a diquark $[q^\prime q^{\prime \prime}]$ consisting of two light quarks 
($q^\prime,\, q^{\prime \prime} = u,\, d,\, s$), the spin-0 diquarks are believed to be more tightly 
bound than the spin-1, and this hyperfine splitting has implications for the spectroscopy. 
For the heavy-light diquarks, such as~$[cq]$ or~$[bq]$, this splitting is suppressed by $1/m_c$ for the 
$[cq]$-diquark or by $1/m_b$ for the $[bq]$-diquark, and hence both spin configurations are treated at par. 
Thus, the constituents of the hidden-charm pentaquarks in the compact diquark model are~$[cq]_{\bar 3}$, 
$\bar c_{\bar 3}$, and~$[q^\prime q^{\prime\prime}]_{\bar 3}$, which make up a color singlet. 
%  $ P_c=\epsilon_{\alpha\,\beta\,\gamma}\, [cq]_\alpha \, \bar c_\beta\, [q^\prime q^{\prime \prime}]_\gamma$. 
However, it is still a three-body problem to solve and there are several dynamical possibilities 
to model their interconnections. We follow here the intuitive picture in which the heavier components 
form a nucleus and the lighter one is in an orbit around this nucleus, as it is energetically 
easier to excite light degrees of freedom. Among the three constituents, the light 
diquark~$[q^\prime q^{\prime\prime}]_{\bar 3}$ is the lightest. Moreover, its spin-parity quantum 
numbers are fixed due to heavy-quark symmetry constraints. So, we keep the light diquark as emerging 
intact in the $b$-baryon decays, and put it in the orbit for the $P$-wave states, with the heavier 
components, carrying a charm quark or a charm antiquark, acting as a nucleus of (an almost) static 
color source. This is shown in fig.~\ref{ali:fig-pentaquark-model-2} and the color and flavor 
structure in this approach can be written as 
 [$[\bar c_{\bar 3} (c q)_{\bar 3}]_3 (q^\prime q^{\prime\prime})_{\bar 3}$]. 
Such a description is closer to the doubly-charm tetraquarks, with the quark content 
[$(\bar c \bar c)_3 (q^\prime q^{\prime\prime})_{\bar 3}$], and doubly-charm antibaryons, 
having [$(\bar c \bar c)_3 \bar q_{\bar 3}$], as all three systems have two charm (anti)quarks 
at the center~\cite{Manohar:1992nd,Esposito:2013fma,Luo:2017eub,Karliner:2017qjm,Eichten:2017ffp,%
Francis:2016hui,Bicudo:2017szl,Junnarkar:2017sey,Mehen:2017nrh,Czarnecki:2017vco,Maiani:2019cwl,Maiani:2019lpu}. 
We note that a colored diquark-triquark template for pentaquarks 
$\bar\theta\, \delta \equiv \left [ \bar Q \left ( q_1 q_2 \right )_{\bar 3} \right ]_3 \left ( Q q_3 \right )_{\bar 3}$, 
through the color-triplet binding mechanism, was first used by Lebed~\cite{Lebed:2015tna}.
There is also the third possibility to introduce a model for the internal structure 
of a pentaquark~\cite{Ali:2016dkf,Ali:2017ebb} in which the light and heavy diquarks 
bind first into a tetraquark in a color-triplet~$\tt 3$ state which further interacts 
with the charm antiquark, 
[$[(c q)_{\bar 3} (q^\prime q^{\prime\prime})_{\bar 3}]_3 \bar c_{\bar 3}$]. 
Which of the three internal structures discussed  provides a realistic template for the bound charm-anticharm pentaquarks
remains to be seen.

In this paper we use the $L$ - $S$ scheme for describing the angular structure of pentaquarks. 
The spin of the heavy $[c q]$-diquark is $\mathbf{S}_{hd} = \mathbf{S}_c + \mathbf{S}_q$ with 
the allowed values $S_{hd} = 0,\, 1$, where $\mathbf{S}_c$ and $\mathbf{S}_q$ are the spins of 
the charm- and light $u$-, $d$- or $s$-quark, respectively. Combining the heavy diquark spin with 
the spin~$\mathbf{S}_{\bar c}$ of the charm antiquark, we get the spin of the doubly-heavy triquark, 
$\mathbf{S}_t = \mathbf{S}_{hd} + \mathbf{S}_{\bar c}$, having the values $S_t = 1/2,\, 3/2$. 
The spin of the light $[q^\prime q^{\prime\prime}]$-diquark is 
$\mathbf{S}_{ld} = \mathbf{S}_{q^\prime} + \mathbf{S}_{q^{\prime\prime}}$, where $S_{ld} = 0,\, 1$. 
The total spin of the pentaquark is the sum of the triquark and the light-diquark spins, 
$\mathbf{S} = \mathbf{S}_t + \mathbf{S}_{ld}$ with the set of values $S_t = 1/2,\, 3/2,\, 5/2$. 
In the case of the orbitally excited pentaquarks, the orbital angular momentum~$\mathbf{L}$
of the pentaquark is the sum of two terms:~$\mathbf{L}_t$, arising from the triquark system, 
and $\mathbf{L_{ld}}$, which determines the relative motion of the light diquark around the 
doubly heavy triquark. The total orbital angular momentum~$\mathbf{L}$ of the pentaquark 
is then obtained with the help of the momentum sum rules from quantum mechanics, i.\,e., 
it takes a value in the range $L = |L_t - L_{ld}|, \ldots, L_t + L_{ld}$. 
After~$\mathbf{L}$ and~$\mathbf{S}$ are constructed, the total angular momentum~$\mathbf{J}$ 
of the pentaquark follows from $\mathbf{J} = \mathbf{L} + \mathbf{S}$.
The parity of the states depends on the orbital angular momentum as $P = (-1)^{L+1}$. It is negative 
for the $S$-wave states and positive for the states with one unit of the orbital angular momentum.

Mass estimates are worked out in an effective Hamiltonian approach, which apart from the constituent 
quark and diquark masses, includes dominant spin-spin, spin-orbit, orbital, and tensor interactions. 
The effective Hamiltonian for the $S$-wave states in this formulation is the same as the one used  
earlier in the analysis of the 2015 LHCb data on pentaquarks~\cite{Ali:2016dkf,Ali:2017ebb}, but 
the $P$-wave (and higher orbital angular) states are packaged differently. In addition to this, 
we incorporate the tensor interaction in the effective Hamiltonian, which affect the $P$-states, 
and work out the detailed mass spectrum in this framework, using the parameters fixed from earlier 
studies of baryons and hidden-charm tetraquarks. The resulting pentaquark spectrum is very rich. 
However, imposing the spin conservation in the heavy-quark symmetry limit, we argue that only that 
part of the pentaquark spectrum is expected to be observed in $\Lambda_b$-decays, in which the 
pentaquarks have a ``good'' light diquark, i.\,e., having spin $S_{ld} = 0$. This reduces the number 
of anticipated pentaquark states in $\Lambda_b$-decays greatly. The same is also true for the decays 
of the $\Xi_b^{0,-}$-baryons which are from the same $SU (3)_F$-triplet as the $\Lambda_b$-baryon. 
The other weakly decaying $b$-baryon,~$\Omega_b$, in which the light diquark has the spin-1, belongs 
to the $SU (3)_F$-sextet. Its decays will yield pentaquarks from the $SU (3)_F$-decuplets, i.\,e., 
have a spin-1 light diquark ($S_{ld} = 1$) as a constituent~\cite{Li:2015gta,Ali:2016dkf}. 
Some of the unflavored pentaquark states are estimated to lie below the $J/\psi\, p$ 
(even below $\eta_c\, p$) threshold and they will decay via the $c \bar c$-annihilation 
into light hadrons or a lepton pair ($e^+ e^-$ and $\mu^+ \mu^-$), giving rise to states 
narrower than the observed peaks, which can also be searched in the $P_c^+ \to p\, \ell^+ \ell^-$ 
modes in $\Lambda_b \to P_c\, K^-$ decays. The same holds for the decays of some other pentaquark 
states, produced in the decays of the $\Xi_b$- and $\Omega_b$-baryons, in which case the corresponding 
pentaquarks with one, two, or three strange quarks will give rise to narrow $\Lambda^0\, \ell^+ \ell^-$, 
$\Xi^{\prime 0}\, \ell^+ \ell^-$, and $\Omega^-\, \ell^+ \ell^-$ final states, respectively.

We also note that a detailed spectrum of the tetraquark and pentaquark states has also been 
worked out in the dynamical diquark model~\cite{Giron:2019bcs}, using Born-Oppenheimer potentials 
calculated numerically on the lattice. Likewise, the $P_c^+$-states have been studied in different 
color-bindings, such as 
$\left \{ \left ( c \bar c \right ) \left ( \left ( u d \right ) u \right );\, L = 0 \right \}$~\cite{Weng:2019ynv}
and as $\left ( c \bar c \right ) \left ( u u d \right )$ hadrocharmonia~\cite{Eides:2019tgv}.

This paper is organized as follows. In section~\ref{sec:Triquark-Diquark}, we introduce 
the doubly-heavy triquark~--- light diquark model of pentaquark, and define the state 
vectors having a total angular momentum quantum number~$J$ by 
$\left | S_{hd}, S_t, L_t; S_{ld}, L_{ld}; S, L \right\rangle_J$, 
where~$S_{hd}$, $S_t$, and~$S_{ld}$ are the spins of the heavy diquark, doubly-heavy triquark 
and light diquark, respectively. The corresponding sets of the $S$-wave state vectors 
(with $L_t = L_{ld} = L = 0$) with the ``good'' ($S_{ld} = 0$) and ``bad'' ($S_{ld} = 1$) 
light diquarks are presented in tables~\ref{tab:S-wave-pentaquarks-good-ld} 
and~\ref{tab:S-wave-pentaquarks-bad-ld}, respectively. State vectors of the $P$-wave pentaquarks 
with the ground-state triquark ($L_t = 0$, $L_{ld} = L = 1$) and ``good'' light diquark with 
the spin $S_{ld} = 0$ (``bad'' light diquark with the spin $S_{ld} = 1$), are given in 
table~\ref{tab:P-wave-pentaquarks-good-ld} (table~\ref{tab:P-wave-pentaquarks-bad-ld}). 
In section~\ref{sec:Effective-Hamiltonian}, we give the effective Hamiltonian used to work 
out the pentaquark mass spectrum. In section~\ref{sec:Mass-Formulae}, analytical expressions 
for effective Hamiltonian matrix elements are presented taking into account the dominant 
spin-spin interactions in the heavy and light diquarks and in the hidden-charm triquark. 
For the $P$-wave states, additional contributions and mixings due to the orbital, spin-orbit 
and tensor interactions are included. In all the cases, we diagonalize the mass matrices and 
the analytical equations for all the $S$- and $P$-wave pentaquark masses are presented. 
Section~\ref{sec:mass-predictions} contains the values of various input parameters, 
with the constituent diquark masses given in table~\ref{tab:diquark-masses} and the 
spin-spin couplings,~$\mathcal{K}_{\bar Q Q^\prime}$ and~$(\mathcal{K}_{Q Q^\prime})_{\bar 3}$, 
extracted from spectra of mesons and baryons, are given in table~\ref{tab:spin-spin-couplings}.
Our predictions for the unflavored pentaquark masses having the quark flavors 
$(\bar c [cq] [q^\prime q^{\prime \prime}])$, with~$q$, $q^\prime$, and~$q^{\prime \prime}$ 
being $u$- and $d$-quarks, assuming the isospin symmetry, are given 
in table~\ref{tab:masses-predictions-cbar-cq-qq}. Masses of the hidden-charm strange pentaquarks are 
shown in tables~\ref{tab:masses-predictions-cbar-cs-qq}~---~\ref{tab:masses-predictions-cbar-cs-ss}. 
In this section, a comparison of our approach with the dynamical diquark model~\cite{Giron:2019bcs} 
is also presented. We continue with a discussion of dominant decay channels of ground-state pentaquarks 
containing a scalar light diquark in section~\ref{sec:pentaquark-widths} and conclude 
in section~\ref{sec:conclusions}. In appendix~\ref{sec:spin-spin-corrections}, 
corrections due to the spin-spin interactions between the constituents of the doubly-heavy triquark 
and the light diquark are detailed; in appendix~\ref{sec:mass-derivations}, cumbersome mass 
derivations for the $P$-wave pentaquark states are presented, and in appendix~\ref{sec:chi2-Omega-c} 
the details and results of the $\chi^2$-analysis of the orbitally-excited $\Omega_c^*$-baryons 
are shown.

\section{Doubly-heavy triquark~--- light diquark model of pentaquark}
\label{sec:Triquark-Diquark}
\begin{figure}
\centerline{\includegraphics[scale=0.4]{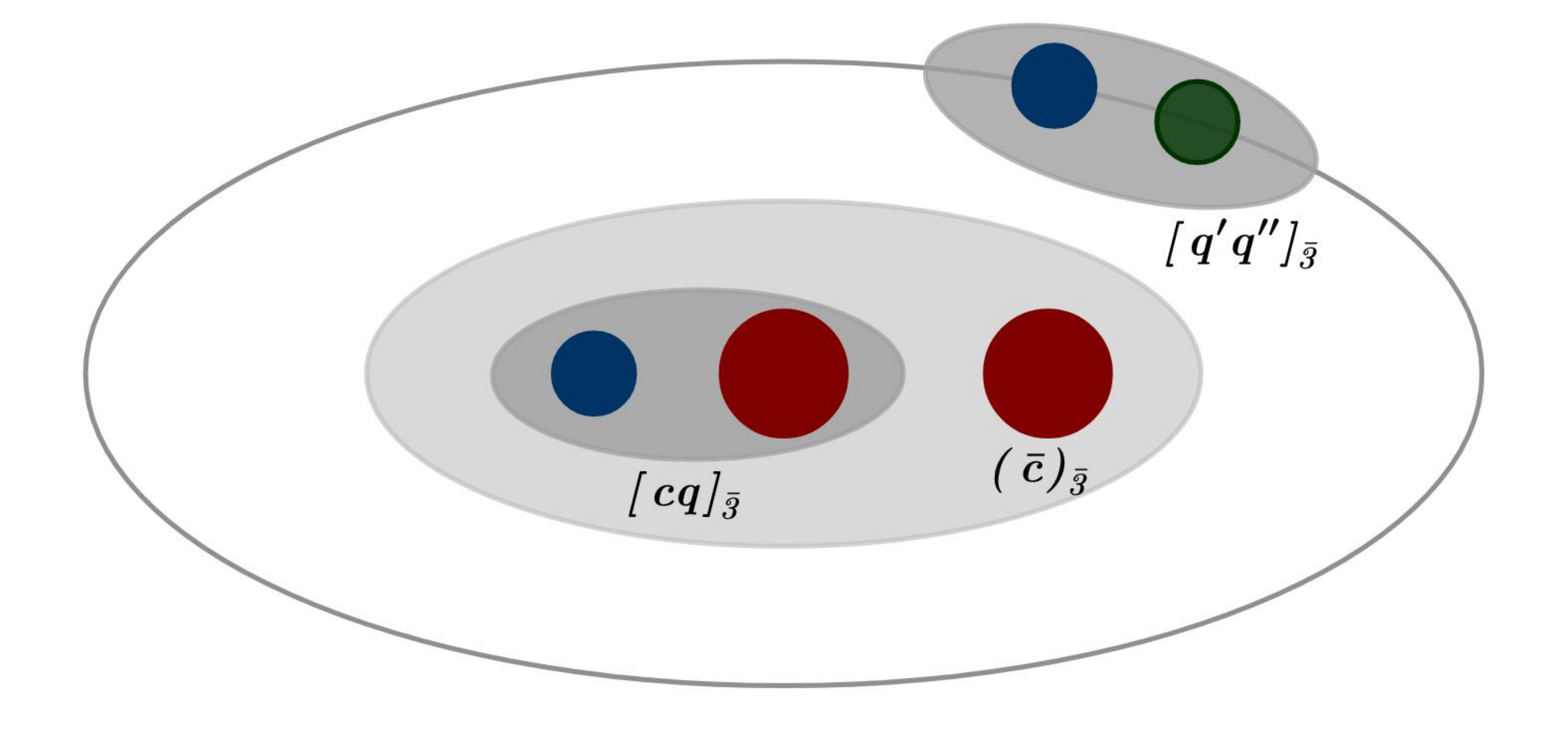}}
\vspace*{-4mm}
\caption{
% \footnotesize 
A picture of pentaquarks in the diquark model involving the heavy diquark 
$[c q]_{\bar 3}$ and charm antiquark $(\bar c)_{\bar 3}$, which form 
the triquark system, which combines with the light diquark $[q^\prime q^{\prime\prime}]_{\bar 3}$ 
to make a color singlet pentaquark. The subscripts indicate that all three 
constituents are color anti-triplets;~$q$, $q^\prime$, and~$q^{\prime\prime}$ 
are light quarks each of which can be~$u$-, $d$-, or $s$-quark.
}
\label{ali:fig-pentaquark-model-2}
\end{figure}

In the pentaquark picture shown in fig.~\ref{ali:fig-pentaquark-model-2}, 
there are two color flux tubes, with the first stretched between the charm 
diquark and the charm antiquark from which the diquark is in the color-antitriplet 
state~$\bar 3$. With the antiquark being also a color antitriplet state~$\bar 3$, 
their product is decomposed into two irreducible representations, 
$\bar 3 \times \bar 3 = 3 + \bar 6$, from which the color triplet~$3$ is kept. 
This color-triplet triquark makes a color-singlet bound state with the light diquark 
through the second flux tube in much the same way as the quark and diquark bind 
in an ordinary baryon. This approach differs from the one~\cite{Ali:2016dkf,Ali:2017ebb} 
used previously to get the mass spectrum of the $S$- and $P$-wave pentaquark states with 
the spin-parities $J^P = 1/2^\pm$, $3/2^-$, and~$5/2^\pm$, where 
the two diquarks are assumed to be organized first into a color-triplet tetraquark state 
which afterwards interacts with the charm antiquark.

The double-charm triquark system $[ [cq] \bar c ]_3$ is expected to be dynamically 
similar to the double-charm antidiquark, $\{ \bar c \bar c \}_3$, as far as their color 
and masses are concerned. The role of the latter, forming an almost static source, 
has been discussed at great length in a number of papers for the doubly-heavy anti-baryons, 
such as $\bar\Xi_{cc} (= \bar c \bar c \bar q)$, and tetraquarks $T (\bar c \bar c q q^\prime)$~\cite{% 
Manohar:1992nd,Karliner:2017qjm,Eichten:2017ffp,Francis:2016hui,Bicudo:2017szl,Junnarkar:2017sey,Mehen:2017nrh}. 
The doubly-heavy triquark and the doubly-heavy diquark also differ in that the former has 
a light quark and the latter none. 
However, under the assumption that the diquark $[cq]_{\bar 3}$ is bound, and the dynamics 
is essentially determined by the color configuration and masses, the two are expected 
not to deviate from each other, at least as a first approximation.
With this in mind, we first detail the doubly-heavy triquark~--- light diquark picture, 
modify the effective Hamiltonian for the $S$-wave pentaquarks~\cite{Ali:2016dkf,Ali:2017ebb} 
by keeping the most relevant terms for the mass determination, and then extend it for 
the $P$-states by including the orbital, spin-orbit and tensor interactions between the 
hidden-charm triquark and light diquark. 

The effective Hamiltonian for the ground-state pentaquarks is described in terms of two 
constituent diquark masses of the heavy diquark~$m_{[cq]} \equiv m_{hd}$ and of the light 
one~$m_{[q^{\prime}q^{\prime\prime}]} \equiv m_{ld}$, the spin-spin interactions between 
the quarks in each diquark shell, and spin-spin interactions between the diquarks. 
To these are added the constituent mass~$m_c$ of the charm antiquark and its spin-spin 
interactions with each of the diquarks.

Using the quantum numbers introduced above, we specify the complete orthogonal set 
of basis vectors for the hidden-charm pentaquark states. As already stated, we define 
the state vectors having a total angular momentum quantum number~$J$ by 
$$
\left | S_{hd}, S_t, L_t; S_{ld}, L_{ld}; S, L \right\rangle_J . 
$$
%
%where~$S_{hd}$, $S_t$, and~$S_{ld}$ are the spins of the heavy diquark, 
%doubly-heavy triquark and light diquark, respectively. 
The corresponding sets 
of the $S$-wave state vectors (with $L_t = L_{ld} = L = 0$) with the ``good'' 
($S_{ld} = 0$) and ``bad'' ($S_{ld} = 1$) light diquark are presented in 
tables~\ref{tab:S-wave-pentaquarks-good-ld} and~\ref{tab:S-wave-pentaquarks-bad-ld}, 
respectively.
\begin{table}[tb] 
\begin{minipage}[ct]{0.45\textwidth}
\begin{center}
\begin{tabular}{|l|l|} 
\hline\hline 
  $J^P$ & $\left | S_{hd}, S_t, L_t; S_{ld}, L_{ld};   S, L \right\rangle_J$  \\ 
\hline\hline
$1/2^-$ &      $\left | 0, 1/2,   0;   0,   0; 1/2, 0 \right\rangle_{1/2}$ \\ \hline 
$1/2^-$ &      $\left | 1, 1/2,   0;   0,   0; 1/2, 0 \right\rangle_{1/2}$ \\
$3/2^-$ &      $\left | 1, 3/2,   0;   0,   0; 3/2, 0 \right\rangle_{3/2}$ \\  
\hline\hline 
\end{tabular}
\end{center}
\caption{
Spin-parity~$J^P$ and state vectors of the $S$-wave pentaquarks
containing the ``good'' light diquark with the spin $S_{ld} = 0$.
The horizontal line demarcates the spin~$S_{hd}$ of the heavy diquark.
}
\label{tab:S-wave-pentaquarks-good-ld}
\end{minipage}
\qquad\qquad
\begin{minipage}[ct]{0.45\textwidth}
\begin{center}
\begin{tabular}{|l|l|} 
\hline\hline 
  $J^P$ & $\left | S_{hd}, S_t, L_t; S_{ld}, L_{ld};   S, L \right\rangle_J$  \\ 
\hline\hline
$1/2^-$ &      $\left | 0, 1/2,   0;   1,   0; 1/2, 0 \right\rangle_{1/2}$ \\
$3/2^-$ &      $\left | 0, 1/2,   0;   1,   0; 3/2, 0 \right\rangle_{3/2}$ \\ \hline 
$1/2^-$ &      $\left | 1, 1/2,   0;   1,   0; 1/2, 0 \right\rangle_{1/2}$ \\
$3/2^-$ &      $\left | 1, 1/2,   0;   1,   0; 3/2, 0 \right\rangle_{3/2}$ \\
$1/2^-$ &      $\left | 1, 3/2,   0;   1,   0; 1/2, 0 \right\rangle_{1/2}$ \\
$3/2^-$ &      $\left | 1, 3/2,   0;   1,   0; 3/2, 0 \right\rangle_{3/2}$ \\
$5/2^-$ &      $\left | 1, 3/2,   0;   1,   0; 5/2, 0 \right\rangle_{5/2}$ \\
\hline\hline 
\end{tabular}
\end{center}
\caption{
Spin-parity~$J^P$ and state vectors of the $S$-wave pentaquarks 
containing the ``bad'' light diquark with the spin $S_{ld} = 1$.
The horizontal line demarcates the spin~$S_{hd}$ of the heavy diquark.
}
\label{tab:S-wave-pentaquarks-bad-ld}
\end{minipage}
\end{table}
Among the overall ten states with a fixed light-quark flavor content, there are five 
states with $J^P = 1/2^-$, four states with $J^P = 3/2^-$ and one with $J^P = 5/2^-$.  
In view of the propensity of these states, we note that in the heavy-quark symmetry limit,
the production of the states with the quantum numbers shown in table~\ref{tab:S-wave-pentaquarks-bad-ld} 
is forbidden in $\Lambda_b$- or $\Xi_b$-decays. Nevertheless, they are present 
in the pentaquark mass spectrum, and they are reachable in other processes.

For the orbitally excited states, one needs to specify which part of the pentaquark 
wave function is excited. In the triquark-diquark template used here, and shown in 
fig.~\ref{ali:fig-pentaquark-model-2}, the heavy triquark state consists of the charm 
diquark and charm antiquark. Since both are heavy, the triquark is an (almost) static 
system and, hence, $L_t = 0$ is its most probable orbital quantum number. 
Thus, the orbital excitation is generated by the light diquark (i.\,e., $L = L_{ld}$). 
With this, we list the lowest-lying orbitally excited states ($L = 1$) with the ``good'' ($S_{ld} = 0$) 
and the ``bad'' ($S_{ld} = 1$) light diquark in tables~\ref{tab:P-wave-pentaquarks-good-ld}
and~\ref{tab:P-wave-pentaquarks-bad-ld}, respectively. In total, there are 25 pentaquark 
states with a fixed light-quark flavor content which are divided into nine states with 
$J^P = 1/2^+$, ten with $J^P = 3/2^+$, five with $J^P = 5/2^+$ and one with $J^P = 7/2^+$.

\begin{table}[tb] 
\begin{minipage}[ct]{0.45\textwidth}
\begin{center}
\begin{tabular}{|l|l|} 
\hline\hline 
  $J^P$ & $\left | S_{hd}, S_t, L_t; S_{ld}, L_{ld};  S, L \right\rangle_J$    \\ 
\hline\hline 
$1/2^+$ &      $\left | 0, 1/2,   0;   0,   1; 1/2, 1 \right\rangle_{1/2}$ \\ 
$3/2^+$ &      $\left | 0, 1/2,   0;   0,   1; 1/2, 1 \right\rangle_{3/2}$ \\ \hline 
$1/2^+$ &      $\left | 1, 1/2,   0;   0,   1; 1/2, 1 \right\rangle_{1/2}$ \\
$3/2^+$ &      $\left | 1, 1/2,   0;   0,   1; 1/2, 1 \right\rangle_{3/2}$ \\
$1/2^+$ &      $\left | 1, 3/2,   0;   0,   1; 3/2, 1 \right\rangle_{1/2}$ \\
$3/2^+$ &      $\left | 1, 3/2,   0;   0,   1; 3/2, 1 \right\rangle_{3/2}$ \\
$5/2^+$ &      $\left | 1, 3/2,   0;   0,   1; 3/2, 1 \right\rangle_{5/2}$ \\ 
\hline\hline 
\end{tabular}
\end{center}
\caption{
Spin-parity~$J^P$ and state vectors of the $P$-wave pentaquarks with the ground-state 
triquark ($L_t = 0$) and ``good'' light diquark with the spin $S_{ld} = 0$.
The horizontal line demarcates the spin~$S_{hd}$ of the heavy diquark.
}
\label{tab:P-wave-pentaquarks-good-ld}
\end{minipage}
\qquad\qquad 
\begin{minipage}[ct]{0.45\textwidth}
\begin{center}
\begin{tabular}{|l|l|} 
\hline\hline 
  $J^P$ & $\left | S_{hd}, S_t, L_t; S_{ld}, L_{ld};  S, L \right\rangle_J$    \\ 
\hline\hline 
$1/2^+$ &      $\left | 0, 1/2,   0;   1,   1; 1/2, 1 \right\rangle_{1/2}$ \\
$3/2^+$ &      $\left | 0, 1/2,   0;   1,   1; 1/2, 1 \right\rangle_{3/2}$ \\
$1/2^+$ &      $\left | 0, 1/2,   0;   1,   1; 3/2, 1 \right\rangle_{1/2}$ \\
$3/2^+$ &      $\left | 0, 1/2,   0;   1,   1; 3/2, 1 \right\rangle_{3/2}$ \\
$5/2^+$ &      $\left | 0, 1/2,   0;   1,   1; 3/2, 1 \right\rangle_{5/2}$ \\ \hline 
$1/2^+$ &      $\left | 1, 1/2,   0;   1,   1; 1/2, 1 \right\rangle_{1/2}$ \\
$3/2^+$ &      $\left | 1, 1/2,   0;   1,   1; 1/2, 1 \right\rangle_{3/2}$ \\
$1/2^+$ &      $\left | 1, 1/2,   0;   1,   1; 3/2, 1 \right\rangle_{1/2}$ \\
$3/2^+$ &      $\left | 1, 1/2,   0;   1,   1; 3/2, 1 \right\rangle_{3/2}$ \\
$5/2^+$ &      $\left | 1, 1/2,   0;   1,   1; 3/2, 1 \right\rangle_{5/2}$ \\
$1/2^+$ &      $\left | 1, 3/2,   0;   1,   1; 1/2, 1 \right\rangle_{1/2}$ \\
$3/2^+$ &      $\left | 1, 3/2,   0;   1,   1; 1/2, 1 \right\rangle_{3/2}$ \\
$1/2^+$ &      $\left | 1, 3/2,   0;   1,   1; 3/2, 1 \right\rangle_{1/2}$ \\
$3/2^+$ &      $\left | 1, 3/2,   0;   1,   1; 3/2, 1 \right\rangle_{3/2}$ \\
$5/2^+$ &      $\left | 1, 3/2,   0;   1,   1; 3/2, 1 \right\rangle_{5/2}$ \\
$3/2^+$ &      $\left | 1, 3/2,   0;   1,   1; 5/2, 1 \right\rangle_{3/2}$ \\
$5/2^+$ &      $\left | 1, 3/2,   0;   1,   1; 5/2, 1 \right\rangle_{5/2}$ \\
$7/2^+$ &      $\left | 1, 3/2,   0;   1,   1; 5/2, 1 \right\rangle_{7/2}$ \\
\hline\hline 
\end{tabular}
\end{center}
\caption{
Spin-parity~$J^P$ and state vectors of the $P$-wave pentaquarks with the ground-state 
triquark ($L_t = 0$) and ``bad'' light diquark with the spin $S_{ld} = 1$.
The horizontal line demarcates the spin~$S_{hd}$ of the heavy diquark.
}
\label{tab:P-wave-pentaquarks-bad-ld}
\end{minipage}
\end{table}

The pentaquark spectrum emerging from the underlying diquark picture is very rich. 
Contrasting it with the current experimental situation, with only three known hidden-charm 
pentaquark states~$P_c (4312)^\pm$, $P_c(4440)^\pm$, and~$P_c (4457)^\pm$ observed 
in $\Lambda_b$-baryon decays, some selection rules have to be applied to restrict the number 
of observable pentaquarks. Since in the heavy-quark symmetry limit, the spin of the light quarks 
is conserved in heavy-baryon decays, and $S_{ld} = 0$ in $\Lambda_b$-baryon, the states with 
the light-diquark spin $S_{ld} = 1$ are suppressed. This means that we need to consider only 
the states in tables~\ref{tab:S-wave-pentaquarks-good-ld} and~\ref{tab:P-wave-pentaquarks-good-ld}, 
for the $S$- and $P$-wave pentaquarks, respectively. This reduces the number of states 
to three ($S$-wave) and seven ($P$-wave). The states with $S_{ld} = 1$ can, however, be produced 
in $\Omega_b$-baryon decays or in prompt production processes.

A related issue is the presence of two identical light quarks in the Fock space of a hidden-charm pentaquark. 
Among the light quark~$q$ present in the triquark, and the light quarks~$q^\prime$ and~$q^{\prime\prime}$ 
in the diquark, there are two identical quarks. Treated as free quarks, they are subject 
to Pauli's exclusion principle, affecting the resulting spectrum of hadrons. This, for example, would 
be the case in the hadroquarkonium picture of the hidden-charm pentaquarks. However, if the $q^\prime q^{\prime\prime}$-pair 
is bound in a diquark, which is a boson, then this exclusion does not apply so long as
the diquark is not broken into its fermionic constituents.  We illustrate this situation by
focusing on the discovery mode $\Lambda_b \to P_c^+ K^-$. This decay is induced at the quark level by the  
$b \left [ u d \right ] \to c\, \bar c\, s \left [ u d \right ] + \left ( u \bar u \right )_{\rm vac}$
transition. In this, the $\left [ u d \right ]$-diquark, which has well-defined quantum numbers (spin 
$S_{[ud]} = 0$ and isospin $I_{[ud]} = 0$), is a boson. In the heavy-quark symmetry limit, assumed 
in our paper, the light degrees of freedom remain conserved, and the light $\left [ u d \right ]$-diquark  
emerges in tact from the decay. Indeed, this imposes restrictions on the quantum numbers of final states, 
as we have emphasized earlier. If this symmetry is broken, say, by $1/m_b$ terms, then the $u$- 
and $d$-quarks in the $\left [ u d \right ]$-diquark are no longer bound, and in the example here, 
one has two identical quarks, the $u$-quark produced from the vacuum excitation and the $u$-quark 
from the break-up of the spectator diquark, and the Pauli's exclusion principle would apply. 
In the heavy-quark symmetry limit, we have the spin-$1/2$ $u$-quark from the vacuum excitation, 
and the spin-0 $\left [ u d \right ]$-diquark, and they are distinctly different building blocks. 
In this case, no exclusion restrictions from identical quarks apply, which is what we have used 
in building the pentaquark spectrum.

\section{Effective Hamiltonian for pentaquark spectrum}
\label{sec:Effective-Hamiltonian}
We calculate the mass spectrum of pentaquark under the assumption that their 
underlying structure is given by $\bar c$, $[cq]$, and $[q^\prime q^{\prime \prime}]$, 
with~$q$, $q^\prime$, and~$q^{\prime \prime}$ being any of the light $u$-, $d$-, 
and $s$-quarks. We also assume the isospin symmetry among the states. For this, 
we extend the effective Hamiltonian proposed for the tetraquark spectroscopy 
by Maiani et al.~\cite{Maiani:2014aja}. The effective Hamiltonian 
for the $S$-wave pentaquark mass spectrum can be written as follows:  
\begin{equation}
H^{(L = 0)} = H_t + H_{ld} . 
\label{eq:Hamiltonian-S-wave} 
\end{equation}
The first term in the Hamiltonian~(\ref{eq:Hamiltonian-S-wave})  
is related with the color triquark: 
\begin{equation}
H_t = m_c + m_{hd} + 
2 \, (\mathcal{K}_{cq})_{\bar 3} \, (\mathbf{S}_c \cdot \mathbf{S}_q) + 
2 \, \mathcal{K}_{\bar c q} \left ( \mathbf{S}_{\bar c} \cdot \mathbf{S}_q \right ) +    
2 \, \mathcal{K}_{\bar c c} \left ( \mathbf{S}_{\bar c} \cdot \mathbf{S}_c \right ) ,
\label{eq:H-triquark}
\end{equation}
where~$m_c$ and~$m_{hd}$ are the constituent masses of the charm antiquark and charm 
diquark, respectively. The last three terms describe the spin-spin interactions in the 
charm diquark and between the diquark constituents and the charm antiquark. Among 
the three spin-spin couplings~$(\mathcal{K}_{cq})_{\bar 3}$, $\mathcal{K}_{\bar c q}$, 
and~$\mathcal{K}_{\bar c c}$, the spin-spin interaction inside the diquark 
$(\mathcal{K}_{cq})_{\bar 3}$ is argued to be the dominant one~\cite{Maiani:2014aja}.     

The second term~$H_{ld}$ in the Hamiltonian~(\ref{eq:Hamiltonian-S-wave}) contains 
the operators responsible for the spin-spin interaction in the light diquark and 
its interaction with the triquark: 
\begin{equation}
% \begin{array}{rclrclrcl} 
H_{ld} = m_{ld} + 
2 \, (\mathcal{K}_{q^\prime q^{\prime \prime}})_{\bar 3} \, 
(\mathbf{S}_{q^\prime} \cdot \mathbf{S}_{q^{\prime\prime}}) + 
H_{SS}^{t-ld} , 
\label{eq:H-tri-diquark-interaction}
% \nonumber \\ 
\end{equation}
where $m_{ld}$ is the constituent mass of the light diquark, consisting 
of the light quarks~$q^\prime$ and~$q^{\prime\prime}$, and the contribution 
of the spin-spin interaction in this diquark to the pentaquark mass is 
determined by the coupling~$(\mathcal{K}_{q^\prime q^{\prime\prime}})_{\bar 3}$. 
The last term in~$H_{ld}$ is responsible for all the possible spin-spin 
interactions between the constituents of the light diquark and doubly-heavy triquark: 
\begin{eqnarray}
H_{SS}^{t-ld} & = & 
2 \, (\tilde{\mathcal{K}}_{c q^\prime})_{\bar 3} \, (\mathbf{S}_c \cdot \mathbf{S}_{q^\prime}) + 
2 \, (\tilde{\mathcal{K}}_{q q^\prime})_{\bar 3} \, (\mathbf{S}_q \cdot \mathbf{S}_{q^\prime}) + 
2 \, \tilde{\mathcal{K}}_{\bar c q^\prime} \, (\mathbf{S}_{\bar c} \cdot \mathbf{S}_{q^\prime})  
\nonumber \\ 
& + & 
2 \, (\tilde{\mathcal{K}}_{c q^{\prime\prime}})_{\bar 3} \, (\mathbf{S}_c \cdot \mathbf{S}_{q^{\prime\prime}}) + 
2 \, (\tilde{\mathcal{K}}_{q q^{\prime\prime}})_{\bar 3} \, (\mathbf{S}_q \cdot \mathbf{S}_{q^{\prime\prime}}) + 
2 \, \tilde{\mathcal{K}}_{\bar c q^{\prime\prime}} \, (\mathbf{S}_{\bar c} \cdot \mathbf{S}_{q^{\prime\prime}}) . 
\label{eq:H-SS-t-d}
\end{eqnarray}
Here, the coefficients with the tilde differ from the ones introduced above as the 
former show the strengths of the spin-spin interactions inside the compact objects 
like diquarks and triquark while the later ones determine the strengths between the 
constituents of two objects, the heavy triquark and light diquark, and are strongly 
suppressed. The rationale of this is specific to the diquark model, in which the 
hadronic size of the diquarks is much smaller or compared to the overall size of the
multiquark hadrons. Thus, only local spin-spin interactions (within a diquark or triquark) 
are allowed. This then accounts for all possible spin-spin interactions and completes 
the content of the effective Hamiltonian~(\ref{eq:Hamiltonian-S-wave}) for the masses 
of the ground-state pentaquarks. A comment is in order: Despite the fact that this picture, 
which is physically more motivated, differs form the one used in~\cite{Ali:2016dkf,Ali:2017ebb}, 
for the analysis and predictions of the ground-state pentaquarks, actually all the terms 
in the effective Hamiltonian~(\ref{eq:Hamiltonian-S-wave}) are the same.

The general form of the effective Hamiltonian for the orbitally-excited pentaquark 
mass spectrum can be written as follows:  
\begin{equation}
H = H^{(L = 0)} + H_L + H_T . 
\label{eq:Hamiltonian-general} 
\end{equation}
In addition to the spin-spin interactions introduced for the ground-state pentaquarks
described above, the extended effective Hamiltonian~(\ref{eq:Hamiltonian-general})  
includes terms explicitly dependent on the internal orbital angular momentum~$L_t$ 
of the hidden-charm triquark and orbital momentum~$L_{ld}$ of the light diquark 
relative to the triquark system. The corresponding Hamiltonian is called~$H_L$ 
in~(\ref{eq:Hamiltonian-general}). The terms relevant for the tensor interactions 
in each subsystem specified above are subsumed in~$H_T$ in~(\ref{eq:Hamiltonian-general}). 
As already stated, we assume that the triquark state is an $S$-wave (i.\,e., $L_t = 0$). 
Thus, $L = L_{ld}$. Inclusion of the triquark orbital angular momentum~$L_t$ will be 
commented later on. The terms in~$H_L$ which contain the orbital angular momentum 
operator~$\mathbf{L}$ are as follows: 
\begin{equation}
H_L = 2 A_t    \left ( \mathbf{S}_t    \cdot \mathbf{L} \right ) +  
      2 A_{ld} \left ( \mathbf{S}_{ld} \cdot \mathbf{L} \right ) + 
      \frac{1}{2} \, B \, \mathbf{L}^2 ,  
\label{eq:Hamiltonian-L-pentaquark} 
\end{equation}
where quantities~$A_t$, $A_{ld}$, and~$B$ parametrize the strengths of the triquark 
spin-orbit, light-diquark spin-orbit and orbital momentum couplings, respectively.
The last term in~(\ref{eq:Hamiltonian-general}) represents the tensor 
interaction among the heavy triquark and light diquark: 
\begin{equation}
H_T =  
b \left [ 
3 \, \frac{({\bf S}_t \cdot {\bf R}) \, ({\bf S}_{ld} \cdot {\bf R})}{R^2} 
   - ({\bf S}_t \cdot {\bf S}_{ld})  
\right ] , 
\label{eq:Hamiltonian-tensor-pentaquark}
\end{equation}
where ${\bf R}$ determines the position of the light diquark relative 
to the heavy triquark and~$b$ is the strength of the tensor interaction.  

The inclusion of a possible excitation in the triquark system ($L_t \neq 0$) 
results into additional terms in the effective Hamiltonian: 
\begin{equation}
\Delta H = \Delta H_L + \Delta H_T . 
\label{eq:Hamiltonian-triquark-addon} 
\end{equation}
The structure of~$\Delta H_L$ which includes the angular momentum 
operator~$\mathbf{L}_t$ in the triquark system is similar 
to~(\ref{eq:Hamiltonian-L-pentaquark}): 
\begin{equation}
\Delta H_L = 2 A_{\bar c} \left ( \mathbf{S}_{\bar c} \cdot \mathbf{L}_t \right ) +  
             2 A_{hd} \left ( \mathbf{S}_{hd} \cdot \mathbf{L}_t \right ) + 
             \frac{1}{2} \, B_t \, \mathbf{L}_t^2 ,  
\label{eq:Hamiltonian-Delta-L-pentaquark} 
\end{equation}
where quantities~$A_{\bar c}$, $A_{hd}$, and~$B_t$ parametrize the strengths 
of the charm antiquark and diquark spin-orbit and triquark orbital momentum 
couplings, respectively. As the triquark contains charm quark and antiquark, 
one expects a suppression of~$A_{\bar c}$, $A_{hd}$, and~$B_t$ by the 
$c$-quark mass in comparison with the parameters~$A_t$, $A_{ld}$, and~$B$ 
entering~(\ref{eq:Hamiltonian-L-pentaquark}). In the numerical calculations, 
we set them to zero.

The second term in~(\ref{eq:Hamiltonian-triquark-addon}) represents the 
tensor interaction among the charm diquark and antiquark in the triquark: 
\begin{equation}
\Delta H_T =  
b_t \left [ 
3 \, \frac{({\bf S}_{\bar c} \cdot {\bf r}) \, ({\bf S}_{hd} \cdot {\bf r})}{r^2} 
   - ({\bf S}_{\bar c} \cdot {\bf S}_{hd})  
\right ] , 
\label{eq:Hamiltonian-tensor-triquark}
\end{equation}
where ${\bf r}$ determines the position of the charm diquark relative 
to the charm antiquark and~$b_t$ is the strength of this interaction.  
In the phenomenological analysis of the hidden-charm orbitally excited 
$Y$-tetraquarks, it was shown~\cite{Ali:2017wsf} that a similar strength 
entering the tensor interaction between the charm diquark and antidiquark 
is consistent with zero. Following~\cite{Ali:2017wsf}, it is reasonable 
to assume that~$b_t$ in~(\ref{eq:Hamiltonian-tensor-triquark}) is also 
small, and hence we neglect this interaction in the analysis reported 
below by setting $\Delta H = 0$.

Returning to the discussion of the tensor contribution, the expression in 
eq.~(\ref{eq:Hamiltonian-tensor-pentaquark}) is of the following general form: 
\begin{eqnarray}
Q ({\bf S}_1, {\bf S}_2) & = & 
3 \left ( {\bf S}_1 \cdot {\bf n} \right ) 
  \left ( {\bf S}_2 \cdot {\bf n} \right ) - 
  \left ( {\bf S}_1 \cdot {\bf S}_2 \right ) = 
% \nonumber \\ 
% & = & 
3 S_{1 i} S_{2 j} \, N_{ij} ,  
\label{eq:tensor-general}
\end{eqnarray}
where~${\bf S}_1$ and~${\bf S}_2$ are the spins of two particles and  
${\bf n} = {\bf R}/R$ is the unit vector in the direction of~${\bf R}$.
%  from one particle to the other, i.\,e. ${\bf r}_{AB} = {\bf r}_B - {\bf r}_A$. 
%
This notation was used by us previously~\cite{Ali:2017wsf} in the analysis 
of the orbitally excited ($L = 1$) $\Omega_c$-baryons and $Y$-tetraquarks. 
In both cases~${\bf S}_1$ was identified with the spin of the diquark,~${\bf S}_Q$, 
and~${\bf S}_2$ is either the spin of the charm quark,~${\bf S}_c$, in 
the $\Omega_c$-baryon, or the spin of the antidiquark,~${\bf S}_{\bar Q}$, 
in the $Y$-tetraquark. 
  
The scalar operator~$Q ({\bf S}_1, {\bf S}_2)$ in eq.~(\ref{eq:tensor-general}) 
is expressed as the convolution of spins with the tensor operator: 
\begin{equation}
N_{ij} = n_{i} n_{j} - \frac{1}{3} \, \delta_{ij} .
\label{eq:N-def}
\end{equation}
For further applications, we need the matrix elements of the operator~$N_{ij}$ 
between the states with the same fixed value~$L$ of the orbital angular momentum 
operator~${\bf L}$ which can be obtained with the help of the identity from 
Landau and Lifshitz~\cite{Landau:1977}:
\begin{equation}
\langle N_{ij} \rangle = - \frac{1}{(2 L - 1) (2 L + 3)} \left [ 
L_i L_j + L_j L_i - \frac{2}{3} \, \delta_{ij} \, L \left ( L + 1 \right ) 
\right ] .     
\label{eq:LL-identity}
\end{equation}
It is obvious that this matrix element is trivial when the two-particle 
quantum state is in the $S$-wave, $L = 0$.   

In terms of~$Q ({\bf S}_1, {\bf S}_2)$, eq.~(\ref{eq:Hamiltonian-tensor-pentaquark}) 
for $L = 1$ can be expressed as follows 
\begin{equation}
H_T =  b \, \langle Q ({\bf S}_t, {\bf S}_{ld}) \rangle = 
- \frac{3}{5} \, b \left\langle 
\left ( {\bf L} \cdot {\bf S}_t \right )  
\left ( {\bf L} \cdot {\bf S}_{ld} \right ) + 
\left ( {\bf L} \cdot {\bf S}_{ld} \right )  
\left ( {\bf L} \cdot {\bf S}_t \right ) - 
\frac{4}{3} \left ( {\bf S}_t \cdot {\bf S}_{ld} \right ) 
\right\rangle .  
\label{eq:tensor-Q-operators}
\end{equation}
and similarly $\Delta H_T = b_t \, Q ({\bf S}_{hd}, {\bf S}_{\bar c})$, if $b_t \neq 0$.  
In eq.~(\ref{eq:tensor-Q-operators}), the mixed-spin $Q$-operator can be reexpressed 
in terms of the single spin ones:  % ~(\ref{eq:Q-L-1}):  
\begin{equation}
Q ({\bf S}_t, {\bf S}_{ld}) = Q ({\bf S}_{ld}, {\bf S}_t) = \frac{1}{2} \left [  
Q ({\bf S}, {\bf S}) - Q ({\bf S}_t, {\bf S}_t) - Q ({\bf S}_{ld}, {\bf S}_{ld}) 
\right ] , 
\label{eq:Q-relation}
\end{equation}
where ${\bf S} = {\bf S}_t + {\bf S}_{ld}$. To calculate the single-spin $Q$-operators 
inside the brackets, it is necessary to account for appropriate commutation relations of 
the components of~${\bf L}$ and~${\bf S}_X$, where ${\bf S}_X = {\bf S}_t$, ${\bf S}_{ld}$, 
and~${\bf S}$ are the spins of the triquark, light-diquark, and the total spin of the 
pentaquark, respectively. Setting $L = 1$, one finds easily~\cite{Karliner:2017kfm}:
\begin{equation}
\left\langle Q ({\bf S}_X, {\bf S}_X) \right\rangle = 
- \frac{3}{5} \, \left\langle 
2 \left ( {\bf L} \cdot {\bf S}_X \right )^2 + 
\left ( {\bf L} \cdot {\bf S}_X \right ) - 
\frac{4}{3} \left ( {\bf S}_X \cdot {\bf S}_X \right ) 
\right\rangle . 
\label{eq:Q-L-1}
\end{equation}

The matrix elements can be computed directly by applying the operators 
$\left ( {\bf L} \cdot {\bf S}_X \right )$ to the products of states 
corresponding to the individual spins and angular momenta. 
More effectively, one can use Wigner $6j$-symbols, as is customary for analogous 
cases in atomic and nuclear physics. An explicit example of their application 
to the analysis of the orbitally-excited $\Omega_c$-baryons and $Y$-tetraquarks 
is presented in the appendix of ref.~\cite{Ali:2017wsf}. 
So, to get the mass predictions one needs to specify a concrete basis of the 
orbitally-excited pentaquark states in which the mass predictions can be done.

\section{Mass formulae for pentaquark spectrum}
\label{sec:Mass-Formulae} 

\subsection[$S$-wave pentaquarks]{\boldmath $S$-wave pentaquarks}
\label{sec:Mass-Formulae-S-wave} 

With the basis vectors of the pentaquark states chosen, one can derive analytical 
expressions for calculating the pentaquark spectrum. They are the matrix elements 
of the effective Hamiltonian presented above in sec.~\ref{sec:Effective-Hamiltonian}. 
Note that this is simpler for the Model~II by Maiani et al~\cite{Maiani:2014aja}, 
but becomes more involved in the Model~I~\cite{Maiani:2004vq}, where additional 
mixings between the spins of (anti)quarks in compact shells are included. 
As the later couplings are suppressed in comparison with the spin-spin interactions 
inside the shells, we neglect the later mixings and restrict ourselves with the Model~II.

Restricting ourselves to the light $u$- and $d$-quarks, and assuming isospin symmetry, 
the universal contribution entering all the pentaquark states is defined as $M_0$, 
which is the sum of the constituent masses of the heavy and light diquarks 
and charm antiquark: 
\begin{equation} 
M_0 \equiv m_{hd} + m_{ld} + m_c . 
\label{eq:M0-def}
\end{equation} 
Apart from this, there are two terms in the effective Hamiltonian explicitly 
related with the spins of the diquarks [see eqs.~(\ref{eq:H-triquark}) 
and~(\ref{eq:H-tri-diquark-interaction})]: 
\begin{eqnarray} 
&& 
{}_J \langle S_{hd}, S_t, L_t; S_{ld}, L_{ld}; S, L | \, 
2 \, (\mathcal{K}_{cq})_{\bar 3} \, (\mathbf{S}_c \cdot \mathbf{S}_q) \,    
| S_{hd}, S_t, L_t; S_{ld}, L_{ld}; S, L \rangle_J 
\label{eq:hd-spin-contribution} \\ 
&& \hspace{17mm}
= (\mathcal{K}_{cq})_{\bar 3} 
\left [ S_{hd} \left ( S_{hd} + 1 \right ) - \frac{3}{2} \right ] = 
\frac{1}{2} \, (\mathcal{K}_{cq})_{\bar 3} \times 
\left \{ 
\begin{array}{cc} 
- 3, & S_{hd} = 0, \\ 
  1, & S_{hd} = 1,   
\end{array}
\right. 
\nonumber   
\end{eqnarray}
\begin{eqnarray} 
&& 
{}_J \langle S_{hd}, S_t, L_t; S_{ld}, L_{ld}; S, L | \, 
2 \, (\mathcal{K}_{q^\prime q^{\prime\prime}})_{\bar 3} \, 
(\mathbf{S}_{q^\prime} \cdot \mathbf{S}_{q^{\prime\prime}}) \,   
| S_{hd}, S_t, L_t; S_{ld}, L_{ld}; S, L \rangle_J 
\label{eq:ld-spin-contribution} \\ 
&& \hspace{17mm} = 
(\mathcal{K}_{q^\prime q^{\prime\prime}})_{\bar 3} 
\left [ S_{ld} \left ( S_{ld} + 1 \right ) - \frac{3}{2} \right ] = 
\frac{1}{2} \, (\mathcal{K}_{q^\prime q^{\prime\prime}})_{\bar 3} \times 
\left \{ 
\begin{array}{cc} 
- 3, & S_{ld} = 0, \\ 
  1, & S_{ld} = 1.    
\end{array}
\right.  
\nonumber 
\end{eqnarray}
The two terms left in eq.~(\ref{eq:H-triquark}) are responsible for the contributions  
of the spin-spin interactions between the charm antiquark and the quarks inside 
the heavy diquark, which together form the triquark state. To calculate their impact, 
it is convenient to use the Wigner $6j$-symbols, which allows us to describe 
a recoupling between three angular momenta, say~$j_1$, $j_2$ and~$j_3$, which 
are combined into the state with the total momentum~$J$. Following the angular 
momentum sum rules in quantum mechanics, one combines two of the three momenta 
first and then adds the third momentum. Thus, there are three possibilities: 
in the the first, the momenta~$j_1$ and~$j_2$ couple together in the state 
with the angular momentum~$j_{12}$, which after combining with~$j_3$ results 
in the state with total angular momentum~$J$; in the second, the momenta~$j_1$ 
and~$j_3$ produce the state with the momentum~$j_{13}$ and it is combined 
with~$j_2$ to get the state with the momentum~$J$; and in the third, the momenta~$j_2$ 
and~$j_3$ produce the state with the momentum~$j_{23}$ which then combines with~$j_1$ 
to obtain the state with the momentum~$J$. Each way of summing up the momenta results 
in its own set of basis vectors. If, for example, the momentum~$j_2$ is coupled 
initially to~$j_1$ and one wants to recouple it to~$j_3$, there are two extra 
momenta~$j_{12}$ and~$j_{23}$, each of which has its own set of basis vectors, 
$| (j_1, \, j_2)_{j_{12}}, \, j_3; \, J \rangle$ and $| j_1, \, (j_2, \, j_3)_{j_{23}}; \, J \rangle$. 
The Wigner $6j$-symbols describe the transformation from one basis to the other 
as follows~\cite{Edmonds:1957}:    
\begin{equation}
| j_1,\, (j_2, \, j_3)_{j_{23}}; \, J \rangle = \sum_{j_{12}} 
(-1)^{j_1 + j_2 + j_3 + J} \sqrt{(2j_{12} + 1)\, (2j_{23}+1)} 
\left \{ 
\begin{array}{rrr} 
j_1 & j_2 & j_{12} \\ 
j_3 &   J & j_{23} 
\end{array} 
\right \} 
| (j_1, \, j_2)_{j_{12}}, \, j_3; \, J \rangle .  
\label{eq:transform-6j}
\end{equation}
Here, Wigner $6j$-symbols are represented by the curly brackets. 
To find the matrix elements of the spin-spin operators  
$2 \, \mathcal{K}_{\bar c q} \left ( \mathbf{S}_{\bar c} \cdot \mathbf{S}_q \right )$    
and 
$2 \, \mathcal{K}_{\bar c c} \left ( \mathbf{S}_{\bar c} \cdot \mathbf{S}_c \right )$ 
in eq.~(\ref{eq:H-triquark}) with the help of the Wigner $6j$-symbols, 
it is convenient to rewrite the pentaquark vector state 
$| S_{hd}, S_t, L_t; S_{ld}, L_{ld}; S, L \rangle_J$ as 
$| S_{\bar c},\, (S_q, \, S_c)_{S_{hd}}; \, S_t \rangle$ and 
$| S_{\bar c},\, (S_c, \, S_q)_{S_{hd}}; \, S_t \rangle$, respectively, 
(here, all the unnecessary fixed quantum numbers are omitted from the vector 
and the charm antiquark spin, $S_{\bar c}$, is shown explicitly) 
and make the angular momenta recoupling as in eq.~(\ref{eq:transform-6j}): 
\begin{eqnarray}
&& 
| S_{\bar c},\, (S_q, \, S_c)_{S_{hd}}; \, S_t \rangle = \sum_{S_{\bar c q}} 
(-1)^{S_{\bar c} + S_q + S_c + S_t} \sqrt{(2 S_{\bar c q} + 1)\, (2 S_{hd} + 1)} 
\nonumber \\ 
&& \hspace*{29mm}
\times \left \{ 
\begin{array}{rrr} 
S_{\bar c} & S_q & S_{\bar c q} \\ 
S_c        & S_t &       S_{hd} 
\end{array} 
\right \} 
| (S_{\bar c},\, S_q)_{S_{\bar c q}}, \, S_c; \, S_t \rangle ,   
\label{eq:transform-Shd-Scbarq} \\ 
&& 
| S_{\bar c},\, (S_c, \, S_q)_{S_{hd}}; \, S_t \rangle = \sum_{S_{\bar c c}} 
(-1)^{S_{\bar c} + S_q + S_c + S_t} \sqrt{(2 S_{\bar c c} + 1)\, (2 S_{hd} + 1)} 
\nonumber \\ 
&& \hspace*{29mm}
\times \left \{ 
\begin{array}{rrr} 
S_{\bar c} & S_c & S_{\bar c c} \\ 
S_q        & S_t &       S_{hd} 
\end{array} 
\right \} 
| (S_{\bar c},\, S_c)_{S_{\bar c c}}, \, S_q; \, S_t \rangle .    
\label{eq:transform-Shd-Scbarc}  
\end{eqnarray}
The required matrix elements can be written in the form: 
\begin{eqnarray} 
&& 
{}_J \langle S_{hd}^\prime, S_t, L_t; S_{ld}, L_{ld}; S, L | \, 
2 \, \mathcal{K}_{\bar c q} \, 
(\mathbf{S}_{\bar c} \cdot \mathbf{S}_q) \,   
| S_{hd}, S_t, L_t; S_{ld}, L_{ld}; S, L \rangle_J  
\nonumber \\ 
&& \hspace{7mm} 
= \mathcal{K}_{\bar c q} \, (- 1)^{2 S_t + 1} 
\sqrt{(2 S_{hd} + 1)\, (2 S_{hd}^\prime + 1)} \, 
\sum_{S_{\bar c q}} \left ( 2 S_{\bar c q} + 1 \right ) 
\left [ S_{\bar c q} \left ( S_{\bar c q} + 1 \right ) - \frac{3}{2} \right ]   
\nonumber \\ 
&& \hspace{7mm} \times 
\left \{ 
\begin{array}{rrr} 
1/2 & 1/2 & S_{\bar c q} \\ 
1/2 & S_t &       S_{hd} 
\end{array} 
\right \} 
\left \{ 
\begin{array}{rrr} 
1/2 & 1/2 &  S_{\bar c q} \\ 
1/2 & S_t & S_{hd}^\prime 
\end{array} 
\right \} , 
\label{eq:cbarq-spin-contribution}  
\end{eqnarray} 
\begin{eqnarray} 
&& 
{}_J \langle S_{hd}^\prime, S_t, L_t; S_{ld}, L_{ld}; S, L | \, 
2 \, \mathcal{K}_{\bar c c} \, 
(\mathbf{S}_{\bar c} \cdot \mathbf{S}_c) \,   
| S_{hd}, S_t, L_t; S_{ld}, L_{ld}; S, L \rangle_J  
\nonumber \\ 
&& \hspace{7mm} 
= \mathcal{K}_{\bar c c} \, (- 1)^{2 S_t + 1} 
\sqrt{(2 S_{hd} + 1)\, (2 S_{hd}^\prime + 1)} \, 
\sum_{S_{\bar c c}} \left ( 2 S_{\bar c c} + 1 \right ) 
\left [ S_{\bar c c} \left ( S_{\bar c c} + 1 \right ) - \frac{3}{2} \right ]   
\nonumber \\ 
&& \hspace{7mm} \times 
\left \{ 
\begin{array}{rrr} 
1/2 & 1/2 & S_{\bar c c} \\ 
1/2 & S_t &       S_{hd} 
\end{array} 
\right \} 
\left \{ 
\begin{array}{rrr} 
1/2 & 1/2 &  S_{\bar c c} \\ 
1/2 & S_t & S_{hd}^\prime 
\end{array} 
\right \} , 
\label{eq:cbarc-spin-contribution}   
\end{eqnarray}
where spins of all the quarks are replaced by their numerical values, 
$S_{\bar c} = S_c = S_q = 1/2$.  
Up to the factors~$\mathcal{K}_{\bar c q}$ and~$\mathcal{K}_{\bar c c}$, 
both equations coincide, so the contribution of these two terms entering 
the effective Hamiltonian to the mass formulae is proportional to the sum 
of the coupling strengths, $\mathcal{K}_{\bar c q} + \mathcal{K}_{\bar c c}$.

The matrix elements of the operators in eq.~(\ref{eq:H-SS-t-d}) have in general 
six different coupling strengths. For the pentaquark having the light $u$- and $d$-quarks 
only, isospin symmetry results into relations among the couplings. This simplifies 
the spin-spin interaction~(\ref{eq:H-SS-t-d}) substantially. For the pentaquarks 
having one or more strange quarks, some of the relations no longer hold, as 
$SU(3)_F$-symmetry is broken. Details of the matrix element calculation in this case 
are presented in appendix~\ref{sec:spin-spin-corrections}. This contribution will be 
neglected in the numerical analysis as we are working in the Model~II~\cite{Maiani:2014aja}.

We apply the above formalism for the calculation of the mass spectrum beginning 
from the pentaquark states from table~\ref{tab:S-wave-pentaquarks-good-ld}.  
From three states presented there, two states with $J^P = 1/2^-$ mix 
due to the spin-spin interaction of the charm antiquark and the heavy diquark, 
and the third one with $J^P = 3/2^-$ remains unmixed. For the later state, 
the mass~$m^{S0}_3$ (here, the superscript denotes the $S$-wave pentaquark 
with the ``good'' light diquark, $S_{ld} = 0$) is the average of the effective 
Hamiltonian over this state: 
\begin{equation}  
m^{S0}_3 = M_0 + 
\frac{1}{2} \, (\mathcal{K}_{c q})_{\bar 3} - 
\frac{3}{2} \, (\mathcal{K}_{q^\prime q^{\prime\prime}})_{\bar 3} + 
\frac{1}{2} \left ( \mathcal{K}_{\bar c q} + \mathcal{K}_{\bar c c} \right ) , 
\label{eq:mass-S0-3} 
\end{equation}
where~$M_0$ is defined in~(\ref{eq:M0-def}). The former two states with $J^P = 1/2^-$, 
after sandwiching the effective Hamiltonian, yield the following $(2 \times 2)$ mass matrix: 
\begin{equation}  
M^{S0}_{J = 1/2} = M_0 - 
\frac{1}{2} \, (\mathcal{K}_{c q})_{\bar 3} - 
\frac{3}{2} \, (\mathcal{K}_{q^\prime q^{\prime\prime}})_{\bar 3} - 
(\mathcal{K}_{c q})_{\bar 3} 
\left ( 
\begin{array}{cc} 
 1 &  0 \\
 0 & -1
\end{array}
\right ) 
+ \frac{1}{2} \left ( \mathcal{K}_{\bar c q} + \mathcal{K}_{\bar c c} \right )  
\left ( 
\begin{array}{cc} 
       0 & \sqrt 3 \\
 \sqrt 3 &      -2
\end{array}
\right ) . 
\label{eq:MM-S0-1/2} 
\end{equation}
Diagonalizing this matrix yields the masses of the other two states:
\begin{eqnarray} 
m^{S0}_1 = M_0 - 
\frac{1}{4} \left ( \mathcal{K}_{\bar c q} + \mathcal{K}_{\bar c c} \right )  
\left [ 2 + r_{hd} + 3 r_{ld} + 2 \sqrt{3 + (1 - r_{hd})^2} \right ] ,  
\label{eq:mass-S0-1} \\ 
m^{S0}_2 = M_0 - 
\frac{1}{4} \left ( \mathcal{K}_{\bar c q} + \mathcal{K}_{\bar c c} \right )  
\left [ 2 + r_{hd} + 3 r_{ld} - 2 \sqrt{3 + (1 - r_{hd})^2} \right ] ,  
\label{eq:mass-S0-2} 
\end{eqnarray}
where two ratios~$r_{hd}$ and~$r_{ld}$ of the couplings are defined as:  
\begin{equation}
r_{hd} \equiv \frac{2 (\mathcal{K}_{c q})_{\bar 3}}{
               \mathcal{K}_{\bar c q} + \mathcal{K}_{\bar c c}} , 
\qquad  
r_{ld} \equiv \frac{2 (\mathcal{K}_{q^\prime q^{\prime\prime}})_{\bar 3}}{ 
               \mathcal{K}_{\bar c q} + \mathcal{K}_{\bar c c}} . 
\label{eq:rhd-rld-def}
\end{equation} 
The averaged mass of the states with masses given in~(\ref{eq:mass-S0-1}) 
and~(\ref{eq:mass-S0-2}) has the following value: 
\begin{equation}
\bar m^{S0}_{12} = \frac{1}{2} \left [ m^{S0}_1 + m^{S0}_2 \right ] = 
M_0 - \frac{1}{4} \left ( \mathcal{K}_{\bar c q} + \mathcal{K}_{\bar c c} \right )  
\left ( 2 + r_{hd} + 3 r_{ld} \right ) ,  
\label{eq:mass-S0-12av} 
\end{equation}
and the mass of the third state lies above  by an amount  %  $\Delta m^{S0}_{3,\overline{12}}$
\begin{equation}
\Delta m^{S0}_{3,\overline{12}} = m^{S0}_3 - \bar m^{S0}_{12} = 
\frac{1}{2} \left ( \mathcal{K}_{\bar c q} + \mathcal{K}_{\bar c c} \right )  
\left ( 2 + r_{hd} \right ) .   
\label{eq:delta-m-S0-3-12av} 
\end{equation}

We continue calculations of the mass spectrum taking the $S$-wave pentaquark states 
with the ``bad'' light diquark from table~\ref{tab:S-wave-pentaquarks-bad-ld}.  
From the seven states presented there, two states with $J^P = 1/2^-$ and the other 
two with $J^P = 3/2^-$, both pairs having the triquark spin $S_t = 1/2$, mix
due to the spin-spin interaction of the charm antiquark and the heavy diquark. 
The other three states with the triquark spin $S_t = 3/2$ remain unmixed due 
to this interaction but can mix through the spin-spin interactions between 
(anti)quarks in the two separated shells~--- the heavy triquark and light diquark. 
As mentioned earlier, such types of spin-spin interactions are suppressed and 
neglected in this analysis. For the later three states, enumerated according to their 
entries in table~\ref{tab:S-wave-pentaquarks-bad-ld}, the masses~$m^{S1}_{5,6,7}$ 
are the averages of the effective Hamiltonian over the states considered (here, the 
superscript denotes the $S$-wave pentaquark with the ``bad'' light diquark, $S_{ld} = 1$): 
\begin{eqnarray}  
m^{S1}_{5, 6, 7} = M_0 + 
\frac{1}{2} \, (\mathcal{K}_{c q})_{\bar 3} + 
\frac{1}{2} \, (\mathcal{K}_{q^\prime q^{\prime\prime}})_{\bar 3} + 
\frac{1}{2} \left ( \mathcal{K}_{\bar c q} + \mathcal{K}_{\bar c c} \right ) . 
\label{eq:mass-S1-567} 
\end{eqnarray}
% 
%  where~$M_0$ is the sum of the masses of the pentaquark constituents~(\ref{eq:M0-def}). 
So, one can see that taking into account only the spin-spin interactions inside 
the light diquark and heavy triquark, these states are degenerated in mass. 

The pairs of states with $J^P = 1/2^-$ and $J^P = 3/2^-$, after sandwiching 
the effective Hamiltonian, yield identical $(2 \times 2)$ mass matrices: 
\begin{equation}  
M^{S1}_{J = 1/2, 3/2} = M_0 - 
\frac{1}{2} \, (\mathcal{K}_{c q})_{\bar 3} + 
\frac{1}{2} \, (\mathcal{K}_{q^\prime q^{\prime\prime}})_{\bar 3} - 
(\mathcal{K}_{c q})_{\bar 3} 
\left ( 
\begin{array}{cc} 
 1 &  0 \\
 0 & -1
\end{array}
\right ) 
+ \frac{1}{2} \left ( \mathcal{K}_{\bar c q} + \mathcal{K}_{\bar c c} \right )  
\left ( 
\begin{array}{cc} 
       0 & \sqrt 3 \\
 \sqrt 3 &      -2
\end{array}
\right ) . 
\label{eq:MM-S1-1/2-3/2} 
\end{equation}
Diagonalizing the matrix, the states, being twice degenerate, have the masses: 
\begin{eqnarray} 
m^{S1}_{1,3} = M_0 - 
\frac{1}{4} \left ( \mathcal{K}_{\bar c q} + \mathcal{K}_{\bar c c} \right )  
\left [ 2 + r_{hd} - r_{ld} + 2 \sqrt{3 + (1 - r_{hd})^2} \right ] ,  
\label{eq:mass-S1-1-3} \\ 
m^{S1}_{2,4} = M_0 - 
\frac{1}{4} \left ( \mathcal{K}_{\bar c q} + \mathcal{K}_{\bar c c} \right )  
\left [ 2 + r_{hd} - r_{ld} - 2 \sqrt{3 + (1 - r_{hd})^2} \right ] ,  
\label{eq:mass-S1-2-4} 
\end{eqnarray}
where~$r_{hd}$ and~$r_{ld}$ are defined in eq.~(\ref{eq:rhd-rld-def}). 
The averaged masses of the states with masses in~(\ref{eq:mass-S1-1-3}) 
and~(\ref{eq:mass-S1-2-4}) have the following value: 
\begin{eqnarray}
\bar m^{S1}_{12} = \frac{1}{2} \left [ m^{S1}_1 + m^{S1}_2 \right ] = 
M_0 - \frac{1}{4} \left ( \mathcal{K}_{\bar c q} + \mathcal{K}_{\bar c c} \right )  
\left ( 2 + r_{hd} - r_{ld} \right ) ,  
\label{eq:mass-S1-12av} \\ 
\bar m^{S1}_{34} = \frac{1}{2} \left [ m^{S1}_3 + m^{S1}_4 \right ] = 
M_0 - \frac{1}{4} \left ( \mathcal{K}_{\bar c q} + \mathcal{K}_{\bar c c} \right )  
\left ( 2 + r_{hd} - r_{ld} \right ) ,  
\label{eq:mass-S1-34av} 
\end{eqnarray}
and the mass gap between these average values and the mass~$\bar m^{S0}_{12}$~(\ref{eq:mass-S0-12av}) 
is completely determined by the strength of the spin-spin interaction in the light diquark:  
\begin{equation}
\Delta\bar m^{S1, S0}_{12} = 
\bar m^{S1}_{12} - \bar m^{S0}_{12} = 
\left ( \mathcal{K}_{\bar c q} + \mathcal{K}_{\bar c c} \right ) r_{ld} = 
2 (\mathcal{K}_{q^\prime q^{\prime\prime}})_{\bar 3} .   
\label{eq:delta-m-S1-S0} 
\end{equation}
The degeneracy is lifted if the spin-spin interactions between the triquark 
and the light-diquark constituents are taken into account as suggested by 
Maiani et al. in the Model~I~\cite{Maiani:2004vq}.

\subsection[$P$-wave pentaquarks]{\boldmath $P$-wave pentaquarks}
\label{ssec:Mass-Formulae-P-wave} 
As it was discussed earlier in sec.~\ref{sec:Effective-Hamiltonian}, one needs 
to include the terms dependent on the internal angular momenta of the pentaquark 
system which can be written as additional terms~(\ref{eq:Hamiltonian-L-pentaquark}) 
and~(\ref{eq:Hamiltonian-tensor-pentaquark}) in the pentaquark effective Hamiltonian. 

Let the vector states $| S_{hd}, S_t, L_t; S_{ld}, L_{ld}; S, L \rangle_J$, appearing 
in tables~\ref{tab:P-wave-pentaquarks-good-ld} and~\ref{tab:P-wave-pentaquarks-bad-ld}, 
be denoted as $| L,\, (S_t, \, S_{ld})_S; \, J \rangle$ and transformed to the state 
$| (L,\, S_t)_{J_t}, \, S_{ld}; \, J \rangle$. 
After the transformation~(\ref{eq:transform-6j}) is applied, we get:  
\begin{equation}
| L, \, (S_t, \, S_{ld})_S; \, J \rangle = \sum_{J_t} 
(-1)^{S_t + S_{ld} + L + J} \sqrt{(2 S + 1)\, (2 J_t + 1)} 
\left \{ 
\begin{array}{ccc} 
     L & S_t & J_t \\ 
S_{ld} &   J &   S 
\end{array} 
\right \} 
| (L, \, S_t)_{J_t}, \, S_{ld}; \, J \rangle .  
\label{eq:transform-spin-orbit-L-St}
\end{equation}
The required matrix elements can be written in the form: 
\begin{eqnarray} 
&& 
{}_J \langle S_{hd}, S_t, L_t; S_{ld}, L_{ld}; S^\prime, L | \, 
2 \, A_t \, (\mathbf{L} \cdot \mathbf{S}_t) \,   
| S_{hd}, S_t, L_t; S_{ld}, L_{ld}; S, L \rangle_J 
\nonumber \\ 
&& \hspace{17mm}  
= A_t \, (-1)^{2 S_t + 2 J} \sqrt{(2 S + 1) \, (2 S^\prime + 1)} \, 
\sum_{J_t} \left ( 2 J_t + 1 \right ) 
\nonumber \\ 
&& \hspace{17mm} \times 
\left [ 
J_t \left ( J_t + 1 \right ) - S_t \left ( S_t + 1 \right ) - 2 
\right ]   
\left \{ 
\begin{array}{ccc} 
     1 & S_t & J_t \\ 
S_{ld} &   J &   S 
\end{array} 
\right \} 
\left \{ 
\begin{array}{ccc} 
     1 & S_t &      J_t \\ 
S_{ld} &   J & S^\prime 
\end{array} 
\right \} , 
\label{eq:L-St-spin-orbit-contribution}  
\end{eqnarray}
where the value of the orbital angular momentum, $L = 1$, is substituted.  
 
Next, we denote the state $| S_{hd}, S_t, L_t; S_{ld}, L_{ld}; S, L \rangle_J$   
as $| L, \, (S_{ld}, \, S_t)_S; \, J \rangle$ and transform it to the state 
$| (L, \, S_{ld})_{J_{ld}}, \, S_t; \, J \rangle$.   
After the transformation~(\ref{eq:transform-6j}) is applied, we get:  
\begin{equation}
| L, \, (S_{ld}, \, S_t)_S; \, J \rangle = \sum_{J_{ld}} 
(-1)^{S_t + S_{ld} + L + J} \sqrt{(2 S + 1)\, (2 J_{ld} + 1)} 
\left \{ 
\begin{array}{ccc} 
  L & S_{ld} & J_{ld} \\ 
S_t &      J &      S 
\end{array} 
\right \} 
| (L, \, S_{ld})_{J_{ld}}, \, S_t; \, J \rangle .  
\label{eq:transform-spin-orbit-L-Sld}
\end{equation}
With this, the matrix elements can be written as: 
\begin{eqnarray} 
&& 
{}_J \langle S_{hd}, S_t, L_t; S_{ld}, L_{ld}; S^\prime, L | \, 
2 \, A_{ld} \, (\mathbf{L} \cdot \mathbf{S}_{ld}) \,   
| S_{hd}, S_t, L_t; S_{ld}, L_{ld}; S, L \rangle_J 
\nonumber \\ 
&& \hspace{17mm}  
= A_{ld} \, (-1)^{2 S_t + 2 J} \sqrt{(2 S + 1) \, (2 S^\prime + 1)} \, 
\sum_{J_{ld}} \left ( 2 J_{ld} + 1 \right ) 
\nonumber \\ 
&& \hspace{17mm} \times 
\left [ 
J_{ld} \left ( J_{ld} + 1 \right ) - S_{ld} \left ( S_{ld} + 1 \right ) - 2 
\right ]   
\left \{ 
\begin{array}{ccc} 
  1 & S_{ld} & J_{ld} \\ 
S_t &      J &      S 
\end{array} 
\right \} 
\left \{ 
\begin{array}{ccc} 
  1 & S_{ld} &   J_{ld} \\ 
S_t &      J & S^\prime 
\end{array} 
\right \} , 
\label{eq:L-Sld-spin-orbit-contribution}  
\end{eqnarray}
where again the value of the orbital angular momentum, $L = 1$, is substituted.  
 
We calculate the mass spectrum of the $P$-states, starting with the pentaquark 
states with the ``good'' light diquark from table~\ref{tab:P-wave-pentaquarks-good-ld}.  
From the seven states presented there, two states with $J^P = 1/2^+$ and the other two with 
$J^P = 3/2^+$, both pairs having the triquark spin $S_t = 1/2$, mix due to the spin-spin 
interaction of the charm antiquark and heavy diquark. The other three states with the triquark 
spin $S_t = 3/2$ remain unmixed due to this interaction but can mix through the spin-spin 
interactions between the (anti)quarks entering two separated shells~--- the heavy triquark 
and light diquark, as also discussed earlier for the $S$-wave states. As mentioned earlier, 
such types of spin-spin interactions are suppressed and neglected in this analysis.
From the terms entering the angular-momentum dependent parts~(\ref{eq:Hamiltonian-L-pentaquark}) 
and~(\ref{eq:Hamiltonian-tensor-pentaquark}) of the effective Hamiltonian, one obtains the 
spin-orbit, orbital, and tensor contributions to matrix elements. For the three states, enumerated 
as the fifth, sixth and seventh according to their rows in table~\ref{tab:P-wave-pentaquarks-good-ld},
the masses~$m^{P0}_{5,6,7}$ are the matrix elements of the effective Hamiltonian over these states
(here, the superscript denotes the $P$-wave pentaquark with the ``good'' light diquark, $S_{ld} = 0$): 
\begin{eqnarray}  
&& 
m^{P0}_5 = M_0 + 
\frac{1}{2} \, (\mathcal{K}_{c q})_{\bar 3} - 
\frac{3}{2} \, (\mathcal{K}_{q^\prime q^{\prime\prime}})_{\bar 3} + 
\frac{1}{2} \left ( \mathcal{K}_{\bar c q} + \mathcal{K}_{\bar c c} \right ) 
+ B - 5 A_t , 
\label{eq:mass-P0-5} \\ 
&& 
m^{P0}_6 = m^{P0}_5 + 3 A_t , 
\qquad  
m^{P0}_7 = m^{P0}_5 + 8 A_t .  
\nonumber 
\end{eqnarray}
% 
%where~$M_0$ is the sum of the masses of the pentaquark constituents~(\ref{eq:M0-def}). 
These states are non-degenerate due to the triquark spin-orbit interaction, i.\,e., $A_t \neq 0 $.

The pair of states with $J^P = 1/2^-$, after sandwiching the effective 
Hamiltonian, yields the following $(2 \times 2)$ mass matrix: 
\begin{equation}  
M^{P0}_{J = 1/2} = M_0 - 
\frac{1}{2} \, (\mathcal{K}_{c q})_{\bar 3} - 
\frac{3}{2} \, (\mathcal{K}_{q^\prime q^{\prime\prime}})_{\bar 3} - 
(\mathcal{K}_{c q})_{\bar 3} 
\left ( 
\begin{array}{rr} 
 1 &  0 \\
 0 & -1
\end{array}
\right ) 
+ \frac{1}{2} \left ( \mathcal{K}_{\bar c q} + \mathcal{K}_{\bar c c} \right )  
\left ( 
\begin{array}{cc} 
       0 & \sqrt 3 \\
 \sqrt 3 &      -2
\end{array}
\right ) + B - 2 A_t ,  
\label{eq:MM-P0-1/2} 
\end{equation}
while the masses of the pair with $J^P = 3/2^-$ are determined by the matrix:  
\begin{equation}  
M^{P0}_{J = 3/2} = M_0 - 
\frac{1}{2} \, (\mathcal{K}_{c q})_{\bar 3} - 
\frac{3}{2} \, (\mathcal{K}_{q^\prime q^{\prime\prime}})_{\bar 3} - 
(\mathcal{K}_{c q})_{\bar 3} 
\left ( 
\begin{array}{rr} 
 1 &  0 \\
 0 & -1
\end{array}
\right ) 
+ \frac{1}{2} \left ( \mathcal{K}_{\bar c q} + \mathcal{K}_{\bar c c} \right )  
\left ( 
\begin{array}{cc} 
       0 & \sqrt 3 \\
 \sqrt 3 &      -2
\end{array}
\right ) + B + A_t .   
\label{eq:MM-P0-3/2} 
\end{equation}
Diagonalizing these matrices, we get the masses: 
\begin{eqnarray} 
&& 
m^{P0}_1 = M_0 - 
\frac{1}{4} \left ( \mathcal{K}_{\bar c q} + \mathcal{K}_{\bar c c} \right )  
\left [ 2 + r_{hd} + 3 r_{ld} + 2 \sqrt{3 + (1 - r_{hd})^2} \right ] + B - 2 A_t ,  
\label{eq:mass-P0-1} \\ 
&& 
m^{P0}_2 = M_0 - 
\frac{1}{4} \left ( \mathcal{K}_{\bar c q} + \mathcal{K}_{\bar c c} \right )  
\left [ 2 + r_{hd} + 3 r_{ld} - 2 \sqrt{3 + (1 - r_{hd})^2} \right ] + B - 2 A_t ,  
\label{eq:mass-P0-2} \\ 
&& 
m^{P0}_{3,4} = m^{P0}_{1,2} + 3 A_t ,  
\label{eq:mass-P0-3-4}  
\end{eqnarray}
where~$r_{hd}$ and $r_{ld}$ are defined in eq.~(\ref{eq:rhd-rld-def}).

We continue the calculations of the mass spectrum for the pentaquark states 
with the ``bad'' light diquark from table~\ref{tab:P-wave-pentaquarks-bad-ld}.  
From the eighteen states presented, there are six states with $J^P = 1/2^+$, seven 
with $J^P = 3/2^+$, four with $J^P = 5/2^+$, and the last one with $J^P = 7/2^+$. 
Except for the $J^P = 7/2^+$ state, all the others mix due to the spin-spin interactions,
but also because of the spin-orbit and tensor terms~(\ref{eq:Hamiltonian-L-pentaquark}) 
and~(\ref{eq:Hamiltonian-tensor-pentaquark}) in the effective Hamiltonian. For the last 
state, enumerated as the eighteenth according to table~\ref{tab:P-wave-pentaquarks-bad-ld}, 
the mass~$m^{P1}_{18}$ is the average of the effective Hamiltonian over this state (here, 
the superscript denotes the $P$-wave pentaquark with the ``bad'' light diquark, $S_{ld} = 1$): 
\begin{equation}  
m^{P1}_{18} = M_0 + 
\frac{1}{2} \, (\mathcal{K}_{c q})_{\bar 3} + 
\frac{1}{2} \, (\mathcal{K}_{q^\prime q^{\prime\prime}})_{\bar 3} + 
\frac{1}{2} \left ( \mathcal{K}_{\bar c q} + \mathcal{K}_{\bar c c} \right ) 
+ B + 3 A_t + 2 A_{ld} - \frac{3}{5} \, b . 
\label{eq:mass-P1-18}  
\end{equation}
% 
%  where~$M_0$ is the sum of the masses of the pentaquark constituents~(\ref{eq:M0-def}). 

As mentioned earlier, there are four states with spin-parity $J^P = 5/2^+$ which 
can be divided into two pairs according to their mixing mechanism: the fifth and 
the tenth states in table~\ref{tab:P-wave-pentaquarks-bad-ld} mix due to the 
spin-spin interaction while for the fifteenth and the seventeenth states, their 
mixing is determined by the spin-orbit and tensor interactions. We start from 
the first of the mentioned two states. Their mass matrix is as follows: 
\begin{eqnarray}  
&& 
M^{P1}_{J = 5/2} = M_0 - 
\frac{1}{2} \, (\mathcal{K}_{c q})_{\bar 3} + 
\frac{1}{2} \, (\mathcal{K}_{q^\prime q^{\prime\prime}})_{\bar 3} - 
(\mathcal{K}_{c q})_{\bar 3} 
\left ( 
\begin{array}{rr} 
 1 &  0 \\
 0 & -1
\end{array}
\right ) 
+ \frac{1}{2} \left ( \mathcal{K}_{\bar c q} + \mathcal{K}_{\bar c c} \right )  
\left ( 
\begin{array}{cc} 
       0 & \sqrt 3 \\
 \sqrt 3 &      -2
\end{array}
\right ) 
\nonumber \\ 
&& \hspace*{19mm}
+ B + A_t + 2 A_{ld} - \frac{1}{5} \, b .   
\label{eq:MM-P1-5/2-ss} 
\end{eqnarray}
Diagonalizing the matrix, we obtain the masses: 
\begin{equation}  
m^{P1}_{14} = M_0 - 
\frac{1}{4} \left ( \mathcal{K}_{\bar c q} + \mathcal{K}_{\bar c c} \right )  
\left [ 2 + r_{hd} - r_{ld} + 2 \sqrt{3 + (1 - r_{hd})^2} \right ] + 
B + A_t + 2 A_{ld} - \frac{1}{5} \, b ,  
\label{eq:mass-P1-14} 
\end{equation}  
\begin{equation}  
m^{P1}_{15} = M_0 - 
\frac{1}{4} \left ( \mathcal{K}_{\bar c q} + \mathcal{K}_{\bar c c} \right )  
\left [ 2 + r_{hd} - r_{ld} - 2 \sqrt{3 + (1 - r_{hd})^2} \right ] + 
B + A_t + 2 A_{ld} - \frac{1}{5} \, b .  
\label{eq:mass-P1-15} 
\end{equation}
% 
% where $r_{hd} = 2 (\mathcal{K}_{c q})_{\bar 3}/( \mathcal{K}_{\bar c q} + \mathcal{K}_{\bar c c})$ 
% and $r_{ld} = 2 (\mathcal{K}_{q^\prime q^{\prime\prime}})_{\bar 3}/( \mathcal{K}_{\bar c q} + \mathcal{K}_{\bar c c})$.   

For the remaining two states, the mass matrix has the form: 
\begin{eqnarray}  
&& 
\tilde M^{P1}_{J = 5/2} = M_0 + 
\frac{1}{2} \, (\mathcal{K}_{c q})_{\bar 3} + 
\frac{1}{2} \, (\mathcal{K}_{q^\prime q^{\prime\prime}})_{\bar 3} 
+ \frac{1}{2} \left ( \mathcal{K}_{\bar c q} + \mathcal{K}_{\bar c c} \right )  
+ B 
\label{eq:MM-P1-5/2-so} \\ 
&& \hspace*{11mm}
+ \frac{1}{5} \, A_t 
\left ( 
\begin{array}{cc} 
          11 & 2 \sqrt{21} \\
 2 \sqrt{21} &          -6
\end{array}
\right ) 
+ \frac{2}{5} \, A_{ld} 
\left ( 
\begin{array}{cc} 
         2 & \sqrt{21} \\
 \sqrt{21} &        -2
\end{array}
\right ) 
- \frac{1}{250} \, b 
\left ( 
\begin{array}{cc} 
          584 & 15 \sqrt{21} \\
 15 \sqrt{21} &           24
\end{array}
\right ) .   
\nonumber 
\end{eqnarray}
So, the masses of the pentaquarks are two eigenvalues of this matrix: 
\begin{eqnarray} 
&& 
m^{P1}_{16} = M_0 + 
\frac{1}{4} \left ( \mathcal{K}_{\bar c q} + \mathcal{K}_{\bar c c} \right )  
\left [ 2 + r_{hd} + r_{ld} \right ] + 
B + \frac{1}{2} \, A_t - \frac{152}{125} \, b 
\label{eq:mass-P1-16} \\ 
&& \hspace*{17mm}
- \frac{1}{50} \, 
\sqrt{\left ( 85 A_t + 40 A_{ld} - 56 b \right )^2 + 
      21 \left ( 20 A_t + 20 A_{ld} - 3 b \right )^2} ,  
\nonumber \\ 
&& 
m^{P1}_{17} = M_0 + 
\frac{1}{4} \left ( \mathcal{K}_{\bar c q} + \mathcal{K}_{\bar c c} \right )  
\left [ 2 + r_{hd} + r_{ld} \right ] + 
B + \frac{1}{2} \, A_t - \frac{152}{125} \, b 
\label{eq:mass-P1-17} \\ 
&& \hspace*{17mm}
+ \frac{1}{50} \, 
\sqrt{\left ( 85 A_t + 40 A_{ld} - 56 b \right )^2 + 
      21 \left ( 20 A_t + 20 A_{ld} - 3 b \right )^2} .   
\nonumber  
\end{eqnarray}

Further, there are seven states with the spin-parity $J^P = 3/2^+$ which can be separated 
according to the triquark spin $S_t = 1/2$ (four states) and $S_t = 3/2$ (three states) as 
shown in table~\ref{tab:P-wave-pentaquarks-bad-ld}. The former four states  mix under 
the spin-spin, spin-orbit and tensor interactions while three states with $S_t = 3/2$ 
mix due to the spin-orbit and tensor interactions only. The details of the mass derivation 
can be found in appendix~\ref{sec:mass-derivations}; here we present explicit equations 
for the masses only:  
\begin{equation}  
m^{P1}_{11,12,13} = M_0 + 
% \frac{1}{2} \, (\mathcal{K}_{c q})_{\bar 3} + 
% \frac{1}{2} \, (\mathcal{K}_{q^\prime q^{\prime\prime}})_{\bar 3} + 
\frac{1}{4} \left ( \mathcal{K}_{\bar c q} + \mathcal{K}_{\bar c c} \right )  
\left ( 2 + r_{hd} + r_{ld} \right ) 
+ B - \frac{4}{3} \left ( A_t + A_{ld} \right ) 
- \frac{3622}{1125} \, b + \lambda_{3,2,1} , 
\label{eq:mass-P1-11-12-13}  
\end{equation}
where $\lambda_{1,2,3}$ are determined in eq.~(\ref{eq:lambda-1-2-3}). 
                                                                        
To derive the expressions for the masses of the four states with the spin-parity 
$J^P = 3/2^+$ and the triquark spin $S_t = 1/2$, one needs to find the eigenvalues 
of a non-diagonal symmetric $(4 \times 4)$ matrix. The explicit form of the mass matrix 
and the details of calculations can be found in appendix~\ref{sec:mass-derivations}.  
The set of masses is as follows: 
\begin{equation}  
m^{P1}_7 = M_0 + B - \frac{A_t}{2} - \frac{14}{15} \, b - 
\frac{1}{4} \left ( \mathcal{K}_{\bar c q} + \mathcal{K}_{\bar c c} \right )  
\left ( 2 + r_{hd} - r_{ld} + 2 \sqrt{3 + (1 - r_{hd})^2} \right ) - \mu^{(3/2)} , 
\label{eq:mass-P1-7} 
\end{equation} 
\begin{equation} 
m^{P1}_8 = M_0 + B - \frac{A_t}{2} - \frac{14}{15} \, b - 
\frac{1}{4} \left ( \mathcal{K}_{\bar c q} + \mathcal{K}_{\bar c c} \right )  
\left ( 2 + r_{hd} - r_{ld} + 2 \sqrt{3 + (1 - r_{hd})^2} \right ) + \mu^{(3/2)} , 
\label{eq:mass-P1-8}   
\end{equation} 
\begin{equation} 
m^{P1}_9 = M_0 + B - \frac{A_t}{2} - \frac{14}{15} \, b - 
\frac{1}{4} \left ( \mathcal{K}_{\bar c q} + \mathcal{K}_{\bar c c} \right )  
\left ( 2 + r_{hd} - r_{ld} - 2 \sqrt{3 + (1 - r_{hd})^2} \right ) - \mu^{(3/2)} , 
\label{eq:mass-P1-9}   
\end{equation} 
\begin{equation} 
m^{P1}_{10} = M_0 + B - \frac{A_t}{2} - \frac{14}{15} \, b - 
\frac{1}{4} \left ( \mathcal{K}_{\bar c q} + \mathcal{K}_{\bar c c} \right )  
\left ( 2 + r_{hd} - r_{ld} - 2 \sqrt{3 + (1 - r_{hd})^2} \right ) + \mu^{(3/2)} ,   
\label{eq:mass-P1-10}   
\end{equation} 
where
\begin{equation} 
\mu^{(3/2)} = \frac{1}{30} \, 
%  \frac{1}{15 \left ( \mathcal{K}_{\bar c q} + \mathcal{K}_{\bar c c} \right )} \, 
\sqrt{\left ( 5 A_t + 40 A_{ld} - 12 b \right )^2 + 5 \left ( 20 A_t + 20 A_{ld} + 3 b \right )^2} . 
\label{eq:mu-3/2-def} 
\end{equation}

Finally, there are six states with the spin-parity $J^P = 1/2^+$ which can be divided 
according to the triquark spin $S_t = 1/2$ (four states) and $S_t = 3/2$ (two states) 
as shown in table~\ref{tab:P-wave-pentaquarks-bad-ld}. The four states mix under 
the spin-spin, spin-orbit and tensor interactions while the two states with $S_t = 3/2$ 
 mix through the spin-orbit and tensor interactions only. The mass matrix of the 
later two states is as follows: 
\begin{eqnarray}  
&& 
\tilde M^{P1}_{J = 1/2} = M_0 + 
\frac{1}{2} \, (\mathcal{K}_{c q})_{\bar 3} + 
\frac{1}{2} \, (\mathcal{K}_{q^\prime q^{\prime\prime}})_{\bar 3} 
+ \frac{1}{2} \left ( \mathcal{K}_{\bar c q} + \mathcal{K}_{\bar c c} \right )  
+ B - \frac{4}{3} \, b 
\label{eq:MM-P1-1/2-so} \\ 
&& \hspace*{11mm}
- \frac{1}{3} \, A_t 
\left ( 
\begin{array}{cc} 
        10 & -2 \sqrt 5 \\
-2 \sqrt 5 &         11
\end{array}
\right ) 
+ \frac{2}{3} \, A_{ld} 
\left ( 
\begin{array}{cc} 
       2 & \sqrt 5 \\
 \sqrt 5 &      -2
\end{array}
\right ) 
+ \frac{7}{2 \sqrt 5} \, b 
\left ( 
\begin{array}{cc} 
 0 & 1 \\
 1 & 0
\end{array}
\right ) .   
\nonumber 
\end{eqnarray}
The pentaquark masses are the two eigenvalues of this matrix: 
\begin{eqnarray} 
&& 
m^{P1}_5 = M_0 + 
\frac{1}{4} \left ( \mathcal{K}_{\bar c q} + \mathcal{K}_{\bar c c} \right )  
\left [ 2 + r_{hd} + r_{ld} \right ] + 
B - \frac{7}{2} \, A_t - \frac{4}{3} \, b 
\label{eq:mass-P1-5} \\ 
&& \hspace*{17mm}
- \frac{1}{6 \sqrt 5} \, 
\sqrt{5 \left ( A_t + 8 A_{ld} \right )^2 + 
      \left ( 20 A_t + 20 A_{ld} + 21 b \right )^2} ,  
\nonumber \\ 
&& 
m^{P1}_6 = M_0 + 
\frac{1}{4} \left ( \mathcal{K}_{\bar c q} + \mathcal{K}_{\bar c c} \right )  
\left [ 2 + r_{hd} + r_{ld} \right ] + 
B - \frac{7}{2} \, A_t - \frac{4}{3} \, b 
\label{eq:mass-P1-6} \\ 
&& \hspace*{17mm}
+ \frac{1}{6 \sqrt 5} \, 
\sqrt{5 \left ( A_t + 8 A_{ld} \right )^2 + 
      \left ( 20 A_t + 20 A_{ld} + 21 b \right )^2} .   
\nonumber  
\end{eqnarray}

For the last four states with the spin-parity $J^P = 1/2^+$ and triquark spin $S_t = 1/2$, 
the mass matrix is again a non-diagonal symmetric $(4 \times 4)$ matrix. It is written 
explicitly in appendix~\ref{sec:mass-derivations} and details of calculations of its 
eigenvalues can be also found there. The resulting set of masses is presented below:   
\begin{eqnarray} 
&& 
m^{P1}_1 = M_0 + B - \frac{A_t}{2} - 3 A_{ld} - \frac{31}{30} \, b 
\nonumber \\ 
&& \hspace*{17mm}
- \frac{1}{4} \left ( \mathcal{K}_{\bar c q} + \mathcal{K}_{\bar c c} \right )  
\left ( 2 + r_{hd} - r_{ld} + 2 \sqrt{3 + (1 - r_{hd})^2} \right ) - \mu^{(1/2)} , 
\label{eq:mass-P1-1} \\  
&& 
m^{P1}_2 = M_0 + B - \frac{A_t}{2} - 3 A_{ld} - \frac{31}{30} \, b 
\nonumber \\ 
&& \hspace*{17mm}
- \frac{1}{4} \left ( \mathcal{K}_{\bar c q} + \mathcal{K}_{\bar c c} \right )  
\left ( 2 + r_{hd} - r_{ld} + 2 \sqrt{3 + (1 - r_{hd})^2} \right ) + \mu^{(1/2)} , 
\label{eq:mass-P1-2} \\  
&& 
m^{P1}_3 = M_0 + B - \frac{A_t}{2} - 3 A_{ld} - \frac{31}{30} \, b 
\nonumber \\ 
&& \hspace*{17mm}
- \frac{1}{4} \left ( \mathcal{K}_{\bar c q} + \mathcal{K}_{\bar c c} \right )  
\left ( 2 + r_{hd} - r_{ld} - 2 \sqrt{3 + (1 - r_{hd})^2} \right ) - \mu^{(1/2)} , 
\label{eq:mass-P1-3} \\  
&& 
m^{P1}_4 = M_0 + B - \frac{A_t}{2} - 3 A_{ld} - \frac{31}{30} \, b 
\nonumber \\ 
&& \hspace*{17mm}
- \frac{1}{4} \left ( \mathcal{K}_{\bar c q} + \mathcal{K}_{\bar c c} \right )  
\left ( 2 + r_{hd} - r_{ld} - 2 \sqrt{3 + (1 - r_{hd})^2} \right ) + \mu^{(1/2)} ,  
\label{eq:mass-P1-4}   
\end{eqnarray}
where
\begin{equation} 
\mu^{(1/2)} = \frac{1}{30} \, 
%  \frac{1}{15 \left ( \mathcal{K}_{\bar c q} + \mathcal{K}_{\bar c c} \right )} \, 
\sqrt{25 \left ( 7 A_t + 2 A_{ld} + 3 b \right )^2 + 2 \left ( 20 A_t + 20 A_{ld} + 21 b \right )^2} . 
\label{eq:mu-1/2-def} 
\end{equation}

With these expressions, the analytical calculations of the pentaquark mass 
spectrum in Model~II is done and we present the numerical estimates of the 
pentaquark masses in the next section.

\section{Hidden-charm pentaquark mass predictions}                      
\label{sec:mass-predictions}

\subsection{Input parameters} 
\label{ssec:input-parameters}

Working within the Constituent Quark-Diquark Model~\cite{Maiani:2014aja}, input parameters 
are the masses of the constituents, charm quark and two diquarks, spin-spin couplings, 
and other parameters related with the orbital or radial excitations. In the present paper 
we analyse the ground-state pentaquarks and their first orbital excitations only. 
To estimate the constituent quark masses, there are two possibilities: either extract them 
from the masses of known mesons or from baryons~\cite{Ali:2019roi}. Quark masses obtained from 
the baryon spectrum are larger by typically $50$~MeV~\cite{Ali:2019roi}. This can be exemplified 
by the charm quark mass, which is estimated as $m_c^m = 1667$~MeV from the $D$-meson spectrum, 
as opposed to $m_c^b = 1710$~MeV from the charm baryon masses, yielding a mass difference 
of~$43$~MeV~\cite{Ali:2019roi}. In particular, with the~$m_c^b$ value as an input, predictions 
for the charm baryon masses were obtained~\cite{Karliner:2014gca} which differ from the 
experimentally observed masses~\cite{Tanabashi:2018oca} by about $10$~MeV, which may be viewed 
as an error on the input charm-quark mass. We use~$m_c^b$ in the numerical analysis.  
Based on the same arguments, we accept $m_q^b = (362 \pm 10)$~MeV and $m_s^b = (540 \pm 10)$~MeV
as the light and strange quark masses, respectively~\cite{Ali:2019roi}.

The diquark masses are presented in table~\ref{tab:diquark-masses}. In contrast to quarks, 
which have  spin $S_q = 1/2$, diquarks, being composite objects, have
two possible spin configurations from which the antisymmetric one corresponding to the 
diquark spin $S = 0$ is energetically more favorable. Both configurations are allowed 
if the flavors of the quarks are different, indicated by right brackets, $[\ldots]$, 
in table~\ref{tab:diquark-masses} (for $S = 0$), and by curly brackets, $\{\ldots\}$, 
(for $S = 1$). For the diquarks, having two quarks of the same flavor, only the spin-symmetric 
configuration is allowed by Bose statistics, as indicated for the case 
of the diquark with two $s$-quarks in table~\ref{tab:diquark-masses}.

In the limit of the exact isospin symmetry, the light $u$- and $d$-quarks are mass-degenerate
and denoted by~$q$ in table~\ref{tab:diquark-masses}. For the lightest diquark, $[u d]$, and its 
$SU(3)_F$-symmetry partners,~$[s u]$ and~$[s d]$, the masses are taken from~\cite{Karliner:2018bms} 
as well as the mass of the double strange diquark,~$\{ s s \}$. The masses of the charm  
diquarks,~$[c q]$ and~$[c s]$, are borrowed from~\cite{Ali:2019roi}. To get the diquark-mass error 
estimates, one should compare baryon masses predicted within this model and in experiment.
Based on the analysis presented in~\cite{Karliner:2014gca}, in which the predicted and measured 
mass differences for unflavored (i.\,e., having $u$- or $d$-quarks), strange and charm baryons 
do not exceed $15$~MeV, we take it as a measure of the uncertainty in diquark masses.      
\begin{table}[tb] 
\begin{center}
\begin{tabular}{|cccccc|} 
\hline
   Diquark content & $[u d]$ & $[s q]$ & $\{s s\}$ & $[c q]$ & $[c s]$ \\ \hline 
Diquark mass (MeV) &     576 &     800 &      1099 &    1976 &    2105 \\ \hline   
%   $S = 0$ diquarks &          576 &   800 &       &  1976 &  2105 \\   
%   $S = 1$ diquarks &          776 &   933 &  1099 &  ???? &  ???? \\   
%         Difference &          200 &   133 &       &   ??? &   ??? \\ \hline   
\end{tabular}
\end{center}
\caption{
Constituent diquark masses (in MeV) used for calculating
the hidden-charm pentaquark mass spectrum. Here,~$q^{(\prime)}$ 
denotes~$u$- and $d$-quarks and isospin symmetry is assumed. 
The brackets $[\ldots]$ and $\{\ldots\}$ indicate the antisymmetric ($S = 0$) 
and symmetric ($S = 1$) spin configurations of the quarks 
in the diquark, respectively. (Taken from~\cite{Karliner:2018bms} 
for light diquarks and from~\cite{Ali:2019roi} for the heavy-light ones).
}
\label{tab:diquark-masses}
\end{table} 

The spin-spin couplings,~$\mathcal{K}_{\bar Q Q^\prime}$ and~$(\mathcal{K}_{Q Q^\prime})_{\bar 3}$, 
extracted from the spectra of mesons and baryons, respectively, are presented 
in table~\ref{tab:spin-spin-couplings}. These values are taken from~\cite{Ali:2019roi}
except for $(\mathcal{K}_{c c})_{\bar 3}$, which is from~\cite{Karliner:2014gca}. 
The factor two difference in~$\mathcal{K}_{\bar c c}$ and~$(\mathcal{K}_{c c})_{\bar 3}$ 
reflects the fact that the one-gluon exchange assumption is used to relate these couplings. 
%  This value is close to the one [$(\mathcal{K}_{Q Q^\prime})_{\bar 3} = 59$~MeV] 
%  which was used in the previous analysis~\cite{Ali:2016dkf,Ali:2017ebb}. 
Concerning the $(\mathcal{K}_{c q})_{\bar 3}$-coupling, it follows from table~\ref{tab:spin-spin-couplings} 
that the value $(\mathcal{K}_{c q})_{\bar 3} = 15$~MeV extracted from the charm baryon spectroscopy   
is numerically substantially smaller than all others. In fact, this is the spin-spin interaction 
between the charm quark and the two quarks from a light diquark, but not between the quarks within the charm 
diquark $[cq]$, which is estimated as $(\mathcal{K}_{c q})_{\bar 3} = 67$~MeV~\cite{Maiani:2014aja,Ali:2019roi}. 
The later value was obtained from the Constituent Quark-Diquark Model analysis of the hidden-charm 
exotic mesons in which the underlying structure is assumed to be a charm diquark and charm antidiquark. 
In our numerical estimates, we use the later value.
The spin-spin coupling obtained from the charm baryons can be used to determine 
$(\tilde{\mathcal{K}}_{c q})_{\bar 3}$ in eq.~(\ref{eq:H-SS-t-d}), associated with the spin-spin 
interactions between the charm quark in the doubly-heavy triquark and quarks from the light diquark. 
The same is also true for the spin-spin couplings between the charm and strange quarks, 
for which the value $(\mathcal{K}_{c s})_{\bar 3} = 2$~MeV~\cite{Karliner:2017kfm,Ali:2017wsf}%
\footnote{This value differs by a factor of two from the original estimate presented 
in~\cite{Karliner:2017kfm,Ali:2017wsf} due to the coupling redefinition, i.\,e., 
$(\mathcal{K}_{c s})_{\bar 3} = c/2$.} 
follows from the analysis of the narrow orbitally-excited $\Omega_c^*$-baryons recently 
observed by the LHCb~\cite{Aaij:2017nav} and Belle~\cite{Yelton:2017qxg} collaborations. 
To understand the uncertainty of~$(\mathcal{K}_{c s})_{\bar 3}$, we performed the 
$\chi^2$-analysis based on the masses of these baryons~\cite{Aaij:2017nav,Yelton:2017qxg}, 
which yields $(\mathcal{K}_{c s})_{\bar 3} = (2.01 \pm 0.20)$~MeV. The details of this analysis 
are in appendix~\ref{sec:chi2-Omega-c}. This yields~10\% error on~$(\mathcal{K}_{c s})_{\bar 3}$, 
which we assume as an uncertainty on all the spin-spin couplings. 
\begin{table}[tb] 
\begin{center}
\begin{tabular}{|ccccccc|} 
\hline
                      Couplings (MeV) & $q q^\prime$ & $s q$ & $s s$ & $c q$ & $c s$ & $c c$ \\ \hline 
      $\mathcal{K}_{\bar Q Q^\prime}$ &          318 &   200 &   103 &    70 &    72 &   113 \\   
$(\mathcal{K}_{Q Q^\prime})_{\bar 3}$ &           98 &    59 &    23 &    15 &    50 &    57 \\ \hline  
%                                 Ratio &          3.2 &   3.4 &   4.5 &   4.7 &   1.6 &   2.0 \\ \hline   
\end{tabular} 
\end{center} 
\caption{
Spin-spin couplings,~$\mathcal{K}_{\bar Q Q^\prime}$ 
and~$(\mathcal{K}_{Q Q^\prime})_{\bar 3}$, extracted from the spectra 
of mesons and baryons, respectively (borrowed from~\cite{Ali:2019roi} 
and for $c c$ from~\cite{Karliner:2014gca}). Here,~$q^{(\prime)}$ 
denotes~$u$- and $d$-quarks and isospin symmetry is assumed. 
}
\label{tab:spin-spin-couplings}
\end{table} 
The pentaquark masses involve the ratios of the couplings~(\ref{eq:rhd-rld-def}) 
which for the input values are evaluated as: $r_{hd} = 0.73$ and~$r_{ld} = 1.07$. 
According to the Model~II, these ratios should be of order one which explicitly 
demonstrates that the strengths of all possible spin-spin interactions inside 
the diquarks and triquark are approximately the same.

In estimating the $P$-wave pentaquark mass spectrum, values of the couplings 
in the orbital angular momentum term and the spin-orbit ones are required. 
In the previous  analysis~\cite{Ali:2016dkf,Ali:2017ebb}, the spin-orbit 
coupling was taken to be $A_{\mathcal{P}} = 52$~MeV, which follows from the 
analysis of the vector hidden-charm $Y$-tetraquarks. Updated analysis of these 
states in~\cite{Ali:2017wsf}, which takes into account in addition the tensor interaction 
between the diquark and antidiquark, results in the smaller value, $a_Y = (22 \pm 3)$~MeV. 
A similar analysis~\cite{Karliner:2017kfm,Ali:2017wsf}, performed for the orbitally-excited 
$\Omega_c$-baryons, yields even smaller values,\footnote{
The original values ($a_1 = 26.95$~MeV and $a_2 = 25.75$~MeV) are twice the ones 
in the text because of different definitions in the effective Hamiltonian.}
$a_1 = (13.45 \pm 0.13)$~MeV and $a_2 = (12.94 \pm 0.36)$~MeV, for the spin-orbit 
interactions of the $\{s s\}$-diquark and charm antiquark, respectively. 
The uncertainties in these values result from the $\chi^2$-analysis of the 
$\Omega_c^*$-baryon masses~\cite{Aaij:2017nav,Yelton:2017qxg} and the details 
of this analysis are also presented in appendix~\ref{sec:chi2-Omega-c}. The errors 
in these couplings are typically a few percent and are substantially smaller than 
the relative error from the $Y$-tetraquark spectrum which is approximately~14\%. 
    
For $P$-wave pentaquarks, the spin-orbit couplings are responsible for the interactions 
of the triquark and the light diquark, denoted by~$A_t$ and~$A_{ld}$, respectively, which are, 
in general, different. However, based on the observation that the couplings of the spin-orbit 
interaction in $\Omega_c^*$-baryons are close to each other, $a_1 \simeq a_2 \simeq 13$~MeV, 
one can use the same approximate relation $A_t \simeq A_{ld} = 13$~MeV in the pentaquark system 
but, as we shall show later, this is not supported by the data on the newly observed pentaquark 
states~\cite{Aaij:2019vzc} for the plausible spin-parity assignment, as discussed in~\cite{Ali:2019npk}.

For the mass spectrum of orbitally-excited pentaquarks with the ``bad'' light 
diquark, the tensor coupling~$b$ is required. It was extracted from the spectrum 
of the orbitally-excited $\Omega_c$-baryons and $Y$-tetraquarks, yielding
$b_{\Omega_c} = 13.5$~MeV~\cite{Karliner:2017kfm,Ali:2017wsf}\footnote{
The $\chi^2$-analysis of the $\Omega_c^*$-baryons presented in appendix~\ref{sec:chi2-Omega-c} 
gives slightly smaller value $b_{\Omega_c} = (13.30 \pm 0.48)$~MeV.} 
and $b_Y = (- 136 \pm 6)$~MeV~\cite{Ali:2017wsf}, respectively.
Thus, this coupling is quite different in the charm baryons and in the hidden-charm 
tetraquarks and, moreover, the central values are of opposite signs in these hadrons. 
Here, more experimental input is needed. To be definite, 
we use $b = (13.3 \pm 0.5)$~MeV, which is obtained from the orbitally-excited 
$\Omega_c$-baryons rather than from the $Y$-tetraquarks. 

The last parameter to be specified is the orbital coupling~$B$, several 
estimates of which are available from different hadrons. Some representative   
values are: $B (\bar c c) = 457$~MeV~\cite{Ali:2017wsf}, obtained from charmonia, 
$B (\Omega_c) = 325$~MeV, from $\Omega_c$-baryons, $B (Y) = 362$~MeV 
and $B (Y) = 505$~MeV, from the hidden-charm $Y$-tetraquarks.\footnote{
The values presented are determined by two possible assignments 
of the vector tetraquarks discussed in~\cite{Ali:2017wsf}.} 
In the analysis~\cite{Ali:2016dkf,Ali:2017ebb}, the coupling~$B$ was derived from 
the experimentally observed mass of the $P_c (4450)^+$ having the (preferred) spin-parity 
assignment $J^P = 5/2^+$. The orbital 
coupling was determined from the $P_c (4450)^+$ mass to be $B (P_c) = 220$~MeV 
for the Model~II~\cite{Maiani:2014aja}.
With the observation of three narrow pentaquarks in the updated LHCb data, the experimental 
picture is changed drastically. Following the interpretation of these peaks as suggested 
in~\cite{Ali:2019npk}, which assumes that two resonances are orbitally-excited states and 
the lowest mass peak is the $S$-wave state, it is possible to determine the orbital 
coupling~$B$ phenomenologically, discussed in the next subsection.

\subsection{Predictions for the hidden-charm unflavored pentaquarks} 
\label{ssec:predictions-unflavored}

To get an estimate of the mass spectrum of the ground-state hidden-charm unflavored 
pentaquarks, it is enough to know the diquark and charm quark masses and spin-spin 
couplings, discussed in detail in the previous subsection. 
For the $P$-wave pentaquark masses, the values of the orbital angular momentum coupling~$B$, 
the spin-orbit couplings,~$A_t$ and~$A_{ld}$, and the tensor coupling~$b$ are required. 
Some of these couplings can be determined from the measured masses of the observed 
resonances~\cite{Aaij:2019vzc}. In this subsection we consider two possible assignments, 
both assume that the  heavier states, $P_c (4440)^+$ and $P_c (4457)^+$, are the $P$-wave 
pentaquarks, while the lowest mass $P_c (4312)^+$ can be either $S$- or $P$-wave state, 
and show that the second scenario is physically unacceptable. Assuming that
the $P_c (4312)^+$ pentaquark is an $S$-wave state, we predict the masses and 
$J^P$ quantum numbers of yet unmeasured resonances.  

We discuss first the  pentaquarks $P_c (4440)^+$ and $P_c (4457)^+$. 
As they replace the former narrow $P_c (4450)^+$ state with the preferred spin-parity 
$J^P = 5/2^+$, we tentatively assign this spin-parity to one of the observed 
two states, $P_c (4457)^+$. In this case, the lighter partner, $P_c (4440)^+$, 
most probably has the spin-parity $J^P = 3/2^+$.
In the Model~II by Maiani et al.~\cite{Maiani:2014aja}, the mass splitting 
of the two positive-parity states is related to the spin-orbit coupling~$A_t$ of the triquark: 
\begin{equation} 
M [P_c (4457)^+] - M [P_c (4440)^+] = \left ( 17^{+6.4}_{-4.5} \right )~\mathrm{MeV} = 5 A_t , 
\label{eq:Pc-76-mass-diff}
\end{equation}  
where the error is obtained by adding the experimental errors on the masses 
in quadrature. From~(\ref{eq:Pc-76-mass-diff}), the value of the coefficient~$A_t$ 
follows immediately: 
\begin{equation} 
A_t = \left ( 3.4^{+1.3}_{-0.9} \right )~\mathrm{MeV} .  
\label{eq:At-experiment}
\end{equation}  
It is not surprising that $A_t$ is found numerically small, as the doubly-heavy 
triquark is almost static. 

The third narrow state $P_c (4312)^+$ can have several $J^P$ assignments. Identifying 
it with the $J^P = 3/2^-$ state, one can work out the mass difference 
between this state and the heavier pentaquarks, $P_c (4440)^+$ and $P_c (4457)^+$. 
This is determined by the orbital~$B$ and the triquark spin-orbit~$A_t$ couplings. 
The strength of the latter is already known from the $P_c (4440)^+$ 
and $P_c (4457)^+$ mass splitting~(\ref{eq:At-experiment}), and the mass difference, 
say, between $P_c (4312)^+$ and $P_c (4457)^+$, allows us to read off the strength 
of the orbital interaction: 
\begin{equation}
M [P_c (4457)^+] - M [P_c (4312)^+] = \left ( 145.4^{+4.2}_{-7.1} \right )~\mathrm{MeV} = B + 3 A_t . 
\label{eq:Pc-73-mass-diff}
\end{equation}  
With~$A_t$ from~(\ref{eq:At-experiment}), we get $B = 135$~MeV.   
% 
% \begin{equation} 
% B = \left ( 135.2^{+5.0}_{-8.1} \right )~\mathrm{MeV} .  
% \label{eq:B-splitting-exp}
% \end{equation}  
% 
This is too small in comparison with the strengths of the orbital excitations in other 
hadrons~\cite{Ali:2017wsf}, and, in particular,  $B (\Omega_c) = 325$~MeV, obtained 
from the $\Omega_c^*$-baryons. Moreover, the theoretically predicted masses of the $P_c (4440)^+$ 
and $P_c (4457)^+$ states with the value of~$B \simeq 135$~MeV are found 
to be $\sim 70$~MeV below the experimental values. An error on~$B$ can be obtained by
performing the $\chi^2$-analysis of the experimental data. Assuming that the parameters~$M_0$, 
$B$, and~$A_t$ are free variables and all the spin-spin couplings are fixed, we obtain: 
\begin{equation} 
M_0 = \left ( 4333.9 \pm 3.9 \right )~\mathrm{MeV} , \quad 
B   = \left (  135.2 \pm 4.6 \right )~\mathrm{MeV} , \quad 
A_t = \left (    3.4 \pm 1.1 \right )~\mathrm{MeV} .  
\label{eq:fitting-exp}
\end{equation}  
The best-fit value of~$M_0$ comes out about $72$~MeV higher than the sum of the diquarks' 
and charm quark masses, $M_0 = 4262$~MeV.  

Alternatively, assuming that the spin-spin couplings and the constituent (quark and diquarks) 
masses are known, the strength of~$B$ can also be determined from the masses of $P_c (4440)^+$ 
and $P_c (4457)^+$ only, as follows:  
\begin{equation}
B = \frac{1}{5} \left \{ 3 M [P_c (4440)^+] + 2 M [P_c (4457)^+] \right \} 
- M_0 - \frac{1}{4} \left ( \mathcal{K}_{\bar c q} + \mathcal{K}_{\bar c c} \right ) 
\left ( 2 + r_{hd} - 3 r_{ld} \right ) . 
\label{eq:Pc-76-mass-sum}
\end{equation}  
With the values of the other parameters already assigned,  $B = 207$~MeV  
reproduces the masses of the observed $P_c (4440)^+$ and $P_c (4457)^+$ states,
shown in table~\ref{tab:masses-predictions-cbar-cq-qq} as the last two entries 
situated in the part labeled by $S_{ld} = 0$, $L = 1$ and inserted into the solid boxes.
This value of~$B$ is closer to the estimates in the hidden and open charm hadrons~\cite{Ali:2017wsf}. 
With these parameters, the mass of the third pentaquark $M = 4240$~MeV with $J^P = 3/2^-$, 
also shown in the solid box, is somewhat lower than the mass of the observed $P_c (4312)^+$ 
peak, but is still in the right ball-park. 

The second possibility is to assign the lowest mass state $P_c (4312)^+$ with the one having 
the spin-parity $J^P = 3/2^+$ and mass $M = 4360$~MeV, or with $J^P = 1/2^+$ and $M = 4351$~MeV. 
Both predictions are rather close to the observed pentaquark mass. 
In the case of $J^P = 1/2^+$ assignment for $P_c (4312)^+$, Model~II predicts 
for the mass difference: 
\begin{eqnarray} 
&& 
M [P_c (4440)^+] - M [P_c (4312)^+] = \left ( 128.4^{+4.4}_{-8.4} \right )~\mathrm{MeV} 
\nonumber \\ 
&& \hspace*{19mm}
= \frac{1}{2} \left ( \mathcal{K}_{\bar c q} + \mathcal{K}_{\bar c c} \right ) 
\left ( 2 + r_{hd} - \sqrt{3 + \left ( 1 - r_{hd} \right )^2} \right ) ,
\label{eq:Pc-64-mass-diff}
\end{eqnarray}  
which does not depend on~$B$ and~$A_t$ but is determined by the spin-spin couplings 
from the doubly-heavy triquark. If we fix the spin-spin coupling 
$(\mathcal{K}_{cq})_{\bar 3} = 67$~MeV, eq.~(\ref{eq:Pc-64-mass-diff}) can be solved 
analytically with the roots: 
\begin{equation} 
r_{hd}^{(1)} = 0, 
\qquad 
r_{hd}^{(2)} = \frac{2 \left ( 3 - 2 \rho \right )}{\rho \left ( 2 - \rho\right )} , 
\label{eq:rhd-solutions}
\end{equation} 
where $\rho = \{M [P_c (4440)^+] - M [P_c (4312)^+]\}/(\mathcal{K}_{cq})_{\bar 3}$. 
The first solution is unacceptable, as it implies 
$\mathcal{K}_{\bar c q} + \mathcal{K}_{\bar c c} \to \infty$.
The second solution in~(\ref{eq:rhd-solutions}) yields a negative and, hence, unphysical 
value for~$r_{hd}$, if the ratio~$\rho$ lies in the interval: $3/2 < \rho < 2$. 
With the pentaquark mass difference in~(\ref{eq:Pc-64-mass-diff}) 
and $(\mathcal{K}_{cq})_{\bar 3} = 67$~MeV, one obtains the value $\rho \simeq 1.92$, 
which is in the unphysical interval (the corresponding value 
of the spin-spin coupling ratio $r_{hd}^{(2)} \simeq - 10.4$). The situation does not 
improve if we assign $J^P = 3/2^+$ to the third state. In this case, it has a higher mass  
than the $J^P = 1/2^+$ pentaquark by $3 A_t \simeq 10$~MeV, thereby the 
r.h.s. of~(\ref{eq:Pc-64-mass-diff}) gets an additional factor $- 3 A_t$. There are again 
two solutions as in~(\ref{eq:rhd-solutions}) but in~$r_{hd}^{(2)}$ one should make 
the following replacement: 
$\rho \to \tilde\rho = \{M [P_c (4440)^+] - M [P_c (4312)^+] + 3 A_t\}/(\mathcal{K}_{cq})_{\bar 3}$. 
Numerically $\tilde\rho \simeq 2.07$, and with this value~$r_{hd}^{(2)}$ becomes 
positive but too large, $r_{hd}^{(2)} \simeq 16$, which is again unacceptable. 
So, the assignment of the third state, $P_c(4312)^+$, with the orbitally-excited 
pentaquark is not tenable in the Model~II.   

It should be noted that the assignment of all the three observed states with the $P$-wave pentaquarks 
was also discussed in Ref.~\cite{Stancu:2019qga}. In contrast to the one-gluon-exchange model, 
from which the effective Hamiltonian used in our paper follows, the effective Hamiltonian 
in~\cite{Stancu:2019qga} is based on a confining potential and an attractive flavor-spin interaction,
resulting from the underlying chiral dynamics. Under the assumption that the flavor-spin interaction 
is dominant, the three LHCb states are then identified with the lowest-mass orbitally-excited states. 
Moreover, in this model, the lowest $S$-wave state lies above the $P$-wave states. 
This obviously contradicts the predictions of our model discussed above.

We remark that the assignments we have discussed satisfy the heavy-quark symmetry, 
which means that the spin $S_{ld} = 0$ of the light diquark in $\Lambda_b$-baryon is conserved 
in the decay to the charged pentaquark and $K^-$-meson. That is why, only pentaquarks 
with the ``good'' light diquark are considered. As the heavy-quark symmetry is not 
exact, pentaquarks with $S_{ld} = 1$ can also be produced in the $\Lambda_b$-baryon decay, 
but being suppressed by $1/m_b$, they are rather unlikely.

Masses of the hidden-charm unflavored pentaquarks are presented 
in table~\ref{tab:masses-predictions-cbar-cq-qq},
and compared with the results obtained in~\cite{Ali:2016dkf,Ali:2017ebb}. 
For the $P$-wave pentaquarks, the values of the orbital, spin-orbit and 
tensor couplings are taken from table~\ref{tab:input-P-wave-pentaquarks}.
\begin{table}[tb] 
\begin{center}
\begin{tabular}{|ccc|ccc|} 
\hline
$J^P$ & \; This work \; & \; refs.~\cite{Ali:2016dkf,Ali:2017ebb} \; & 
$J^P$ & \; This work \; & \; refs.~\cite{Ali:2016dkf,Ali:2017ebb} \; \\ 
\hline 
\multicolumn{3}{|c|}{$S_{ld} = 0$, $L = 0$} & \multicolumn{3}{|c|}{$S_{ld} = 1$, $L = 1$} \\
$1/2^-$ & $3830 \pm 34$ & $4086 \pm 42$             & $1/2^+$ & $4144 \pm 37$ & $3970 \pm 50$ \\  
        & $4150 \pm 29$ & $4162 \pm 38$             &         & $4209  \pm 37$& $4174 \pm 44$ \\  
$3/2^-$ & \framebox{$4240 \pm 29$} & $4133 \pm 55$  &         & $4465 \pm 32$ & $4198 \pm 50$ \\ \cline{1-3} 
\multicolumn{3}{|c|}{$S_{ld} = 1$, $L = 0$}         &         & $4530 \pm 32$& $4221 \pm 40$ \\
$1/2^-$ & $4026 \pm 31$ & $4119 \pm 42$             &         & $4564 \pm 33$& $4240 \pm 50$ \\  
        & $4346 \pm 25$ & $4166 \pm 38$             &         & $4663 \pm 32$& $4319 \pm 43$ \\  
        & $4436 \pm 25$ & $4264 \pm 41$             & $3/2^+$ & $4187 \pm 37$ &               \\  
$3/2^-$ & $4026 \pm 31$ & $4072 \pm 40$             &         & $4250 \pm 37$ &               \\  
        & $4346 \pm 25$ & $4300 \pm 40$             &         & $4508 \pm 32$ &               \\  
        & $4436 \pm 25$ & $4342 \pm 40$             &         & $4570 \pm 32$ &               \\  
$5/2^-$ & $4436 \pm 25$ & $4409 \pm 40$             &         & $4511 \pm 33$ &               \\ \cline{1-3} 
\multicolumn{3}{|c|}{$S_{ld} = 0$, $L = 1$}         &         & $4566 \pm 32$&               \\
$1/2^+$ & $4030 \pm 39$ & $4030 \pm 62$             &         & $4656 \pm 32$ &               \\  
        & $4351 \pm 35$ & $4141 \pm 44$             & $5/2^+$ & $4260 \pm 37$ & $4450 \pm 44$ \\  
        & $4430 \pm 35$ & $4217 \pm 40$             &         & $4581 \pm 32$ & $4524 \pm 41$ \\  
$3/2^+$ & $4040 \pm 39$ &                           &         & $4601 \pm 32$& $4678 \pm 44$ \\  
        & $4361 \pm 35$ &                           &         & $4656 \pm 32$& $4720 \pm 44$ \\  
        & \framebox{$4440 \pm 35$} &                & $7/2^+$ & $4672\pm 32$ &               \\  
$5/2^+$ & \framebox{$4457 \pm 35$} & $4510 \pm 57$  &         &      &               \\ \hline 
\end{tabular}
\end{center}
\caption{
Masses of the hidden-charm unflavored pentaquarks (in MeV) and their 
comparison with the results presented in~\cite{Ali:2016dkf,Ali:2017ebb}. 
For the $P$-wave pentaquarks, the values of the orbital, spin-orbit and 
tensor couplings are taken from table~\ref{tab:input-P-wave-pentaquarks}.  
}
\label{tab:masses-predictions-cbar-cq-qq}
\end{table} 
The threshold for the observed pentaquarks in the $P_c^+ \to J/\psi + p$ decay mode 
is $M^{\rm thr}_{J/\psi\, p} = m_{J/\psi} + m_p = 4035.17$~MeV~\cite{Tanabashi:2018oca}. 
With the masses given in table~\ref{tab:masses-predictions-cbar-cq-qq}, 
there are two states with the ``good'' light diquark ($S_{ld} = 0$), the $J^P = 1/2^-$ state 
with the mass $M = 3830$~MeV and the $J^P = 1/2^+$ state with the mass $M = 4030$~MeV, which 
lie below the $M^{\rm thr}_{J/\psi\, p}$ threshold. Also, the third state having $J^P = 3/2^+$, 
with the mass $M = 4040$~MeV, may also lie below~$M^{\rm thr}_{J/\psi\, p}$. 
There are also two states, $J^P = 1/2^-$ and $J^P = 3/2^-$, with the ``bad'' light diquark 
($S_{ld} = 1$), having the degenerate masses $M = 4026$~MeV, which also lie below the 
$M^{\rm thr}_{J/\psi\, p}$ threshold. 
One of these states (with the mass $M = 3830$~MeV) is even below the threshold for the decay 
$P_c^+ \to \eta_c + p$ with $M^{\rm thr}_{\eta_c\, p} = m_{\eta_c} + m_p = 3922$~MeV~\cite{Tanabashi:2018oca}. 
They will decay through the annihilation of the $c \bar c$-pair into light hadrons or a lepton 
pair, and hence will be narrower the $P_c^+$-resonances observed. It is conceivable that 
a dedicated search of narrow states in the LHCb data with improved statistics may reveal the 
existence of these pentaquark states. It is interesting to note that the virtual $J/\psi$-meson 
decay will also lead to an off-shell dilepton pair. Hence, a search for a resonance in the 
$p\, \mu^+ \mu^-$ (as well as $p\, e^+ e^-$) may result into observing narrow structures.
In addition, the other states shown in table~\ref{tab:masses-predictions-cbar-cq-qq} 
are also reachable in the $\Lambda_b \to J/\psi\, p\, K^-$ decay, in particular, our model 
predicts a peak with the mass $M = (4150 \pm 29)$~MeV below the lowest-mass observed pentaquark, 
$P_c (4312)^+$, and the other peak with a mass near $4350$~MeV, being a combination of two 
$J^P = 1/2^+$ and $J^P = 3/2^+$ pentaquarks with close masses. The measured spectrum by the 
LHCb collaboration~\cite{Aaij:2019vzc} has indications for such states, though more data are 
needed for a definite conclusion.

\begin{table}[tb] 
\begin{center}
\begin{tabular}{|cccc|} 
\hline
     $B$     &     $A_t$     &    $A_{ld}$    &      $b$       \\ \hline
$207 \pm 20$ & $3.4 \pm 1.1$ & $13.5 \pm 0.4$ & $13.3 \pm 0.5$ \\ \hline
\end{tabular} 
\end{center} 
\caption{
Input values (in MeV) for the couplings required for the mass 
estimations of the $P$-wave pentaquarks. For the coupling~$B$, 
the error is taken from~(\ref{eq:B-Omega-c-extimate}). 
}
\label{tab:input-P-wave-pentaquarks}
\end{table}

\subsection{Comparisons with the dynamical diquark model} 
\label{ssec:dynamical-diquark-model}

Before presenting the strange pentaquark spectrum, we compare some of our estimates in
the non-strange sector with those obtained in the dynamical diquark model~\cite{Giron:2019bcs}.

The common feature of our model and the dynamical diquark model is the assumption of point-like 
diquarks in the color-antitriplet representation, with the pentaquarks built from a diquark and 
a color-triplet triquark. However, the two approaches differ in their dynamical details and 
in assigning the flavors of the quarks in the diquarks and triquarks.
In our model, we assume that the light diquark $[q^\prime q^{\prime\prime}]_{\bar 3}$ 
is bound to the doubly-heavy triquark $\left [ \bar c_{\bar 3} \left [ c q \right ]_{\bar 3} \right ]_3$ 
and the orbital excitations in the pentaquark are due to the light-diquark orbital excitations
(the doubly-heavy triquark is assumed to be in the $S$-wave). In the dynamical diquark 
model~\cite{Giron:2019bcs}, both the diquark $\left [ c q \right ]_{\bar 3}$ and the triquark 
$\left [ \bar c_{\bar 3} \left [ q^\prime q^{\prime\prime} \right ]_{\bar 3} \right ]_3$ 
are heavy objects, and orbital excitations are between them~\cite{Giron:2019bcs}. 
More importantly, in our analysis we take into account all sizable spin-spin, spin-orbit, orbital 
and tensor interactions between the charm antiquark and the diquarks, except for the spin-spin 
interaction between the constituents of the doubly-heavy triquark and the light diquark, which 
are assumed to be strongly suppressed. All these interactions are omitted in the current
formulation of the dynamical diquark model and left for further study~\cite{Giron:2019bcs}. 
So, a lot of pentaquark states which differ in their masses in our model (see, for example, 
tables~\ref{tab:S-wave-pentaquarks-good-ld} and~\ref{tab:P-wave-pentaquarks-good-ld}) are mass 
degenerate in the dynamical diquark model, which makes a numerical comparison rather difficult. 

We also differ in the $J^P$ assignments of the states. In the dynamical diquark 
model~\cite{Giron:2019bcs}, two possible assignments are suggested: The first one identifies 
the two pentaquark states with larger masses, $P_c (4440)^+$ and $P_c (4457)^+$, with the states 
from the first orbitally-excited multiplet called~$\Sigma_g^+ (1P)$ (see table~7 in~\cite{Giron:2019bcs}). 
With this assignment, there is no place in the mass spectrum for the third narrow state, $P_c (4312)^+$, 
observed recently by the LHCb collaboration. This is not surprising, as also remarked in~\cite{Giron:2019bcs},
since the strength of the orbital excitation in this assignment is about $\sim 400$~MeV for the states 
with the principle quantum number $n = 1$, in accord with the corresponding excitation energy in charmonia. 
The observed mass splitting $\sim 140$~MeV is, therefore, impossible to accommodate with these assignment. 
So, the second possibility discussed in~\cite{Giron:2019bcs} is to put the lowest-mass state, $P_c (4312)^+$, 
in the first orbitally-excited multiplet~$\Sigma_g^+ (1P)$, and in this case the two other states~$P_c (4440)^+$ 
and~$P_c (4457)^+$ are the members of the radially excited multiplet~$\Sigma_g^+ (2S)$. 
The suggested spin-parity of the $P_c (4380)^+$-state is $J^P = 5/2^+$ (this means that the bottom 
of the $\Sigma_g^+ (1P)$-multiplet is fixed to the mass of the $P_c (4312)^+$-pentaquark) while one 
of the~$P_c (4440)^+$- and~$P_c (4457)^+$-states has $J^P = 3/2^-$. 
In our model, all three observed pentaquark states have the principle quantum number $n = 1$ and 
have the indicated parities: $J^P = 3/2^-$ for $P_c (4312)^+$ and $J^P = 3/2^+$ and $J^P = 5/2^+$ 
for~$P_c (4440)^+$ and~$P_c (4457)^+$, respectively. We are able to explain quantitatively the mass 
splitting in the higher mass pentaquarks by the spin-orbit interaction, and the difference between 
the lower- and higher-mass states, to a large extent, by the orbital interaction. The strength 
of the latter interaction was determined as $\sim 200$~MeV, which is smaller by $\sim 100$~MeV 
compared to charmonia, but still acceptable as the size of the pentaquark and, hence, the distance 
from the doubly-heavy triquark to the light diquark, is anticipated to be significantly larger 
than the hadronic size of the charmonium system. 

Thus, despite the similarity in the general approach about the constituents (colored diquark and 
triquark) of the pentaquarks, the two model are different in detail and lead to different spectra. 
Both models require estimates of the diquark-size effects on the spectroscopy, a point also
discussed later in this paper.

\begin{table}[tb] 
\begin{center}
\begin{tabular}{|ccc|ccc|} 
\hline
$J^P$ & \; This work \; & \; refs.~\cite{Ali:2016dkf,Ali:2017ebb} \; & 
$J^P$ & \; This work \; & \; refs.~\cite{Ali:2016dkf,Ali:2017ebb} \; \\ 
\hline 
\multicolumn{3}{|c|}{$S_{ld} = 0$, $L = 0$} & \multicolumn{3}{|c|}{$S_{ld} = 1$, $L = 1$} \\
$1/2^-$ & $3961 \pm 34$ & $4318 \pm 42$             & $1/2^+$ & $4275 \pm 37$ & $4202 \pm 50$ \\  
        & $4292 \pm 29$ & $4392 \pm 38$             &         & $4341 \pm 37$ & $4406 \pm 44$ \\  
$3/2^-$ & $4362 \pm 29$ & $4365 \pm 55$             &         & $4607 \pm 32$ & $4430 \pm 50$ \\ \cline{1-3} 
\multicolumn{3}{|c|}{$S_{ld} = 1$, $L = 0$}         &         & $4672 \pm 32$ & $4453 \pm 40$ \\
$1/2^-$ & $4157 \pm 31$ & $4351 \pm 42$             &         & $4685 \pm 33$ & $4472 \pm 50$ \\  
        & $4488 \pm 25$ & $4398 \pm 38$             &         & $4784 \pm 32$ & $4551 \pm 43$ \\  
        & $4558 \pm 25$ & $4496 \pm 41$             & $3/2^+$ & $4319 \pm 37$ &               \\  
$3/2^-$ & $4157 \pm 31$ & $4304 \pm 55$             &         & $4381 \pm 37$ &               \\  
        & $4488 \pm 25$ & $4532 \pm 40$             &         & $4650 \pm 32$ &               \\  
        & $4558 \pm 25$ & $4574 \pm 40$             &         & $4712 \pm 32$ &               \\  
$5/2^-$ & $4558 \pm 25$ & $4641 \pm 40$             &         & $4633 \pm 33$ &               \\ \cline{1-3} 
\multicolumn{3}{|c|}{$S_{ld} = 0$, $L = 1$}         &         & $4687 \pm 32$ &               \\
$1/2^+$ & $4161 \pm 39$ & $4262 \pm 63$             &         & $4778 \pm 32$ &               \\  
        & $4492 \pm 35$ & $4373 \pm 44$             & $5/2^+$ & $4391 \pm 37$ & $4682 \pm 57$ \\  
        & $4552 \pm 35$ & $4449 \pm 40$             &         & $4723 \pm 32$ & $4756 \pm 41$ \\  
$3/2^+$ & $4171 \pm 39$ &                           &         & $4723 \pm 32$ & $4910 \pm 44$ \\  
        & $4503 \pm 35$ &                           &         & $4777 \pm 32$ & $4952 \pm 44$ \\  
        & $4562 \pm 35$ &                           & $7/2^+$ & $4794 \pm 32$ &               \\  
$5/2^+$ & $4579 \pm 35$ & $4742 \pm 57$             &         &               &               \\ \hline 
\end{tabular}
\end{center}
\caption{
Masses of the hidden-charm strange pentaquarks (in MeV) with a strange-heavy diquark, 
i.\,e. having the structure $(\bar c_{\bar 3} [c s]_{\bar 3} [q q^\prime]_{\bar 3})$, 
where $q^{(\prime)}$ is $u$- or $d$-quark, and their comparison with the results 
presented in~\cite{Ali:2016dkf,Ali:2017ebb}. 
For the $P$-wave pentaquarks, the values of the orbital, spin-orbit and 
tensor couplings are taken from table~\ref{tab:input-P-wave-pentaquarks}.  
}
\label{tab:masses-predictions-cbar-cs-qq}
\end{table} 

\begin{table}[tb] 
\begin{center}
\begin{tabular}{|ccc|ccc|} 
\hline
$J^P$ & \; This work \; & \; refs.~\cite{Ali:2016dkf,Ali:2017ebb} \; & 
$J^P$ & \; This work \; & \; refs.~\cite{Ali:2016dkf,Ali:2017ebb} \; \\ 
\hline 
\multicolumn{3}{|c|}{$S_{ld} = 0$, $L = 0$} & \multicolumn{3}{|c|}{$S_{ld} = 1$, $L = 1$} \\
$1/2^-$ & $4112 \pm 32$ & $4094 \pm 44$             & $1/2^+$ & $4348 \pm 36$ & $3929 \pm 53$ \\  
        & $4433 \pm 26$ & $4132 \pm 43$             &         & $4414 \pm 36$ & $4183 \pm 45$ \\  
$3/2^-$ & \framebox{$4523 \pm 26$} & $4172 \pm 47$  &         & $4669 \pm 32$ & $4159 \pm 53$ \\ \cline{1-3} 
\multicolumn{3}{|c|}{$S_{ld} = 1$, $L = 0$}         &         & $4735 \pm 32$ & $4189 \pm 44$ \\
$1/2^-$ & $4230 \pm 30$ & $4128 \pm 44$             &         & $4768 \pm 32$ & $4201 \pm 53$ \\  
        & $4551 \pm 25$ & $4134 \pm 42$             &         & $4867 \pm 32$ & $4275 \pm 45$ \\  
        & $4641 \pm 25$ & $4220 \pm 43$             & $3/2^+$ & $4392 \pm 36$ &               \\  
$3/2^-$ & $4230 \pm 30$ & $4031 \pm 43$             &         & $4454 \pm 36$ &               \\  
        & $4551 \pm 25$ & $4262 \pm 43$             &         & $4713 \pm 32$ &               \\  
        & $4641 \pm 25$ & $4303 \pm 43$             &         & $4775 \pm 32$ &               \\  
$5/2^-$ & $4641 \pm 25$ & $4370 \pm 43$             &         & $4716 \pm 32$ &               \\ \cline{1-3} 
\multicolumn{3}{|c|}{$S_{ld} = 0$, $L = 1$}         &         & $4770 \pm 32$ &               \\
$1/2^+$ & $4312 \pm 37$ & $4069 \pm 56$             &         & $4861 \pm 32$ &               \\  
        & $4633 \pm 33$ & $4149 \pm 45$             & $5/2^+$ & $4465 \pm 36$ & $4409 \pm 47$ \\  
        & $4713 \pm 33$ & $4187 \pm 44$             &         & $4786 \pm 32$ & $4486 \pm 45$ \\  
$3/2^+$ & $4323 \pm 37$ &                           &         & $4806 \pm 32$ & $4639 \pm 47$ \\  
        & $4643 \pm 33$ &                           &         & $4860 \pm 32$ & $4681 \pm 47$ \\  
        & \framebox{$4723 \pm 33$} &                & $7/2^+$ & $4877 \pm 32$ &               \\  
$5/2^+$ & \framebox{$4740 \pm 33$} & $4549 \pm 51$  &         &               &               \\ \hline 
\end{tabular}
\end{center} 
\caption{
Masses of the hidden-charm strange pentaquarks (in MeV) with a strange-light diquark, 
i.\,e. having the structure $(\bar c_{\bar 3} [c q]_{\bar 3} [s q^\prime]_{\bar 3})$, 
where $q^{(\prime)}$ is $u$- or $d$-quark, and their comparison with the results 
presented in~\cite{Ali:2016dkf,Ali:2017ebb}. 
For the $P$-wave pentaquarks, the values of the orbital, spin-orbit and 
tensor couplings are taken from table~\ref{tab:input-P-wave-pentaquarks}.  
}
\label{tab:masses-predictions-cbar-cq-sq}
\end{table} 

\subsection{Mass predictions for the strange hidden-charm pentaquarks} 
\label{ssec:predictions-strange}

Including the $s$-quark, the hidden-charm pentaquark spectrum becomes much richer.
Depending on the strange content of the pentaquarks, they can be classified  
into singly-, doubly-, and triple-strange pentaquarks and their $SU (3)_F$-multiplets 
are in one to one correspondence with the ordinary and strange baryons~\cite{Ali:2016dkf}. 
The mass estimates given below assume mass-degenerate $u$- and $d$-quarks. The constituent 
masses of the quarks are accepted as specified in subsec.~\ref{ssec:input-parameters}.   
The diquark masses and spin-spin couplings are not $SU (3)_F$-invariant, and their breaking 
is taken into account (see tables~\ref{tab:diquark-masses} and~\ref{tab:spin-spin-couplings}). 
For the $P$-wave pentaquarks, the orbital, spin-orbit and tensor couplings are assumed equal 
for all three light quarks and are taken from table~\ref{tab:input-P-wave-pentaquarks}.   

The masses of the singly-strange pentaquarks are presented 
in tables~\ref{tab:masses-predictions-cbar-cs-qq}
and~\ref{tab:masses-predictions-cbar-cq-sq} and compared with the results 
of~\cite{Ali:2016dkf,Ali:2017ebb}. The difference between the entries in the two tables 
lies in the $s$-quark attachment either to the heavy diquark or to the light diquark, 
i.\,e., the pentaquark structures are $(\bar c_{\bar 3} [c s]_{\bar 3} [q q^\prime]_{\bar 3})$ 
and $(\bar c_{\bar 3} [c q]_{\bar 3} [s q^\prime]_{\bar 3})$, respectively, with~$q^{(\prime)}$ 
being $u$- or $d$-quark. One should also remember that the states presented in 
tables~\ref{tab:masses-predictions-cbar-cs-qq} and~\ref{tab:masses-predictions-cbar-cq-sq}
with the same set of quantum numbers: spin-parity~$J^P$ and isospin~$I$, having strangeness $S = -1$, 
are experimentally indistinguishable and, hence, mix with each other. 
Thus, the mass eigenstates, resulting from this mixing, and the weak interaction eigenstates, 
in which the decay of the $\Xi_b$-baryon would lead dominantly to the states with the internal 
quantum number $(\bar c_{\bar 3} [c q]_{\bar 3} [s q^\prime]_{\bar 3})$, are different. 
In the present analysis, we have not taken into account this mixing.

As for the discovery modes of such pentaquarks in weak decays of bottom baryons, they were 
studied in~\cite{Ali:2016dkf} in the heavy-quark symmetry limit, assuming
$SU (3)_F$-symmetry. Both Cabibbo-allowed $b \to c \bar c s$ (with $\Delta I = 0$, 
$\Delta S = -1$) and Cabibbo-suppressed $b \to c \bar c d$ (with $\Delta I = 1/2$, 
$\Delta S = 0$) transitions were considered. Among the possible modes of the weakly-decaying 
$b$-baryons, the most promising modes in searching for the hidden-charm singly-strange 
pentaquarks at the LHC are: $\Xi_b^- \to P_\Lambda^0\, K^- \to J/\psi\, \Lambda^0\, K^-$ 
and $\Xi_b^{0,-} \to P_\Sigma^{+,0}\, K^- \to J/\psi\, \Sigma^{+,0}\, K^-$, where the symbol 
in the subscript on the pentaquarks denotes its light quark content, expressed as 
the corresponding baryon in the $SU (3)_F$-multiplet. 

Assuming a similarity in the unflavored and strange pentaquark production, we anticipate 
the observation of the strange partners of the known pentaquarks. We have marked the masses 
of these states by enclosing them in solid boxes in table~\ref{tab:masses-predictions-cbar-cq-sq}. 
The threshold for the strange pentaquarks in the $P_\Lambda^0 \to J/\psi\, \Lambda$ decay mode is 
$M^{\rm thr}_{J/\psi\, \Lambda} = m_{J/\psi} + m_\Lambda = 4212.58$~MeV~\cite{Tanabashi:2018oca}. 
With the masses given in table~\ref{tab:masses-predictions-cbar-cq-sq}, 
there is only one state with the ``good'' light diquark ($S_{ld} = 0$), the $J^P = 1/2^-$ with the 
mass $M = (4112 \pm 32)$~MeV, which lies $100$~MeV below the $M^{\rm thr}_{J/\psi\, \Lambda}$ threshold. 
This will result into a state narrower than one should expect due to a strong decay, with 
the $c \bar c$-pair annihilating into light hadrons or a lepton pair. This makes a good case 
for a search of narrow $\Lambda\, \mu^+ \mu^-$ (and $\Lambda\, e^+ e^-$) resonant structure.
The thresholds in other two modes $P_\Sigma^{+,0} \to J/\psi\, \Sigma^{+,0}$ are higher: 
$M^{\rm thr}_{J/\psi\, \Sigma^+} = m_{J/\psi} + m_{\Sigma^+} = 4286.27$~MeV and 
$M^{\rm thr}_{J/\psi\, \Sigma^0} = m_{J/\psi} + m_{\Sigma^0} = 4289.54$~MeV~\cite{Tanabashi:2018oca}, 
and there are two mass-degenerate states with the mass $M = 4230$~MeV and spin-parities 
$J^P = 1/2^-$ and $J^P = 3/2^-$ which are below these thresholds. One should remember that 
these states contain the ``bad'' light diquark and their production in the $\Xi_b$-baryon 
decays is suppressed by heavy quark symmetry.

\begin{table}[tb] 
\begin{center}
\begin{tabular}{|ccc|ccc|} 
\hline
$J^P$ & \; This work \; & \; refs.~\cite{Ali:2016dkf,Ali:2017ebb} \; & 
$J^P$ & \; This work \; & \; refs.~\cite{Ali:2016dkf,Ali:2017ebb} \; \\ 
\hline 
\multicolumn{3}{|c|}{$S_{ld} = 0$, $L = 0$} & \multicolumn{3}{|c|}{$S_{ld} = 1$, $L = 1$} \\
$1/2^-$ & $4243 \pm 32$ & $4326 \pm 44$             & $1/2^+$ & $4480 \pm 37$ & $4161 \pm 53$ \\  
        & $4575 \pm 26$ & $4364 \pm 43$             &         & $4545 \pm 37$ & $4415 \pm 45$ \\  
$3/2^-$ & $4644 \pm 26$ & $4404 \pm 47$             &         & $4811 \pm 32$ & $4391 \pm 53$ \\ \cline{1-3} 
\multicolumn{3}{|c|}{$S_{ld} = 1$, $L = 0$}         &         & $4877 \pm 32$ & $4421 \pm 44$ \\
$1/2^-$ & $4361 \pm 31$ & $4360 \pm 43$             &         & $4890 \pm 32$ & $4433 \pm 52$ \\  
        & $4693 \pm 25$ & $4366 \pm 43$             &         & $4989 \pm 32$ & $4507 \pm 45$ \\  
        & $4762 \pm 25$ & $4452 \pm 43$             & $3/2^+$ & $4523 \pm 37$ &               \\  
$3/2^-$ & $4361 \pm 31$ & $4263 \pm 43$             &         & $4585 \pm 37$ &               \\  
        & $4693 \pm 25$ & $4494 \pm 43$             &         & $4855 \pm 32$ &               \\  
        & $4762 \pm 25$ & $4535 \pm 43$             &         & $4917 \pm 32$ &               \\  
$5/2^-$ & $4762 \pm 25$ & $4602 \pm 43$             &         & $4837 \pm 32$ &               \\ \cline{1-3} 
\multicolumn{3}{|c|}{$S_{ld} = 0$, $L = 1$}         &         & $4892 \pm 32$ &               \\
$1/2^+$ & $4443 \pm 38$ & $4301 \pm 56$             &         & $4982 \pm 32$ &               \\  
        & $4775 \pm 33$ & $4381 \pm 45$             & $5/2^+$ & $4596 \pm 37$ & $4641 \pm 47$ \\  
        & $4834 \pm 33$ & $4419 \pm 44$             &         & $4928 \pm 32$ & $4718 \pm 45$ \\  
$3/2^+$ & $4454 \pm 38$ &                           &         & $4927 \pm 32$ & $4871 \pm 47$ \\  
        & $4785 \pm 33$ &                           &         & $4982 \pm 32$ & $4913 \pm 47$ \\  
        & $4844 \pm 33$ &                           & $7/2^+$ & $4998 \pm 32$ &               \\  
$5/2^+$ & $4861 \pm 33$ & $4781 \pm 51$             &         &               &               \\ \hline 
\end{tabular}
\end{center}
\caption{
Masses of the hidden-charm doubly-strange pentaquarks (in MeV) with both the heavy and light 
strange diquarks, i.\,e. having the structure $(\bar c_{\bar 3} [c s]_{\bar 3} [s q]_{\bar 3})$, 
where $q$ is $u$- or $d$-quark, and their comparison 
with the results presented in~\cite{Ali:2016dkf,Ali:2017ebb}. 
For the $P$-wave pentaquarks, the values of the orbital, spin-orbit and 
tensor couplings are taken from table~\ref{tab:input-P-wave-pentaquarks}.  
}
\label{tab:masses-predictions-cbar-cs-sq}
\end{table} 

\begin{table}[tb] 
\begin{center}
\begin{tabular}{|ccc|ccc|} 
\hline
$J^P$ & \; This work \; & \; refs.~\cite{Ali:2016dkf,Ali:2017ebb} \; & 
$J^P$ & \; This work \; & \; refs.~\cite{Ali:2016dkf,Ali:2017ebb} \; \\ 
\hline 
\multicolumn{3}{|c|}{$S_{ld} = 1$, $L = 0$} & \multicolumn{3}{|c|}{$S_{ld} = 1$, $L = 1$} \\
$1/2^-$ & $4511 \pm 30$ & $4598 \pm 44$             & $3/2^+$ & $4673 \pm 36$ &               \\   
        & $4832 \pm 25$ & $4666 \pm 43$             &         & $4735 \pm 36$ &               \\   
        & $4922 \pm 25$ & $4775 \pm 44$             &         & $4994 \pm 32$ &               \\  
$3/2^-$ & $4511 \pm 30$ & $4577 \pm 43$             &         & $5056 \pm 32$ &               \\  
        & $4832 \pm 25$ & $4810 \pm 43$             &         & $4997 \pm 32$ &               \\  
        & $4922 \pm 25$ & $4851 \pm 43$             &         & $5051 \pm 32$ &               \\
$5/2^-$ & $4922 \pm 25$ & $4918 \pm 47$             &         & $5142 \pm 32$ &               \\ \cline{1-3} 
\multicolumn{3}{|c|}{$S_{ld} = 1$, $L = 1$}         & $5/2^+$ & $4746 \pm 36$ & $4954 \pm 47$ \\ 
$1/2^+$ & $4629 \pm 36$ & $4474 \pm 53$             &         & $5067 \pm 32$ & $5033 \pm 47$ \\  
        & $4695 \pm 36$ & $4653 \pm 45$             &         & $5087 \pm 32$ & $5187 \pm 47$ \\  
        & $4950 \pm 32$ & $4707 \pm 53$             &         & $5141 \pm 32$ & $5228 \pm 47$ \\  
        & $5016 \pm 32$ & $4721 \pm 44$             & $7/2^+$ & $5158 \pm 32$ &               \\  
        & $5049 \pm 32$ & $4748 \pm 52$             &         &               &               \\ 
        & $5148 \pm 32$ & $4830 \pm 45$             &         &               &               \\ \hline 
\end{tabular}
\end{center}
\caption{
Masses of the hidden-charm doubly-strange pentaquarks (in MeV) with the doubly-strange 
light diquark, i.\,e. having the structure $(\bar c_{\bar 3} [c q]_{\bar 3} \{s s\}_{\bar 3})$, 
where $q$ is $u$- or $d$-quark and curly brackets denote the spin-1 light diquark, 
and their comparison with the results presented in~\cite{Ali:2016dkf,Ali:2017ebb}. 
For the $P$-wave pentaquarks, the values of the orbital, spin-orbit and 
tensor couplings are taken from table~\ref{tab:input-P-wave-pentaquarks}.  
}
\label{tab:masses-predictions-cbar-cq-ss}
\end{table} 

The masses of doubly-strange pentaquarks are presented in tables~\ref{tab:masses-predictions-cbar-cs-sq}
and~\ref{tab:masses-predictions-cbar-cq-ss} and compared with the results 
of~\cite{Ali:2016dkf,Ali:2017ebb}. The difference between the entries in these tables is as follows: 
the states from table~\ref{tab:masses-predictions-cbar-cs-sq} contain both the strange heavy 
diquark and the strange light diquark, while those from table~\ref{tab:masses-predictions-cbar-cq-ss} 
have hidden-charm unflavored triquark and doubly-strange light diquark, i.\,e., 
the pentaquark structures are $(\bar c_{\bar 3} [c s]_{\bar 3} [s q]_{\bar 3})$ and 
$(\bar c_{\bar 3} [c q]_{\bar 3} \{s s\}_{\bar 3})$, respectively, with~$q$ being $u$- 
or $d$-quark. The number of states with the doubly-strange light diquark is smaller\footnote{ 
In refs.~\cite{Ali:2016dkf,Ali:2017ebb} similar analysis was done for the hidden-charm 
doubly-strange pentaquarks and mass predictions for the states with the doubly-strange 
``good'' light diquark were presented erroneously. The corresponding entries should be 
excluded from tables~V and~VI (the last values in the columns entitled~$\mathcal{P}_{X_2}$ 
and~$\mathcal{P}_{Y_2}$, respectively) in~\cite{Ali:2016dkf} and from tables~V (the last 
values in the columns entitled~$\mathcal{P}_{X_1}$ and~$\mathcal{P}_{X_3}$),~VI (similar 
values in the columns~$\mathcal{P}_{Y_1}$ and~$\mathcal{P}_{Y_3}$), and~VII (the last 
value in the columns entitled~$\mathcal{P}_{Y_7}$) in~\cite{Ali:2017ebb}.
}   
because such a diquark, being flavor-symmetric, can be only the ``bad'' diquark with 
the spin $S_{ld} = 1$. So, this means that the pentaquarks with the ``good'' light 
diquark from table~\ref{tab:masses-predictions-cbar-cs-sq} remain unmixed while  
the states with the ``bad'' light diquark mix, in general, with the corresponding 
states (with the same set of quantum numbers: spin-parity~$J^P$ and isospin~$I$) from 
table~\ref{tab:masses-predictions-cbar-cq-ss}. In this paper, we neglect this mixing.

Based on the analysis in~\cite{Ali:2016dkf}, we specify the most promising decay mode 
of the $\Omega_b$-baryon for searching of hidden-charm doubly-strange pentaquarks 
at the LHC as $\Omega_b^- \to P_{\Xi_{10}}^0\, K^- \to J/\psi\, \Xi^{\prime 0} \, K^-$.    
The threshold for the doubly-strange pentaquarks in the $P_{\Xi_{10}}^0 \to J/\psi\, \Xi^{\prime 0}$ 
decay mode is $M^{\rm thr}_{J/\psi\, \Xi^{\prime 0}} = m_{J/\psi} + m_{\Xi^{\prime 0}} = 
(4628.70 \pm 0.32)$~MeV~\cite{Tanabashi:2018oca}. 
With the masses given in table~\ref{tab:masses-predictions-cbar-cq-ss}, 
there are two mass degenerate states, having $J^P = 1/2^-$ and $J^P = 3/2^-$,   
with the mass $M = 4511$~MeV, which lie below the $M^{\rm thr}_{J/\psi\, \Xi^{\prime 0}}$ 
threshold. Also, the third orbitally-excited state with the spin-parity $J^P = 1/2^+$ 
and the mass $M = 4629$~MeV could also lie below this threshold. All the other states 
presented in table~\ref{tab:masses-predictions-cbar-cq-ss} can be produced in the 
$\Omega_b$-baryon decays and their observation will herald the era of the as yet 
unknown field of doubly-strange pentaquarks.

The masses of the triple-strange pentaquarks are presented 
in table~\ref{tab:masses-predictions-cbar-cs-ss}. 
As these pentaquarks do not contain $u$- or $d$-quarks, they are isospin singlets. 
All these states have the doubly-strange ``bad'' light diquark only, i.\,e., 
the pentaquark structure is $(\bar c_{\bar 3} [c s]_{\bar 3} \{s s\}_{\bar 3})$. 
They can be produced at the LHC in the weak decays of the $\Omega_b$-baryons, 
in particular, due to the $b \to c \bar c s$ transition which gives rise to the quasi-two-body 
decay mode $\Omega_b^- \to P_{\Omega_{10}}^-\, \phi \to J/\psi\, \Omega^-\, \phi$. 
The threshold for the triple-strange pentaquarks in the $P_{\Omega_{10}}^- \to J/\psi\, \Omega^-$ 
decay is $M^{\rm thr}_{J/\psi\, \Omega^-} = m_{J/\psi} + m_{\Omega^-} = 
(4769.35 \pm 0.29)$~MeV~\cite{Tanabashi:2018oca}. With the masses given in 
table~\ref{tab:masses-predictions-cbar-cs-ss}, there are two mass degenerate states, 
having $J^P = 1/2^-$ and $J^P = 3/2^-$, with the mass $M = (4642 \pm 31)$~MeV, which lie $\sim 100$~MeV 
below the $M^{\rm thr}_{J/\psi\, \Omega^-}$ threshold. Also, the third orbitally-excited 
state with the spin-parity $J^P = 1/2^+$ and the mass $M = (4761 \pm 37)$~MeV could also lie below 
this threshold. All the other states presented in table~\ref{tab:masses-predictions-cbar-cs-ss} 
can be produced in the $\Omega_b$-baryon decays. The other possibility is to produce such 
pentaquarks promptly, but their production is highly suppressed due to their triple-strange content.

\begin{table}[tb] 
\begin{center} 
\begin{tabular}{|cc|cc|} 
\hline
$J^P$ & \; Mass \; & $J^P$ & \; Mass \; \\ 
\hline 
\multicolumn{2}{|c|}{$S_{ld} = 1$, $L = 0$} & \multicolumn{2}{|c|}{$S_{ld} = 1$, $L = 1$} \\
$1/2^-$ & $4642 \pm 31$                            & $3/2^+$ & $4804 \pm 37$           \\   
        & $4974 \pm 25$                            &         & $4866 \pm 37$           \\   
        & $5043 \pm 25$                            &         & $5136 \pm 32$           \\  
$3/2^-$ & $4642 \pm 31$                            &         & $5198 \pm 32$           \\  
        & $4974 \pm 25$                            &         & $5118 \pm 32$           \\  
        & $5043 \pm 25$                            &         & $5173 \pm 32$           \\
$5/2^-$ & $5043 \pm 25$                            &         & $5263 \pm 32$           \\ \cline{1-2} 
\multicolumn{2}{|c|}{$S_{ld} = 1$, $L = 1$}        & $5/2^+$ & $4877 \pm 37$           \\ 
$1/2^+$ & $4761 \pm 37$                            &         & $5209 \pm 32$           \\  
        & $4826 \pm 37$                            &         & $5208 \pm 32$           \\  
        & $5092 \pm 32$                            &         & $5263 \pm 32$           \\  
        & $5158 \pm 32$                            & $7/2^+$ & $5279 \pm 32$           \\  
        & $5171 \pm 32$                            &         &                         \\ 
        & $5270 \pm 32$                            &         &                         \\ \hline 
\end{tabular}
\end{center}
\caption{
Masses of the hidden-charm triple-strange pentaquarks (in MeV) 
containing the strange heavy diquark and doubly-strange light diquark, 
i.\,e. having the structure $(\bar c_{\bar 3} [c s]_{\bar 3} \{s s\}_{\bar 3})$, 
where curly brackets denote the spin-1 light diquark. 
For the $P$-wave pentaquarks, the values of the orbital, spin-orbit and 
tensor couplings are taken from table~\ref{tab:input-P-wave-pentaquarks}.  
}
\label{tab:masses-predictions-cbar-cs-ss}
\end{table} 

\section{Pentaquark decay widths} 
\label{sec:pentaquark-widths}  

In the compact diquark picture, the quarks in a diquark are bound and not free. 
In the present context it means that there is a barrier (or bound-state effect) which 
reduces the probability of the $\bar c$-quark and the charm quark in the $[uc]$-diquark 
to form a charmonium state. This is seen also in the decays of the $X,\, Y,\, Z$ states, 
which are tetraquark candidates in the compact diquark picture~\cite{% 
Maiani:2017kyi,Maiani:2019cwl,Brodsky:2014xia,Lebed:2017min,Giron:2019bcs,Esposito:2018cwh}.
In the case of the hidden-charm tetraquarks, it is the radius of the tetraquark $R_{4q} \sim 1 - 2$~fm 
(or, perhaps somewhat larger) compared to the size of the compact heavy-light diquark, 
typically $R_{Qq} \sim 0.5$~fm, which acts as a barrier to form a $J/\psi$- or $\eta_c$-meson.
For example, it is argued in~\cite{Maiani:2017kyi}, that the relative ratio~$\lambda$ 
is expected to be $\lambda \equiv R_{4q}/R_{Qq} \geq 3$. 
At long distance (but within the confined tetraquark radius), the potential is attractive, which 
changes as the internal structure of the diquark (and anti-diquark) gets resolved. 
The tunneling probability depends on the mass of the quark, with the probability exponentially 
suppressed the heavier the quark is. This can be  expressed through the semi-classical approximation 
of the tunneling amplitude $A_M \sim e^{-\sqrt{2 M E}\, R_{4q}}$, where~$M$ is the quark mass
% ,~$\ell$ is the radius of a tetraquark 
and the energy~$E$ is typically 100~MeV~\cite{Maiani:2017kyi}. This suppresses the formation 
of the charmonium states, leading to a much reduced $P_c \to J/\psi\, p$ decay width.
In all these cases, the size of the diquark~$R_{Qq}$ plays an important role in reducing
the decay widths, though the extent of this suppression may be different in the tetraquark 
and pentaquark cases. To work it out requires a dynamical theory, which is currently not at hand.
As some of the states are orbitally excited states, anticipated in the compact diquark picture, 
the decay widths are further reduced due to the angular momentum barrier.   Diquark-size effects
are also present in the tetraquark and pentaquark mass estimates in the diquark model, and
they have been calculated in a particular case, namely double-bottom tetraquark~\cite{An:2018cln}.

For four quarks~($q$, $q^\prime$, $q^{\prime\prime}$ and~$c$) and the charm antiquark, present
in a pentaquark $P_c$, there are different alternatives for clustering in color neutral states. 
The first possibility is the triquark-diquark alternative: 
\begin{eqnarray} 
\Psi^D_1 = \frac{1}{\sqrt 3} \left [ \frac{1}{\sqrt 2} \, \epsilon_{ijk} \bar c^i 
\left [ \frac{1}{\sqrt 2} \, \epsilon^{jlm} c_l q_m \right ] \right ] 
\left [ \frac{1}{\sqrt 2} \, \epsilon^{knp} q^\prime_n q^{\prime\prime}_p \right ] 
\equiv \left [ \bar c \left [ c q \right ] \right ]  \left [ q^\prime q^{\prime\prime} \right ] ,  
\label{eq:PsiD1-def} \\ 
\Psi^D_2 = \frac{1}{\sqrt 3} \left [ \frac{1}{\sqrt 2} \, \epsilon_{ikj} \bar c^i 
\left [ \frac{1}{\sqrt 2} \, \epsilon^{knp} q^\prime_n q^{\prime\prime}_p \right ] \right ]  
\left [ \frac{1}{\sqrt 2} \, \epsilon^{jlm} c_l q_m \right ] 
\equiv \left [ \bar c \left [ q^\prime q^{\prime\prime} \right ] \right ] \left [ c q \right ] . 
\label{eq:PsiD2-def}
\end{eqnarray} 
From the color algebra point of view, these states are related, $\Psi^D_2 = - \Psi^D_1$, 
but other internal dynamical properties can be different. 
The square parenthesis on the r.h.s. of the second equality in each equation 
denote color antisymmetrization of the constituents.
The color connection of the quarks in  $\Psi^D_1$ is assumed in the present paper, while 
the $\Psi^D_2$ color structure is employed in the dynamical diquark model 
of multiquark exotic hadrons~\cite{Lebed:2015tna,Lebed:2017min,Giron:2019bcs}. 
The other color-singlet combinations are the meson-baryon alternatives: 
\begin{eqnarray} 
&& 
\Psi^H_1 = 
\left ( \frac{1}{\sqrt 3} \, \bar c^i c_i \right )
\left [ \frac{1}{\sqrt 6} \, \epsilon^{jkl} q_j q^\prime_k q^{\prime\prime}_l \right ] 
\equiv \left ( \bar c c \right ) \left [ q q^\prime q^{\prime\prime} \right ] , 
\label{eq:PsiH1-def} \\ 
&& 
\Psi^H_2 = 
\left ( \frac{1}{\sqrt 3} \, \bar c^i q_i \right )
\left [ \frac{1}{\sqrt 6} \, \epsilon^{jkl} c_j q^\prime_k q^{\prime\prime}_l \right ]  
\equiv \left ( \bar c q \right ) \left [ c q^\prime q^{\prime\prime} \right ] , 
\label{eq:PsiH2-def} \\ 
&& 
\Psi^H_3 = 
\left ( \frac{1}{\sqrt 3} \, \bar c^i q^\prime_i \right )
\left [ \frac{1}{\sqrt 6} \, \epsilon^{jkl} c_j q_k q^{\prime\prime}_l \right ]   
\equiv \left ( \bar c q^\prime \right ) \left [ c q q^{\prime\prime} \right ] , 
\label{eq:PsiH3-def} \\  
&& 
\Psi^H_4 = 
\left ( \frac{1}{\sqrt 3} \, \bar c^i q^{\prime\prime}_i \right )
\left [ \frac{1}{\sqrt 6} \, \epsilon^{jkl} c_j q_k q^\prime_l \right ]  
\equiv \left ( \bar c q^{\prime\prime} \right ) \left [ c q q^\prime \right ] .  
\label{eq:PsiH4-def}  
\end{eqnarray} 
Among these four combinations, only $\Psi^H_1$ and $\Psi^H_2$ satisfy the heavy-quark-symmetry 
condition, which implies that the light $\left [ q^\prime q^{\prime\prime} \right ]$-diquark 
is transmitted intact, retaining its spin quantum number, from the $b$-baryon decays to the pentaquark.

It is possible to reduce the states in the color configurations~(\ref{eq:PsiD1-def}) 
or~(\ref{eq:PsiD2-def}) to a linear combination of the states in the 
set~(\ref{eq:PsiH1-def})--(\ref{eq:PsiH4-def}). Keeping the color of the light diquark unchanged, 
the convolution of two Levi-Civita tensors entering the triquark in~(\ref{eq:PsiD1-def}) 
gives the following decomposition: 
\begin{equation} 
\Psi^D_1 = - \frac{\sqrt 3}{2} \left [ \Psi^H_1 + \Psi^H_2 \right ] ,  
\label{eq:PsiD1-PsiH1-PsiH2}
\end{equation}
as mentioned earlier. However, a color reconnection is not enough to reexpress 
a pentaquark operator as a direct product of the meson and baryon operators. As hadrons 
are determined by their spin-parities, one should also project spins of the quarks and 
diquarks onto definite hadronic spin states. This requires to know the Dirac structure 
of the pentaquark operators to undertake the Fierz transformations in the Dirac space. 
This is a rather involved problem, already for the $S$-wave pentaquarks, and requires 
a dedicated study, to which we hope to return in a future publication.    
  
We exemplify it here by considering the three $S$-wave pentaquarks from 
table~\ref{tab:S-wave-pentaquarks-good-ld} with a light ``good'' diquark, following similar
considerations for the hidden-charm tetraquarks~\cite{Maiani:2017kyi}. 
The heavy diquark in them, being also in $S$-wave, can have both spins, $S_{hd} = 0$ and $S_{hd} = 1$. 
We further simplify the problem by assuming that the charm quark and antiquark are non-relativistic 
particles, keeping them as point-like objects at different positions in the Minkowski space-time.
The diquark-diquark-antiquark operators of the first state from table~\ref{tab:S-wave-pentaquarks-good-ld},
containing the heavy-diquark with $S_{hd} = 0$, after color reconnection (the spin structure 
is determined by eq.~(\ref{eq:PsiD1-def})) and making the color indices explicit, are as follows: 
\begin{eqnarray} 
&& 
\Psi_1^{H(1)} (x, y, z) = \frac{1}{3}
\left ( \tilde c^i (x)\, \sigma_2 \right ) 
\left ( c_i (y)\, \sigma_2\, q_k (y) \right ) d_0^k (z) , 
\label{eq:psiH1-1-spin} \\ 
&& 
\Psi_2^{H(1)} (x, y, z) = \frac{1}{3} 
\left ( \tilde c^i (x)\, \sigma_2 \right ) 
\left ( c_k (y)\, \sigma_2\, q_i (y) \right ) d_0^k (z) , 
\label{eq:psiH2-1-spin}  
\end{eqnarray}
where $\tilde c (x)$, $c (y)$ and $q (y)$ are the charm-antiquark, charm-quark and light-quark 
spinors in the non-relativistic limit, respectively,~$i$ and~$k$ are the color indices, the light 
diquark operator is denoted as~$d_0 (z)$, and the usual notation for the product of matrices is 
employed. With the help of the Fierz transform for the Pauli matrices~\cite{Ali:2019roi}: 
\begin{equation} 
\left ( \sigma_2 \right )^\alpha_\beta \left ( \sigma_2 \right )^\gamma_\delta = 
\frac{1}{2} \left [ 
\left ( \sigma_2 \right )^\alpha_\delta \left ( \sigma_2 \right )^\gamma_\beta +  
\left ( \sigma_2 \sigma_a \right )^\alpha_\delta \left ( \sigma_2 \sigma_a \right )^\gamma_\beta   
\right ] , 
\label{eq:sigma-FT-1}
\end{equation}
we reduce the states~(\ref{eq:psiH1-1-spin}) and~(\ref{eq:psiH2-1-spin}) to the forms: 
\begin{equation} 
\Psi_1^{H(1)} (x, y, z) = - \frac{1}{2} \left [ 
\left ( \tilde c (x)\, \sigma_2\, c (y) \right ) 
\left ( q (y) \, \sigma_2\, d_0 (z) \right ) + 
\left ( \tilde c (x)\, \sigma_2\, \boldsymbol\sigma\, c (y) \right ) 
\left ( q (y) \, \sigma_2\, \boldsymbol\sigma\, d_0 (z) \right ) 
\right ] , 
\label{eq:psiH1-1-spin-q} 
\end{equation} 
\begin{equation} 
\Psi_2^{H(1)} (x, y, z) = - \frac{1}{2} \left [ 
\left ( \tilde c (x)\, \sigma_2\, q (y) \right ) 
\left ( c (y)\, \sigma_2\, d_0 (z) \right ) +  
\left ( \tilde c (x)\, \sigma_2\, \boldsymbol\sigma\, q (y)  \right ) 
\left ( c (y)\, \sigma_2\, \boldsymbol\sigma\, d_0 (z) \right )   
\right ] ,  
\label{eq:psiH2-1-spin-q}  
\end{equation}
where the quark-antiquark and quark-diquark combinations are properly normalized in the color space. 

The transition to hadrons needs specifying the  flavors of all the quarks. 
If, as an example, we consider pentaquarks with the same quark content as the observed 
ones, this immediately specifies $q = u$ and $d_0 = \left [ u\, C\, \gamma_5\, d \right ]$, 
being the color-antitriplet scalar state\footnote{
In the diquark, operator~$C$ is the charge-conjugation matrix. 
}.  
Neglecting the position dependence of the operators, the quark operators~(\ref{eq:psiH1-1-spin-q}) 
and~(\ref{eq:psiH2-1-spin-q}) can be rewritten in terms of the hadrons: 
\begin{eqnarray} 
&& 
\Psi_1^{H(1)} = - \frac{i}{\sqrt 2}\, p \left [ 
a\, \eta_c + b \left ( \boldsymbol\sigma\, \mathbf{J}\boldsymbol{/\psi} \right ) 
\right ] , 
\label{eq:psiH1-1-spin-h-uf} \\ 
&& 
\Psi_2^{H(1)} = - \frac{i}{\sqrt 2}\, \Lambda_c^+ \left [ 
A\, \bar D^0 + B \left ( \boldsymbol\sigma\, \mathbf{\bar D^{*0}} \right ) 
\right ] ,  
\label{eq:psiH2-1-spin-h-uf}  
\end{eqnarray}
where~$A$ and~$B$ ($a$ and~$b$) are non-perturbative coefficients associated 
with different barrier penetration amplitudes for different light-quark 
(heavy-quark) spin configurations. They are equal in the limit of naive Fierz 
coupling. As mentioned in subsec.~\ref{ssec:predictions-unflavored}, 
the predicted mass of this state, $M = (3830 \pm 34)$~MeV, is below the $J/\psi\, p$ 
and $\eta_c\, p$ thresholds, so its decay to these states is forbidden by 
phase space, and it will decay through the charm-anticharm pair annihilation, 
having a width much smaller than the widths of the newly observed pentaquarks 
in table~\ref{tab:LHCb-data-2019}. The decays of this pentaquark to the 
$\Lambda_c^+\, \bar D^0$ and $\Lambda_c^+\, \bar D^{*0}$ final states 
are also forbidden by phase space, as these channels have higher thresholds 
than the proton-charmonium modes. 

Along similar lines, assigning $q = s$, but keeping the light diquark~$d_0$
as above, we get the first state with the internal spins $S_{hd} = S_{ld} = 0$ from 
table~\ref{tab:masses-predictions-cbar-cq-sq}. In terms of the physical hadrons, 
the quark operators~(\ref{eq:psiH1-1-spin-q}) and~(\ref{eq:psiH2-1-spin-q}) can 
be written as follows: 
\begin{eqnarray} 
&& 
\Psi_1^{H(1)} = - \frac{i}{\sqrt 2}\, \Lambda \left [ 
a_s\, \eta_c + b_s \left ( \boldsymbol\sigma\, \mathbf{J}\boldsymbol{/\psi} \right ) 
\right ] , 
\label{eq:psiH1-1-spin-h-1s} \\ 
&& 
\Psi_2^{H(1)} = - \frac{i}{\sqrt 2}\, \Lambda_c^+ \left [ 
A_s\, \bar D_s^0 + B_s \left ( \boldsymbol\sigma\, \mathbf{\bar D_s^{*0}} \right ) 
\right ] ,  
\label{eq:psiH2-1-spin-h-1s}  
\end{eqnarray}
where~$A_s$ and~$B_s$ ($a_s$ and~$b_s$) are non-perturbative coefficients associated 
with barrier penetration amplitudes. The predicted mass of this state, $M = (4112 \pm 32)$~MeV, 
(see table~\ref{tab:masses-predictions-cbar-cq-sq} in subsec.~\ref{ssec:predictions-strange}) 
is below the $J/\psi\, \Lambda$ and $\eta_c\, \Lambda$ thresholds, so its decay to these states 
is forbidden by  phase space and it will decay through the charm-anticharm pair annihilation.  
As the thresholds of $\Lambda_c^+\, \bar D_s^0$ and $\Lambda_c^+\, \bar D_s^{*0}$ are higher, 
the decays of this pentaquark to these final states are also forbidden by phase space.

In the same approximation, the diquark-diquark-antiquark operators of the other states from 
table~\ref{tab:S-wave-pentaquarks-good-ld}, containing the heavy-diquark with $S_{hd} = 1$, 
after color reconnection (the spin structure is determined by eq.~(\ref{eq:PsiD1-def})) 
and making the color indices explicit, are as follows: 
\begin{eqnarray} 
&& 
\boldsymbol\Psi_1^{H(2)} (x, y, z) = \frac{1}{3}
\left ( \tilde c^i (x)\, \sigma_2 \right ) 
\left ( c_i (y)\, \sigma_2\, \boldsymbol\sigma\, q_k (y) \right ) d_0^k (z) , 
\label{eq:psiH1-2-spin} \\ 
&& 
\boldsymbol\Psi_2^{H(2)} (x, y, z) = \frac{1}{3} 
\left ( \tilde c^i (x)\, \sigma_2 \right ) 
\left ( c_k (y)\, \sigma_2\, \boldsymbol\sigma\, q_i (y) \right ) d_0^k (z) . 
\label{eq:psiH2-2-spin}  
\end{eqnarray}
These operators, being the direct product of a spinor and a vector, are required 
to be decomposed into two irreducible representations, corresponding to the spin $J = 1/2$ 
and $J = 3/2$ pentaquarks. The projection onto these states is  done 
after Fierz transform is performed, for which  one uses  
the relation~\cite{Ali:2019roi}: 
\begin{equation} 
\left ( \sigma_2 \right )^\alpha_\beta \left ( \sigma_2\, \sigma_a \right )^\gamma_\delta = 
\frac{1}{2} \left [ 
\left ( \sigma_2 \right )^\alpha_\delta \left ( \sigma_2\, \sigma_a \right )^\gamma_\beta +  
\left ( \sigma_2\, \sigma_a \right )^\alpha_\delta \left ( \sigma_2 \right )^\gamma_\beta +  
i\, \varepsilon_{abc} 
\left ( \sigma_2 \sigma_b \right )^\alpha_\delta \left ( \sigma_2 \sigma_c \right )^\gamma_\beta   
\right ] .  
\label{eq:sigma-FT-2}
\end{equation}
With this, the states~(\ref{eq:psiH1-2-spin}) and~(\ref{eq:psiH2-2-spin}) are reduced to: 
\begin{eqnarray} 
&& 
\boldsymbol\Psi_1^{H(2)} (x, y, z) = \frac{1}{2} \left \{ 
\left ( \tilde c (x)\, \sigma_2\, \boldsymbol\sigma\, c (y) \right ) 
\left ( q (y) \, \sigma_2\, d_0 (z) \right ) + 
\left ( \tilde c (x)\, \sigma_2\, c (y) \right ) 
\left ( q (y) \, \sigma_2\, \boldsymbol\sigma\, d_0 (z) \right ) 
\right. 
\nonumber \\ 
&& \hspace*{25mm} 
+ i \left. \left [ 
\left ( \tilde c (x)\, \sigma_2\, \boldsymbol\sigma\, c (y) \right ) \times 
\left ( q (y) \, \sigma_2\, \boldsymbol\sigma\, d_0 (z) \right ) 
\right ] 
\right \} , 
\label{eq:psiH1-2-spin-q} \\ 
&& 
\boldsymbol\Psi_2^{H(2)} (x, y, z) = - \frac{1}{2} \left \{  
\left ( \tilde c (x)\, \sigma_2\, \boldsymbol\sigma\, q (y) \right ) 
\left ( c (y)\, \sigma_2\, d_0 (z) \right ) +  
\left ( \tilde c (x)\, \sigma_2\, q (y)  \right ) 
\left ( c (y)\, \sigma_2\, \boldsymbol\sigma\, d_0 (z) \right ) 
\right.   
\nonumber \\ 
&& \hspace*{31mm} 
+ i \left. \left [ 
\left ( \tilde c (x)\, \sigma_2\, \boldsymbol\sigma\, q (y) \right ) \times 
\left ( c (y)\, \sigma_2\, \boldsymbol\sigma\, d_0 (z) \right ) 
\right ] \right \} ,  
\label{eq:psiH2-2-spin-q}  
\end{eqnarray}
where the quark-antiquark and quark-diquark combinations are properly normalized in the color space. 

Neglecting as above the position dependence, the quark operators~(\ref{eq:psiH1-2-spin-q}) 
and~(\ref{eq:psiH2-2-spin-q}) of the pentaquarks with the light quark $q = u$ and light diquark 
$d_0 = \left [ u\, C\, \gamma_5\, d \right ]$ can be rewritten in terms of the hadronic states: 
\begin{eqnarray} 
&& 
\boldsymbol\Psi_1^{H(2)} = \frac{i}{\sqrt 2}\, p \left \{ 
a'\, \eta_c\, \boldsymbol\sigma + b'\, \mathbf{J}\boldsymbol{/\psi} - 
i c' \left [ \boldsymbol\sigma \times \mathbf{J}\boldsymbol{/\psi} \right ] 
\right \} , 
\label{eq:psiH1-2-spin-h-uf} \\ 
&& 
\boldsymbol\Psi_2^{H(2)} = - \frac{i}{\sqrt 2}\, \Lambda_c^+ \left \{ 
A'\, \bar D^0\, \boldsymbol\sigma + B'\, \mathbf{\bar D^{*0}} - 
i C' \left [ \boldsymbol\sigma \times \mathbf{\bar D^{*0}} \right ]  
\right \} ,  
\label{eq:psiH2-2-spin-h-uf}  
\end{eqnarray}
where~$A'$, $B'$ and~$C'$ ($a'$, $b'$ and~$c'$) are non-perturbative coefficients 
associated with different barrier penetration amplitudes. They are equal in the limit 
of naive Fierz coupling. These operators are the linear combination of two operators 
corresponding to baryons with spins $J = 1/2$ and $J = 3/2$. They can be separated 
by using the equation: 
\begin{equation} 
\boldsymbol\Psi_{1,2}^{H(2)} = \left [ 
\boldsymbol\Psi_{1,2}^{H(2)} - 
\frac{1}{3} \left ( \boldsymbol\Psi_{1,2}^{H(2)}\, \boldsymbol\sigma \right ) 
\boldsymbol\sigma \right ] + 
\frac{1}{3} \left ( \boldsymbol\Psi_{1,2}^{H(2)}\, \boldsymbol\sigma \right ) 
\boldsymbol\sigma = 
\boldsymbol\Psi_{1,2}^{H(3/2)} + \Psi_{1,2}^{H(1/2)}\, \boldsymbol\sigma , 
\label{eq:1/2-3/2-separation}  
\end{equation}
with the following result: 
\begin{eqnarray} 
&& 
\Psi_1^{H(1/2)} = \frac{i}{\sqrt 2}\, p \left \{ 
a'\, \eta_c + \frac{b'}{3} \left ( \boldsymbol\sigma\, \mathbf{J}\boldsymbol{/\psi} \right )  
\right \} , 
\label{eq:psiH1-1/2-spin-h-uf} \\ 
&& 
\boldsymbol\Psi_1^{H(3/2)} = \frac{i \sqrt 2}{3}\, p \left \{ 
b'\, \mathbf{J}\boldsymbol{/\psi} - 2 i c' \left [ \boldsymbol\sigma \times \mathbf{J}\boldsymbol{/\psi} \right ] 
\right \} , 
\label{eq:psiH1-3/2-spin-h-uf} \\ 
&& 
\Psi_2^{H(1/2)} = - \frac{i}{\sqrt 2}\, \Lambda_c^+ \left \{ 
A'\, \bar D^0 + \frac{B'}{3}\, \left ( \boldsymbol\sigma\, \mathbf{\bar D^{*0}} \right )  
\right \} ,  
\label{eq:psiH2-1/2-spin-h-uf} \\ 
&& 
\boldsymbol\Psi_2^{H(3/2)} = - \frac{i \sqrt 2}{3}\, \Lambda_c^+ \left \{ 
B'\, \mathbf{\bar D^{*0}} - 2 i C' \left [ \boldsymbol\sigma \times \mathbf{\bar D^{*0}} \right ]  
\right \} .   
\label{eq:psiH2-3/2-spin-h-uf}  
\end{eqnarray}
This analysis shows that the unflavored pentaquark with the spin-parity $J^P = 3/2^-$
is mainly decaying either to $J/\psi\, p$ final state, in which it was observed due 
to our assignment, or to $\Lambda_c^+\, \bar D^{*0}$ state. 

If instead we assign $q = s$, but keeping the light diquark~$d_0$ as above, 
we get two states with spins $S_{hd} = 1$ from table~\ref{tab:masses-predictions-cbar-cq-sq} 
and the corresponding operators are as follows: 
\begin{eqnarray} 
&& 
\Psi_1^{H(1/2)} = \frac{i}{\sqrt 2}\, \Lambda \left \{ 
a'_s\, \eta_c + \frac{b'_s}{3} \left ( \boldsymbol\sigma\, \mathbf{J}\boldsymbol{/\psi} \right )  
\right \} , 
\label{eq:psiH1-1/2-spin-h-s} \\ 
&& 
 \boldsymbol\Psi_1^{H(3/2)} = \frac{i \sqrt 2}{3}\, \Lambda \left \{ 
b'_s\, \mathbf{J}\boldsymbol{/\psi} - 2 i c'_s 
\left [ \boldsymbol\sigma \times \mathbf{J}\boldsymbol{/\psi} \right ] 
\right \} , 
\label{eq:psiH1-3/2-spin-h-s} \\ 
&& 
\Psi_2^{H(1/2)} = - \frac{i}{\sqrt 2}\, \Lambda_c^+ \left \{ 
A'_s\, \bar D_s^0 + \frac{B'_s}{3}\, \left ( \boldsymbol\sigma\, \mathbf{\bar D_s^{*0}} \right )  
\right \} ,  
\label{eq:psiH2-1/2-spin-h-s} \\ 
&& 
\boldsymbol\Psi_2^{H(3/2)} = - \frac{i \sqrt 2}{3}\, \Lambda_c^+ \left \{ 
B'_s\, \mathbf{\bar D_s^{*0}} - 2 i C'_s \left [ \boldsymbol\sigma \times \mathbf{\bar D_s^{*0}} \right ]  
\right \} .   
\label{eq:psiH2-3/2-spin-h-s}  
\end{eqnarray}
It is easy to recognize that the main decays in which a singly-strange $J^P = 3/2^-$ pentaquark 
should be searched, are the $J/\psi\, \Lambda$ and $\Lambda_c^+\, \bar D_s^{*0}$ final states. 
Analysis of the other pentaquark states can be done along similar lines.

From the foregoings in this section, it is evident that the color representations and color 
reconnections, on one hand, and localization of the diquarks and quarks, on the other, 
are essential to disentangle the dynamics (production and decays) of multiquark hadrons. 
We have worked out some illustrative cases here, in the color-antitriplet diquark representation, 
imposing heavy-quark symmetry constraints in the decays of $b$-baryons.

We end this section with a remark about the composition, localisation, and bindings of the various 
building blocks of a pentaquark in our model. In the compact diquark picture which we follow, the quarks 
in a diquark are bound and not free, and they have a hadronic size, which could be as large as several tenths of a Fermi.
This applies to both the light $[q^\prime q^{\prime\prime}]$-diquark as well as the heavy-light $[q c]$-diquark,
In this respect, 
the compact pentaquark model differs from the hadrocharmonium model~\cite{Eides:2019tgv}, in which 
the light degrees of freedom are considered as a cloud. We implicitly assume that the diquark sizes 
are smaller than that of the triquark, which in turn is smaller than the pentaquark size. So, the 
compact pentaquark model is essentially a three-scale problem. The light diquark is localized at 
some distance from the doubly-heavy triquark and an overlap of their wave functions can be described
within the one-gluon-exchange approach. The pentaquark stabilization is essentially determined by 
the triquark stabilization, which is accomplished by assuming  that there is a barrier/bound-state-effect 
which reduces the probability of the $\bar c$ and the charmed quark in the heavy $[q c]$-diquark 
to annihilate each other. This barrier is essential, as otherwise the~$c$- and $\bar c$-quarks will 
immediately annihilate. The triquark topology is not too different than that of a tetraquark consisting 
of $[q c]$- and $[\bar q^\prime \bar c]$-diquarks, where the tetraquark is stabilized due to a barrier 
between the diquark and the anti-diquark shells~\cite{Maiani:2019cwl,Maiani:2019lpu}.

As a final remark, we note that in the diquark models light diquarks $[q q^\prime]$ are ubiquitous. 
They are present in baryons, such as $\Lambda_b$, and the hidden-charm pentaquarks, discussed in detail 
in our paper, but also in the excited baryons, such as $\Omega_c^*=c\{s s \}$~\cite{Karliner:2017kfm}, 
and in doubly-heavy tetraquarks, such as ${\bar b \bar b} [q q^\prime]$, discussed in a number of papers
on stable tetraquarks~\cite{Karliner:2017qjm,Eichten:2017ffp,Francis:2016hui,Bicudo:2017szl,Junnarkar:2017sey}. 
How their hadronic sizes impact on properties of a given multiquark state depends on an underlying 
dynamical model. In the constituent diquark model, which we use, the size of the light diquark 
influences various parameters of the effective Hamiltonian. We have ignored the differences which emerge 
from intrinsically different embeddings of the diquarks in the baryons, tetraquarks and pentaquarks. 
This is an approximation, as the hadrons in question are not expected to have a universal hadronic size, 
and it limits the precision of the pentaquark masses presented here. A detailed dynamical scheme 
for pentaquarks remains to be developed~--- an issue which we hope to address in the future.

\section{Conclusions} 
\label{sec:conclusions} 

We have presented the mass spectrum of the hidden-charm pentaquark states 
$(c \bar c q q^\prime q^{\prime\prime})$, where~$q$, $q^\prime$, and~$q^{\prime\prime}$ 
are light $u$-, $d$,- and~$s$-quarks, using isospin symmetry for the~$u$- and $d$-quarks. 
In doing this, we have used an effective Hamiltonian, based on a doubly-heavy triquark~--- 
light diquark picture, shown in fig.~\ref{ali:fig-pentaquark-model-2}. Apart from the constituent
diquark and quark masses, the Hamiltonian incorporates dominant spin-spin, spin-orbit, orbital 
and tensor interactions. As the first step, we present analytical expressions for the masses 
of the states having well-defined quantum numbers, making explicit the transformations needed 
in the relations among various angular momentum basis vectors. This formalism is the main 
theoretical contribution of our paper.

For the numerical estimates of the pentaquark masses, we use Model~II by Maiani 
et al.~\cite{Maiani:2014aja}, developed to calculate the~$X$, $Y$, and~$Z$ tetraquark 
mass spectrum in the compact diquark model, in which only the dominant spin-spin couplings 
are taken into account. Since not all parameters in the effective Hamiltonian can be uniquely 
determined, as there are currently only three pentaquark states observed so far, and also their 
spin-parity assignments are not known, we use the results of the known heavy tetraquarks and 
heavy baryons to fix them. Due to the paucity of data on pentaquarks, and the inference on the 
model parameters from other systems, it is difficult to be quantitatively precise.
A remark about the accuracy on our mass estimates due to the neglect of the diquark size effects 
is in order here. While we have neglected this, it has been estimated recently for the double-bottom 
hadrons (baryons and tetraquarks) by taking into account gluon exchanges between the spectator quarks 
and the two heavy quarks in the diquark. Using the non-relativistic potential model 
to estimate the matrix element of the resulting dominant operator leads to a contribution 
$\Delta M_{\Xi_{bbq}} \simeq 30$~MeV, with a similar contribution to the mass shift of the
corresponding tetraquark~\cite{An:2018cln}. The corresponding effect for the hidden-charm tetraquarks 
and pentaquarks is not yet at hand, but is presumably of a similar order of magnitude. Thus, there 
is an intrinsic diquark-size dependent uncertainty in our mass estimates, which we hope can be estimated 
in an improved dynamical diquark framework. In view of this, the masses of the pentaquarks in the 
compact diquark approach presented here, while being in the right ball-park of the current 
experimental measurements, require an improved theoretical framework and more data to be precise.

The resulting pentaquark spectrum is very rich, as the states have both spin-0 and spin-1 light 
diquarks. Since the current experimental effort is dominated by searches in $\Lambda_b$-decays, 
we impose heavy quark symmetry, in which the light-diquark spin in $\Lambda_b$-baryon is a good 
quantum number. This, in turn, implies that only pentaquarks with the spin-0 light diquarks are 
anticipated in $\Lambda_b$-decays. This reduces the spectrum of the observable states greatly, 
yet there are more states present in the spectrum than have been observed, and hence we are not 
able to assign the $J^P$ quantum numbers to the observed states uniquely. We have argued, based 
on theoretical consistency, why some spin-parity assignments for the new pentaquarks can be ruled 
out in the compact diquark model. Yet, by virtue of the fact that both $S$- and $P$-state pentaquarks 
are anticipated in such models, predicting both positive- and negative-parity states, the spectrum 
provides a generic discrimination between the compact pentaquarks and the ones foreseen in hadron 
molecules, in which the observed pentaquarks have all the negative parity. Thus, it is crucial 
to determine the $J^P$ quantum numbers of the observed pentaquarks. Likewise, the remaining states 
in the pentaquark spectrum, whose masses we have presented, should be searched for. We find that  
some of them may lie in mass below their strong thresholds, defined by $J/\psi\, p$ (or $\eta_c\, p$). 
Hence, they will decay via the $c \bar c$-pair annihilation into light hadrons and into a lepton pair.  
Thus, the mode $p\, \mu^+ \mu^-$ may reveal the presence of these additional pentaquark states, 
and we urge the LHCb collaboration to search for them.

The hidden-charm pentaquarks having a strange quark in their valence composition are at present 
an experimentally virgin territory. While they can be produced also promptly, like their non-strange 
counterparts, only search strategies based on weak decays of the $b$-baryons with a strange quark 
are expected to bear fruits. The strange $b$-baryons which decay weakly are $\Xi_b^0\, (b u s)$, 
$\Xi_b^-\, (b d s)$, and $\Omega_b^-\, (b s s)$. Of these, the decays of $\Omega_b^-\, (b s s)$ 
will produce pentaquarks with a spin-1 light diquark, in accordance with the heavy quark symmetry, 
thus they allow to investigate the $SU (3)_F$-decuplet pentaquark sector. However, the production 
of such $b$-baryons is suppressed due to the excitation of the $s \bar s$-pair from the vacuum 
(for producing~$\Xi_b^0$ and~$\Xi_b^-$ in a $b$-quark initiated jet) and twice suppressed 
(for the $\Omega_b^-$-baryon). As an example, the product branching ratio 
${\cal B} (\Omega_b^- \to J/\psi\, \Omega^-)\, {\cal B} (b \to \Omega_b^-) = 
\left ( 2.9^{+1.1}_{-0.8} \right ) \times 10^{-6}$~\cite{Tanabashi:2018oca}. 
This, compared with the product branching ratio
${\cal B} (\Lambda_b^0 \to J/\psi\, \Lambda^0)\, {\cal B} (b \to \Lambda_b^0) = 
(5.8 \pm 0.8) \times 10^{-5}$, is about 20~times less probable. Thus, the discovery of hidden-charm 
strange pentaquarks requires a lot more data, but they remain a firm prediction, as in the compact 
diquark model complete $SU (3)_F$-multiplets are anticipated. Hence, their non-observation 
in experiments would seriously undermine such models. In the foreseeable future, experiments 
at the LHC, in particular LHCb, remain our best hope for their discovery. In the long-term, 
the entire hidden-charm pentaquark sector can be studied at a Tera-$Z$ factory.

\bigskip 

\acknowledgments{
We would like to thank Vladimir Galkin, Marek Karliner, Luciano Maiani, Antonello Polosa, 
Tomasz Skwarnicki,  and Sheldon Stone for helpful discussions.
A.\,A. and A.\,P. acknowledge support and warm hospitality of the Theoretical Physics 
Division at IHEP (Beijing, China) where an initial part of this research was performed,
and Wei Wang for the hospitality at the Laboratory of Particle Physics of the Shanghai 
Jiao-Tong University, where it was completed.  
A.\,P. thanks the Theory Group at DESY for their kind hospitality, where the major 
part of this work was done.  
A.\,P. acknowledges financial support by the German Academic Exchange Service (DAAD),  
by the Russian Foundation for Basic Research and National Natural Science Foundation 
of China for the research project No. 19-52-53041, and by the ``YSU Initiative 
Scientific Research Activity'' (Project No. AAAA-A16-116070610023-3). 
}

\appendix

\section{Corrections due to spin-spin interactions between the constituents 
         of doubly-heavy triquark and light diquark} 
\label{sec:spin-spin-corrections}

The matrix elements of the operators in eq.~(\ref{eq:H-SS-t-d}) have in general 
six different coupling strengths. For the pentaquarks, which contain the light 
$u$- and $d$-quarks only, isospin symmetry among the light quarks can be used 
to reduce the number of couplings: 
%Thus, instead of six different coupling strengths, entering~(\ref{eq:H-SS-t-d}), 
%in general, the isospin symmetry relates three of them:  
$(\tilde{\mathcal{K}}_{c q^\prime})_{\bar 3} = (\tilde{\mathcal{K}}_{c q^{\prime\prime}})_{\bar 3} = (\tilde{\mathcal{K}}_{c q})_{\bar 3}$, 
$(\tilde{\mathcal{K}}_{q q^\prime})_{\bar 3} = (\tilde{\mathcal{K}}_{q q^{\prime\prime}})_{\bar 3} = (\tilde{\mathcal{K}}_{q q})_{\bar 3}$, 
and $\tilde{\mathcal{K}}_{\bar c q^\prime} = \tilde{\mathcal{K}}_{\bar c q^{\prime\prime}} = \tilde{\mathcal{K}}_{\bar c q}$. 
So, the expression for $H_{SS}^{t-ld}$ in the effective Hamiltonian responsible 
for the spin-spin interactions between the (anti)quarks in the diquark and triquark, 
given in eq.~(\ref{eq:H-SS-t-d}), is greatly simplified: 
\begin{eqnarray}
H_{SS}^{t-ld} =  
2 \, \tilde{\mathcal{K}}_{\bar c q} \, (\mathbf{S}_{\bar c} \cdot \mathbf{S}_{ld}) +  
2 \, (\tilde{\mathcal{K}}_{c q})_{\bar 3} \, (\mathbf{S}_c \cdot \mathbf{S}_{ld}) + 
2 \, (\tilde{\mathcal{K}}_{q q})_{\bar 3} \, (\mathbf{S}_q \cdot \mathbf{S}_{ld}) ,  
\label{eq:H-SS-t-d-isospin}
\end{eqnarray}
where the definition of the light-diquark spin operator, 
$\mathbf{S}_{ld} = \mathbf{S}_{q^\prime} + \mathbf{S}_{q^{\prime\prime}}$, 
is used. We start by calculating the matrix elements of the
spin-spin operators from the first term, 
$2 \, \tilde{\mathcal{K}}_{\bar c q} \, (\mathbf{S}_{\bar c} \cdot \mathbf{S}_{ld})$, 
which contains the spins of the charm antiquark and light diquark. 
Again, it is convenient to transform the vector basis. To do so, we
denote the pentaquark state $| S_{hd}, S_t, L_t; S_{ld}, L_{ld}; S, L \rangle_J$ 
as $| S_{ld},\, (S_{\bar c}, \, S_{hd})_{S_t}; \, S \rangle$ and apply 
the transformation~(\ref{eq:transform-6j}):  
\begin{eqnarray}
&& 
| S_{ld},\, (S_{\bar c}, \, S_{hd})_{S_t};\, S \rangle = \sum_{\tilde S} 
(-1)^{S_{ld} + S_{\bar c} + S_{hd} + S} \sqrt{(2 \tilde S + 1)\, (2 S_t + 1)} 
\nonumber \\ 
&& \hspace*{33mm} \times 
\left \{ 
\begin{array}{ccc} 
S_{ld} & S_{\bar c} & \tilde S \\ 
S_{hd} &          S &      S_t 
\end{array} 
\right \} 
| (S_{ld},\, S_{\bar c})_{\tilde S},\, S_{hd}; \, S \rangle .  
\label{eq:transform-spin-cbar-ld}
\end{eqnarray}
The required matrix elements can be written in the form: 
\begin{eqnarray} 
&& 
{}_J \langle S_{hd}, S_t^\prime, L_t; S_{ld}, L_{ld}; S, L | \, 
2 \, \tilde{\mathcal{K}}_{\bar c q} \, 
(\mathbf{S}_{\bar c} \cdot \mathbf{S}_{ld}) \,   
| S_{hd}, S_t, L_t; S_{ld}, L_{ld}; S, L \rangle_J 
\label{eq:cbar-ld-spin-contribution} \\ 
&& \hspace{7mm} 
= \tilde{\mathcal{K}}_{\bar c q} \, (- 1)^{2 S + 1} 
\sqrt{(2 S_t + 1)\, (2 S_t^\prime + 1)} \, 
\sum_{\tilde S} \left ( 2 \tilde S + 1 \right ) 
\nonumber \\ 
&& \hspace{7mm} \times 
\left [ 
\tilde S \left ( \tilde S + 1 \right ) - S_{ld} \left ( S_{ld} + 1 \right ) - \frac{3}{4} 
\right ]   
\left \{ 
\begin{array}{ccc} 
S_{ld} & 1/2 & \tilde S \\ 
S_{hd} &   S &      S_t 
\end{array} 
\right \} 
\left \{ 
\begin{array}{ccc} 
S_{ld} & 1/2 &   \tilde S \\ 
S_{hd} &   S & S_t^\prime 
\end{array} 
\right \} , 
\nonumber  
\end{eqnarray}
where the charm-antiquark spin, $S_{\bar c} = 1/2$, is substituted.   

We calculate now the matrix elements of the last two operators in 
eq.~(\ref{eq:H-SS-t-d-isospin}). Let us start from the operator 
$2 \, (\tilde{\mathcal{K}}_{c q})_{\bar 3} \, (\mathbf{S}_c \cdot \mathbf{S}_{ld})$ 
which works on the wave functions of the charm quark in the heavy diquark 
and light diquark. To find it, we need to couple the heavy and light diquark 
first and then couple the light diquark with the charm quark~--- the 
constituent of heavy-light diquark. This requires a double transformation 
of the original pentaquark vector state $| S_{hd}, S_t, L_t; S_{ld}, L_{ld}; S, L \rangle_J$,  
which we denote as $| S_{ld},\, (S_{hd}, \, S_{\bar c})_{S_t}; \, S \rangle$, 
to the vector state $| (S_{ld},\, S_{hd})_{S_1}, \, S_{\bar c}; \, S \rangle$;
in the second transformation these basis states, for simplicity called 
as $| S_{ld},\, (S_c, \, S_q)_{S_{hd}}; \, S_1 \rangle$, are changed 
to the required ones, $| (S_{ld},\, S_c)_{S_2}, \, S_q; \, S_1 \rangle$.  
Specifically: 
\begin{eqnarray}
&& 
| S_{ld},\, (S_{hd}, \, S_{\bar c})_{S_t};\, S \rangle = \sum_{S_1} 
(-1)^{S_{ld} + S_{hd} + S_{\bar c} + S} \sqrt{(2 S_1 + 1)\, (2 S_t + 1)} 
\nonumber \\ 
&& \hspace{37mm} \times
\left \{ 
\begin{array}{ccc} 
    S_{ld} & S_{hd} & S_1 \\ 
S_{\bar c} &      S & S_t 
\end{array} 
\right \} 
| (S_{ld},\, S_{hd})_{S_1},\, S_{\bar c}; \, S \rangle   
\nonumber \\ 
&& \hspace{7mm} 
= \sum_{S_1} 
(-1)^{S_{ld} + S_{hd} + S_{\bar c} + S} \sqrt{(2 S_1 + 1)\, (2 S_t + 1)} 
\left \{ 
\begin{array}{ccc} 
    S_{ld} & S_{hd} & S_1 \\ 
S_{\bar c} &      S & S_t 
\end{array} 
\right \} 
| S_{ld},\, (S_c, \, S_q)_{S_{hd}}; \, S_1 \rangle 
\nonumber \\ 
&& \hspace{7mm} 
= \sum_{S_1, S_2} 
(-1)^{2 S_{ld} + S_{hd} + S_{\bar c} + S + S_c + S_q + S_1} 
\sqrt{(2 S_1 + 1) \, (2 S_t + 1) \, (2 S_{hd} + 1) \, (2 S_2 + 1)} 
\nonumber \\ 
&& \hspace{17mm} \times 
\left \{ 
\begin{array}{ccc} 
    S_{ld} & S_{hd} & S_1 \\ 
S_{\bar c} &      S & S_t 
\end{array} 
\right \} 
\left \{ 
\begin{array}{ccc} 
S_{ld} & S_c &    S_2 \\ 
   S_q & S_1 & S_{hd} 
\end{array} 
\right \} 
| (S_{ld}, \, S_c)_{S_2}, \, S_q; \, S_1 \rangle . 
\label{eq:transform-spin-c-ld}  
\end{eqnarray}
The matrix elements of interest are as follows: 
\begin{eqnarray} 
&& 
{}_J \langle S_{hd}^\prime, S_t^\prime, L_t; S_{ld}, L_{ld}; S, L | \, 
2 \, (\tilde{\mathcal{K}}_{c q})_{\bar 3} \, 
(\mathbf{S}_c \cdot \mathbf{S}_{ld}) \,   
| S_{hd}, S_t, L_t; S_{ld}, L_{ld}; S, L \rangle_J 
\nonumber \\ 
&& \hspace{7mm} 
= (\tilde{\mathcal{K}}_{c q})_{\bar 3} \, (-1)^{S_{hd} + S_{hd}^\prime + 2 S + 1}  
\sqrt{(2 S_{hd} + 1) \, (2 S_{hd}^\prime + 1) \, (2 S_t + 1) \, (2 S_t^\prime + 1)} \, 
\nonumber \\   
&& \hspace{7mm} \times 
\sum_{S_1, S_2} 
\left ( 2 S_1 + 1 \right ) \left ( 2 S_2 + 1 \right ) 
\left [ 
S_2 \left ( S_2 + 1 \right ) - S_{ld} \left ( S_{ld} + 1 \right ) - \frac{3}{4} 
\right ] 
\nonumber \\   
&& \hspace{7mm} \times 
\left \{ 
\begin{array}{ccc} 
S_{ld} & S_{hd} & S_1 \\ 
   1/2 &      S & S_t 
\end{array} 
\right \} 
\left \{ 
\begin{array}{ccc} 
S_{ld} & S_{hd}^\prime &        S_1 \\ 
   1/2 &             S & S_t^\prime 
\end{array} 
\right \} 
\left \{ 
\begin{array}{ccc} 
S_{ld} & 1/2 &    S_2 \\ 
   1/2 & S_1 & S_{hd} 
\end{array} 
\right \}  
\left \{ 
\begin{array}{ccc} 
S_{ld} & 1/2 &           S_2 \\ 
   1/2 & S_1 & S_{hd}^\prime 
\end{array} 
\right \} , 
\label{eq:c-ld-spin-contribution}   
\end{eqnarray}
where the quark and antiquark spins, $S_c = S_q = S_{\bar c} = 1/2$, are substituted. 
Note that the matrix elements of the last operator in~(\ref{eq:H-SS-t-d-isospin})
$2 \, (\tilde{\mathcal{K}}_{q q})_{\bar 3} \, (\mathbf{S}_q \cdot \mathbf{S}_{ld})$ 
differ from the above matrix elements by the replacement  
$(\tilde{\mathcal{K}}_{c q})_{\bar 3} \to (\tilde{\mathcal{K}}_{q q})_{\bar 3}$ 
and the interchange $S_c = 1/2 \leftrightarrow S_q = 1/2$. It means that, up to 
the coupling strengths, they coincide. So, the contribution of these two operators 
enters the matrix elements with the factor 
$(\tilde{\mathcal{K}}_{c q})_{\bar 3} + (\tilde{\mathcal{K}}_{q q})_{\bar 3}$.

We continue by considering the spin-spin interaction between the strange hidden-charm  
triquark and unflavored (i.\,e., having $u$- and $d$-quarks) light diquark. In this case, 
the expression given in eq.~(\ref{eq:H-SS-t-d}) of the effective Hamiltonian is again 
simplified in the isospin-symmetry limit: 
\begin{eqnarray}
H_{SS}^{t-ld} =  
2 \, \tilde{\mathcal{K}}_{\bar c q} \, (\mathbf{S}_{\bar c} \cdot \mathbf{S}_{ld}) +  
2 \, (\tilde{\mathcal{K}}_{c q})_{\bar 3} \, (\mathbf{S}_c \cdot \mathbf{S}_{ld}) + 
2 \, (\tilde{\mathcal{K}}_{s q})_{\bar 3} \, (\mathbf{S}_s \cdot \mathbf{S}_{ld}) ,  
\label{eq:H-SS-t-strange-d}
\end{eqnarray}
where $\mathbf{S}_s$ and $\mathbf{S}_{ld} = \mathbf{S}_{q^\prime} + \mathbf{S}_{q^{\prime\prime}}$ 
are the spin operators of the $s$-quark in the triquark and the light-diquark, respectively. 
Comparing~(\ref{eq:H-SS-t-strange-d}) with~(\ref{eq:H-SS-t-d-isospin}), 
it is easy to recognize that they differ by the factor entering the last 
terms,~$(\tilde{\mathcal{K}}_{s q})_{\bar 3}$ and~$(\tilde{\mathcal{K}}_{q q})_{\bar 3}$. 
So, the matrix elements~(\ref{eq:cbar-ld-spin-contribution}) and~(\ref{eq:c-ld-spin-contribution}) 
remain unchanged, but to get the contribution of the last operator in~(\ref{eq:H-SS-t-strange-d}) 
one should simply change the coupling: 
$(\tilde{\mathcal{K}}_{q q})_{\bar 3} \to (\tilde{\mathcal{K}}_{s q})_{\bar 3}$. 
Taking into account the statement in the paragraph above, the second and third terms 
in~(\ref{eq:H-SS-t-strange-d}) are equal (modulo couplings) and contribute with the coefficient
$(\tilde{\mathcal{K}}_{c q})_{\bar 3} + (\tilde{\mathcal{K}}_{s q})_{\bar 3}$ 
in the mass formulae.   

The same simple form of the spin-spin interaction is also hold for the hidden-charm pentaquark 
with the strangeness $S = -2$ which consists of the unflavored doubly-heavy triquark and 
light diquark containing two $s$-quarks: 
\begin{eqnarray}
H_{SS}^{t-ld} =  
2 \, \tilde{\mathcal{K}}_{\bar c s} \, (\mathbf{S}_{\bar c} \cdot \mathbf{S}_{ld}) +  
2 \, (\tilde{\mathcal{K}}_{c s})_{\bar 3} \, (\mathbf{S}_c \cdot \mathbf{S}_{ld}) + 
2 \, (\tilde{\mathcal{K}}_{s q})_{\bar 3} \, (\mathbf{S}_q \cdot \mathbf{S}_{ld}) .   
\label{eq:H-SS-t-d-strange-2}
\end{eqnarray}
So, the matrix elements~(\ref{eq:cbar-ld-spin-contribution}) and~(\ref{eq:c-ld-spin-contribution}), 
up to spin-spin couplings, remain unchanged and their contributions to the mass formulae 
are accompanied by the couplings $\tilde{\mathcal{K}}_{\bar c s}$ and 
$(\tilde{\mathcal{K}}_{c s})_{\bar 3} + (\tilde{\mathcal{K}}_{s q})_{\bar 3}$, respectively. 

Similar simple effective Hamiltonian for the spin-spin interactions is present in the 
hidden-charm pentaquarks with the strangeness $S = -3$, which contains three $s$-quarks:  
\begin{eqnarray}
H_{SS}^{t-ld} =  
2 \, \tilde{\mathcal{K}}_{\bar c s} \, (\mathbf{S}_{\bar c} \cdot \mathbf{S}_{ld}) +  
2 \, (\tilde{\mathcal{K}}_{c s})_{\bar 3} \, (\mathbf{S}_c \cdot \mathbf{S}_{ld}) + 
2 \, (\tilde{\mathcal{K}}_{s s})_{\bar 3} \, (\mathbf{S}_s \cdot \mathbf{S}_{ld}) .   
\label{eq:H-SS-t-strange-d-strange-2}
\end{eqnarray}
The matrix elements~(\ref{eq:cbar-ld-spin-contribution}) and~(\ref{eq:c-ld-spin-contribution}) 
can be safely used in this case also and their contributions are entering the mass formulae 
with the couplings $\tilde{\mathcal{K}}_{\bar c s}$ and 
$(\tilde{\mathcal{K}}_{c s})_{\bar 3} + (\tilde{\mathcal{K}}_{s s})_{\bar 3}$, respectively. 
 
The next flavor state which corresponds to the hidden-charm strange pentaquark consists 
of the the unflavored doubly-heavy triquark and strange light diquark. The corresponding 
spin-spin effective Hamiltonian~(\ref{eq:H-SS-t-d}) does not simplified and all the six 
terms, each of which is accompanied by its own coupling, should be considered separately: 
\begin{eqnarray}
H_{SS}^{t-ld} & = &  
2 \, \tilde{\mathcal{K}}_{\bar c q} \, (\mathbf{S}_{\bar c} \cdot \mathbf{S}_{q^\prime}) +  
2 \, (\tilde{\mathcal{K}}_{c q})_{\bar 3} \, (\mathbf{S}_c \cdot \mathbf{S}_{q^\prime}) + 
2 \, (\tilde{\mathcal{K}}_{q q})_{\bar 3} \, (\mathbf{S}_q \cdot \mathbf{S}_{q^\prime})     
\label{eq:H-SS-t-d-strange} \\ 
& + & 
2 \, \tilde{\mathcal{K}}_{\bar c s} \, (\mathbf{S}_{\bar c} \cdot \mathbf{S}_s) +  
2 \, (\tilde{\mathcal{K}}_{c s})_{\bar 3} \, (\mathbf{S}_c \cdot \mathbf{S}_s) + 
2 \, (\tilde{\mathcal{K}}_{s q})_{\bar 3} \, (\mathbf{S}_q \cdot \mathbf{S}_s) ,    
\nonumber 
\end{eqnarray}
where we identify $q^{\prime\prime} = s$ and apply the isospin symmetry to the spin-spin couplings. 

A calculation of these operators requires a recoupling of four angular momenta. 
In general, this transformation scheme is expressed through the $9j$-symbol
which is defined as follows~\cite{Edmonds:1957}: 
\begin{eqnarray} 
&& 
\langle (j_1, j_2)_{j_{12}}, (j_3, j_4)_{j_{34}}, j | (j_1, j_3)_{j_{13}}, (j_2, j_4)_{j_{24}}, j \rangle 
\nonumber \\ 
&& \hspace*{17mm}
= \sqrt{(2 j_{12} + 1) \, (2 j_{34} + 1) \, (2 j_{13} + 1) \, (2 j_{24} + 1)}
\left \{ 
\begin{array}{ccc} 
   j_1 &    j_2 & j_{12} \\ 
   j_3 &    j_4 & j_{34} \\ 
j_{13} & j_{24} &      j  
\end{array}
\right \} .  
\label{eq:9j-symbol-def}
\end{eqnarray}
To demonstrate its necessity, let us consider the first term in~(\ref{eq:H-SS-t-d-strange}). 
In our basis of states the charm antiquark belongs to the doubly-heavy triquark while 
the light $q^\prime$-quark is inside the light diquark. The original pentaquark vector state 
$| S_{hd}, S_t, L_t; S_{ld}, L_{ld}; S, L \rangle_J$, which we denote for simplicity as 
$| (S_{\bar c},\, S_{hd})_{S_t}, (S_{q^\prime},\, S_s)_{S_{ld}}; \, S \rangle$, 
should be transformed to the new state with the recoupling momenta 
$| (S_{\bar c},\, S_{q^\prime})_{S_{\bar c q^\prime}}, \, (S_{hd},\, S_s)_{S_{hd, s}}; \, S \rangle$, 
and the $9j$-symbol is responsible for this transform: 
\begin{eqnarray}
&& \hspace*{-5mm}
| (S_{\bar c},\, S_{hd})_{S_t}, (S_{q^\prime},\, S_s)_{S_{ld}}; \, S \rangle = 
\sum_{S_{\bar c q^\prime}, S_{hd, s}} 
\sqrt{(2 S_t + 1)\, (2 S_{ld} + 1)\, (2 S_{\bar c q^\prime} + 1)\, (2 S_{hd, s} + 1)} 
\nonumber \\
&& \hspace*{37mm} 
\times
\left \{ 
\begin{array}{ccc} 
         S_{\bar c} &    S_{hd} &    S_t \\
       S_{q^\prime} &       S_s & S_{ld} \\ 
S_{\bar c q^\prime} & S_{hd, s} &      S
\end{array} 
\right \} 
| (S_{\bar c},\, S_{q^\prime})_{S_{\bar c q^\prime}}, \, (S_{hd},\, S_s)_{S_{hd, s}}; \, S \rangle .   
\label{eq:transform-spins-cbar-hd-qprime-s}  
\end{eqnarray}
So, the matrix element of the operator 
$2 \, \tilde{\mathcal{K}}_{\bar c q} \, (\mathbf{S}_{\bar c} \cdot \mathbf{S}_{q^\prime})$ 
sandwiched between the states with different triquark~$S_t^{(\prime)}$ and 
light-diquark~$S_{ld}^{(\prime)}$ vector states can be written as a double sum 
(here, we come back to the original notation for the vector state): 
\begin{eqnarray}
&& \hspace{-3mm}
{}_J \langle S_{hd}, S_t^\prime, L_t; S_{ld}^\prime, L_{ld}; S, L |  
2 \, \tilde{\mathcal{K}}_{\bar c q} \, (\mathbf{S}_{\bar c} \cdot \mathbf{S}_{q^\prime}) 
| S_{hd}, S_t, L_t; S_{ld}, L_{ld}; S, L \rangle_J
\nonumber \\ 
&& \hspace{7mm}
= \tilde{\mathcal{K}}_{\bar c q} 
\sqrt{(2 S_t + 1) \, (2 S_t^\prime + 1) \, (2 S_{hd} + 1) \, (2 S_{hd}^\prime + 1)}
\sum_{S_{\bar c q^\prime}, S_{hd, s}} 
(2 S_{\bar c q^\prime} + 1)\, (2 S_{hd, s} + 1)\, 
\nonumber \\ 
&& \hspace{7mm} \times 
\left [ S_{\bar c q^\prime} (S_{\bar c q^\prime} + 1 ) - \frac{3}{2} \right ]  
\left \{ 
\begin{array}{ccc} 
                1/2 &    S_{hd} &    S_t^\prime \\
                1/2 &       1/2 & S_{ld}^\prime \\ 
S_{\bar c q^\prime} & S_{hd, s} &             S
\end{array} 
\right \} 
\left \{ 
\begin{array}{ccc} 
                1/2 &    S_{hd} &    S_t \\
                1/2 &       1/2 & S_{ld} \\ 
S_{\bar c q^\prime} & S_{hd, s} &      S
\end{array} 
\right \} , 
\label{eq:cbar-hd-qprime-s-spin-contribution}  
\end{eqnarray}
where the spins $S_{\bar c} = S_{q^\prime} = S_s = 1/2$ of quarks were subsituted. 
In a similar manner one can get the contribution of the operator  
$2 \, \tilde{\mathcal{K}}_{\bar c s} \, (\mathbf{S}_{\bar c} \cdot \mathbf{S}_{s})$
with the result: 
\begin{eqnarray}
&& \hspace{-3mm}
{}_J \langle S_{hd}, S_t^\prime, L_t; S_{ld}^\prime, L_{ld}; S, L |  
2 \, \tilde{\mathcal{K}}_{\bar c s} \, (\mathbf{S}_{\bar c} \cdot \mathbf{S}_{s}) 
| S_{hd}, S_t, L_t; S_{ld}, L_{ld}; S, L \rangle_J
\nonumber \\ 
&& \hspace{7mm}
= \tilde{\mathcal{K}}_{\bar c s} 
\sqrt{(2 S_t + 1) \, (2 S_t^\prime + 1) \, (2 S_{hd} + 1) \, (2 S_{hd}^\prime + 1)}
\sum_{S_{\bar c s}, S_{hd, q^\prime}} 
(2 S_{\bar c s} + 1)\, (2 S_{hd, q^\prime} + 1)\, 
\nonumber \\ 
&& \hspace{7mm} \times 
\left [ S_{\bar c s} (S_{\bar c s} + 1 ) - \frac{3}{2} \right ]  
\left \{ 
\begin{array}{ccc} 
         1/2 &           S_{hd} &    S_t^\prime \\
         1/2 &              1/2 & S_{ld}^\prime \\ 
S_{\bar c s} & S_{hd, q^\prime} &             S
\end{array} 
\right \} 
\left \{ 
\begin{array}{ccc} 
         1/2 &           S_{hd} &    S_t \\
         1/2 &              1/2 & S_{ld} \\ 
S_{\bar c s} & S_{hd, q^\prime} &      S
\end{array} 
\right \} .  
\label{eq:cbar-hd-s-qprime-spin-contribution}  
\end{eqnarray}
Up to the coupling,~$\tilde{\mathcal{K}}_{\bar c s}$, this equation coincides 
with~(\ref{eq:cbar-hd-qprime-s-spin-contribution}), so their common contribution 
contains the factor $\tilde{\mathcal{K}}_{\bar c q} + \tilde{\mathcal{K}}_{\bar c s}$. 

With the other four operators the situation is a little bit more complicated 
as spin-spin interactions relate quarks from different diquarks, moreover, 
the spin of the heavy diquark is coupled with the spin of the charm antiquark 
inside the triquark system. So, at the first step one needs to decouple the 
heavy diquark from the charm antiquark and  couple it with the light diquark
which generates the tetraquark system. If we denote the vector state as 
$| S_{ld},\, (S_{hd}, \, S_{\bar c})_{S_t}; \, S \rangle$, we need to recouple 
in a way to get the new basis $| (S_{ld},\, S_{hd},)_{S_T} \, S_{\bar c}; \, S \rangle$ 
which is possible to do with the help of the transformation~(\ref{eq:transform-6j}): 
\begin{eqnarray}
&& 
| S_{ld},\, (S_{hd}, \, S_{\bar c})_{S_t};\, S \rangle = \sum_{S_T} 
(-1)^{S_{ld} + S_{hd} + S_{\bar c} + S} \sqrt{(2 S_T + 1)\, (2 S_t + 1)} 
\nonumber \\ 
&& \hspace*{33mm} \times 
\left \{ 
\begin{array}{ccc} 
    S_{ld} & S_{hd} & S_T \\ 
S_{\bar c} &      S & S_t 
\end{array} 
\right \} 
| (S_{ld},\, S_{hd})_{S_T},\, S_{\bar c}; \, S \rangle .   
\label{eq:transform-spin-c-ld-1}  
\end{eqnarray}
At the second step, it is necessary to recouple quark spins inside the tetraquark.   
So, for this transformation the $9j$-symbol~(\ref{eq:9j-symbol-def}) is required.  
To do this, the introduced pentaquark vector state 
$| (S_{ld},\, S_{hd})_{S_T}, \, S_{\bar c}; \, S \rangle$ is more convenient to rewrite 
in the form $| (S_c,\, S_q)_{S_{hd}}, \, (S_{q^\prime},\, S_s)_{S_{ld}}; \, S_T \rangle$
and then perform the transformations: 
\begin{eqnarray}
&& \hspace*{-5mm}
| (S_c,\, S_q)_{S_{hd}}, \, (S_{q^\prime},\, S_s)_{S_{ld}}; \, S_T \rangle = 
\sum_{S_{c q^\prime}, S_{s q}} 
\sqrt{(2 S_{hd} + 1)\, (2 S_{ld} + 1)\, (2 S_{c q^\prime} + 1)\, (2 S_{s q} + 1)} 
\nonumber \\
&& \hspace*{37mm} 
\times
\left \{ 
\begin{array}{ccc} 
           S_c &     S_q & S_{hd} \\
  S_{q^\prime} &     S_s & S_{ld} \\ 
S_{c q^\prime} & S_{s q} &    S_T
\end{array} 
\right \} 
| (S_c,\, S_{q^\prime})_{S_{c q^\prime}}, \, (S_q,\, S_s)_{S_{s q}}; \, S_T \rangle .   
\label{eq:transform-spins-c-q-qprime-s} 
\end{eqnarray}
Now, we can sandwich the operators 
$2 \, (\tilde{\mathcal{K}}_{c q})_{\bar 3} \, (\mathbf{S}_c \cdot \mathbf{S}_{q^\prime})$ and 
$2 \, (\tilde{\mathcal{K}}_{s q})_{\bar 3} \, (\mathbf{S}_s \cdot \mathbf{S}_q)$ 
by the vector states with varying spins of the heavy,~$S_{hd}$, and light,~$S_{ld}$, diquarks 
as well as of the doubly-heavy triquark,~$S_t$. The resulting matrix element of these two 
operators is the triple sum: 
\begin{eqnarray}
&& \hspace{-3mm}
{}_J \langle S_{hd}^\prime, S_t^\prime, L_t; S_{ld}^\prime, L_{ld}; S, L |  
2 \, (\tilde{\mathcal{K}}_{c q})_{\bar 3} \, (\mathbf{S}_c \cdot \mathbf{S}_{q^\prime}) + 
2 \, (\tilde{\mathcal{K}}_{s q})_{\bar 3} \, (\mathbf{S}_s \cdot \mathbf{S}_q) 
| S_{hd}, S_t, L_t; S_{ld}, L_{ld}; S, L \rangle_J
\nonumber \\ 
&& \hspace{5mm}
= \left [ (\tilde{\mathcal{K}}_{c q})_{\bar 3} + (\tilde{\mathcal{K}}_{s q})_{\bar 3} \right ] 
\sqrt{(2 S_t + 1) \, (2 S_t^\prime + 1) \, 
      (2 S_{hd} + 1) \, (2 S_{hd}^\prime + 1) \, 
      (2 S_{ld} + 1) \, (2 S_{ld}^\prime + 1)}
\nonumber \\ 
&& \hspace{5mm}
\times 
(-1)^{S_{hd} + S_{hd}^\prime + S_{ld} + S_{ld}^\prime + 2 S + 1}
\!\!\!\! \sum_{S_T, S_{c q^\prime}, S_{s q}} \!\!\!\! 
(2 S_T + 1)\, (2 S_{c q^\prime} + 1)\, (2 S_{s q} + 1)\, 
\left [ S_{c q^\prime} (S_{c q^\prime} + 1 ) - \frac{3}{2} \right ]  
\nonumber \\ 
&& \hspace{5mm}
\times 
\left \{ 
\begin{array}{ccc} 
 S_{ld}^\prime & S_{hd}^\prime &        S_T \\
           1/2 &             S & S_t^\prime  
\end{array} 
\right \} 
\left \{ 
\begin{array}{ccc} 
 S_{ld} & S_{hd} & S_T \\
    1/2 &      S & S_t  
\end{array} 
\right \} 
\left \{ 
\begin{array}{ccc} 
           1/2 &     1/2 & S_{hd}^\prime \\
           1/2 &     1/2 & S_{ld}^\prime \\ 
S_{c q^\prime} & S_{s q} &           S_T
\end{array} 
\right \} 
\left \{ 
\begin{array}{ccc} 
           1/2 &     1/2 & S_{hd} \\
           1/2 &     1/2 & S_{ld} \\ 
S_{c q^\prime} & S_{s q} &    S_T
\end{array} 
\right \} .  
\label{eq:cbar-c-q-qprime-s-spin-contribution}  
\end{eqnarray}
The matrix elements of the rest two operators  
$2 \, (\tilde{\mathcal{K}}_{c s})_{\bar 3} \, (\mathbf{S}_c \cdot \mathbf{S}_s)$ and 
$2 \, (\tilde{\mathcal{K}}_{q q})_{\bar 3} \, (\mathbf{S}_q \cdot \mathbf{S}_{q^\prime})$ 
are calculated similarly with the following result: 
\begin{eqnarray}
&& \hspace{-3mm}
{}_J \langle S_{hd}^\prime, S_t^\prime, L_t; S_{ld}^\prime, L_{ld}; S, L |  
2 \, (\tilde{\mathcal{K}}_{c s})_{\bar 3} \, (\mathbf{S}_c \cdot \mathbf{S}_s) + 
2 \, (\tilde{\mathcal{K}}_{q q})_{\bar 3} \, (\mathbf{S}_q \cdot \mathbf{S}_{q^\prime}) 
| S_{hd}, S_t, L_t; S_{ld}, L_{ld}; S, L \rangle_J
\nonumber \\ 
&& \hspace{5mm}
= \left [ (\tilde{\mathcal{K}}_{c s})_{\bar 3} + (\tilde{\mathcal{K}}_{q q})_{\bar 3} \right ] 
\sqrt{(2 S_t + 1) \, (2 S_t^\prime + 1) \, 
      (2 S_{hd} + 1) \, (2 S_{hd}^\prime + 1) \, 
      (2 S_{ld} + 1) \, (2 S_{ld}^\prime + 1)}
\nonumber \\ 
&& \hspace{5mm}
\times 
(-1)^{S_{hd} + S_{hd}^\prime + S_{ld} + S_{ld}^\prime + 2 S + 1}
\!\!\!\! \sum_{S_T, S_{c s}, S_{q q^\prime}} \!\!\!\! 
(2 S_T + 1)\, (2 S_{c s} + 1)\, (2 S_{q q^\prime} + 1)\, 
\left [ S_{c s} (S_{c s} + 1 ) - \frac{3}{2} \right ]  
\nonumber \\ 
&& \hspace{5mm}
\times 
\left \{ 
\begin{array}{ccc} 
 S_{ld}^\prime & S_{hd}^\prime &        S_T \\
           1/2 &             S & S_t^\prime  
\end{array} 
\right \} 
\left \{ 
\begin{array}{ccc} 
 S_{ld} & S_{hd} & S_T \\
    1/2 &      S & S_t  
\end{array} 
\right \} 
\left \{ 
\begin{array}{ccc} 
    1/2 &            1/2 & S_{hd}^\prime \\
    1/2 &            1/2 & S_{ld}^\prime \\ 
S_{c s} & S_{q q^\prime} &           S_T
\end{array} 
\right \} 
\left \{ 
\begin{array}{ccc} 
    1/2 &            1/2 & S_{hd} \\
    1/2 &            1/2 & S_{ld} \\ 
S_{c s} & S_{q q^\prime} &    S_T
\end{array} 
\right \} .  
\label{eq:cbar-c-q-s-qprime-spin-contribution}  
\end{eqnarray}
In general, eqs.~(\ref{eq:cbar-c-q-qprime-s-spin-contribution}) 
and~(\ref{eq:cbar-c-q-s-qprime-spin-contribution}) are the matrices which 
mix all the states having the same total spin,~$S$, and the same parity.

The last flavor state which corresponds to the hidden-charm strange pentaquark with the 
strangeness $S = -2$ consists of the the strange doubly-heavy triquark and strange light 
diquark. The corresponding spin-spin effective Hamiltonian~(\ref{eq:H-SS-t-d}) does not get
simplified in this case also, and all the six terms, each of which is accompanied by its 
own coupling, should be considered separately: 
\begin{eqnarray}
H_{SS}^{t-ld} & = &  
2 \, \tilde{\mathcal{K}}_{\bar c q} \, (\mathbf{S}_{\bar c} \cdot \mathbf{S}_q) +  
2 \, (\tilde{\mathcal{K}}_{c q})_{\bar 3} \, (\mathbf{S}_c \cdot \mathbf{S}_q) + 
2 \, (\tilde{\mathcal{K}}_{s q})_{\bar 3} \, (\mathbf{S}_s \cdot \mathbf{S}_q)     
\label{eq:H-SS-t-d-strange-3} \\ 
& + & 
2 \, \tilde{\mathcal{K}}_{\bar c s} \, (\mathbf{S}_{\bar c} \cdot \mathbf{S}_{s^\prime}) +  
2 \, (\tilde{\mathcal{K}}_{c s})_{\bar 3} \, (\mathbf{S}_c \cdot \mathbf{S}_{s^\prime}) + 
2 \, (\tilde{\mathcal{K}}_{s s})_{\bar 3} \, (\mathbf{S}_s \cdot \mathbf{S}_{s^\prime}) ,    
\nonumber 
\end{eqnarray}
where we make the substitution $q = s$, $q^\prime = q$, and $q^{\prime\prime} = s^\prime$. 
The above consideration holds also here, and one can use the 
expressions~(\ref{eq:cbar-hd-s-qprime-spin-contribution}),
(\ref{eq:cbar-c-q-qprime-s-spin-contribution}), 
and~(\ref{eq:cbar-c-q-s-qprime-spin-contribution}) with the obvious replacements 
$(\tilde{\mathcal{K}}_{s q})_{\bar 3} \to (\tilde{\mathcal{K}}_{s s})_{\bar 3}$ 
in~(\ref{eq:cbar-c-q-qprime-s-spin-contribution}) and  
$(\tilde{\mathcal{K}}_{q q})_{\bar 3} \to (\tilde{\mathcal{K}}_{s q})_{\bar 3}$ 
in~(\ref{eq:cbar-c-q-s-qprime-spin-contribution}). 

The contributions discussed in this appendix are relevant for the Model~I by Maiani et al.~\cite{Maiani:2004vq},
where such (subdominant) spin-spin interactions are included in the effective Hamiltonian. 
We presented the formalism here. They are neglected in the numerical analysis given in this paper, 
as we use the Model~II~\cite{Maiani:2014aja} for our mass estimates.

\section{\boldmath Mass derivations for the $P$-wave pentaquark states}
\label{sec:mass-derivations}

\subsection[The states with $J^P = 3/2^+$ and triquark spin $S_t = 3/2$]
           {\boldmath The states with $J^P = 3/2^+$ and triquark spin $S_t = 3/2$} 
\label{ssec:J-3/2-St-3/2}

The mass matrix of the states with the spin-parity $J^P = 3/2^+$ 
and triquark spin $S_t = 1/2$ is as follows: 
\begin{eqnarray}  
&& \hspace*{-3mm}
\tilde M^{P1}_{J = 3/2} = M_0 + 
\frac{1}{2} \, (\mathcal{K}_{c q})_{\bar 3} + 
\frac{1}{2} \, (\mathcal{K}_{q^\prime q^{\prime\prime}})_{\bar 3} 
+ \frac{1}{2} \left ( \mathcal{K}_{\bar c q} + \mathcal{K}_{\bar c c} \right )  
+ B  
\nonumber \\ 
&& \hspace*{17mm}
+ \frac{1}{15} \, A_t 
\left ( 
\begin{array}{ccc} 
         25 & 25 \sqrt 2 &         0 \\
 25 \sqrt 2 &        -22 & 9 \sqrt 6 \\ 
          0 &  9 \sqrt 6 &       -63 
\end{array}
\right ) 
+ \frac{1}{15} \, A_{ld} 
\left ( 
\begin{array}{ccc} 
        -10 & 25 \sqrt 2 &         0 \\
 25 \sqrt 2 &         -8 & 9 \sqrt 6 \\ 
          0 &  9 \sqrt 6 &       -42 
\end{array}
\right ) 
\nonumber \\ 
&& \hspace*{17mm}
+ b 
\left ( 
\begin{array}{ccc} 
       -10/3 &      \sqrt 2/4 &   -3 \sqrt 3/5 \\
   \sqrt 2/4 &      -1256/375 & 81 \sqrt 6/100 \\ 
-3 \sqrt 3/5 & 81 \sqrt 6/100 &       -372/125  
\end{array}
\right ) .   
\label{eq:MM-P1-3/2-so}  
\end{eqnarray}
Before calculating the mass spectrum, we make the three matrices above to be traceless:    
\begin{eqnarray}  
&& \hspace*{-3mm}
\tilde M^{P1}_{J = 3/2} = M_0 + 
\frac{1}{2} \, (\mathcal{K}_{c q})_{\bar 3} + 
\frac{1}{2} \, (\mathcal{K}_{q^\prime q^{\prime\prime}})_{\bar 3} 
+ \frac{1}{2} \left ( \mathcal{K}_{\bar c q} + \mathcal{K}_{\bar c c} \right )  
+ B - \frac{4}{3} \left ( A_t + A_{ld} \right ) 
- \frac{3622}{1125} \, b  
\nonumber \\ 
&& \hspace*{17mm}
+ \frac{1}{15} \, A_t 
\left ( 
\begin{array}{ccc} 
         45 & 25 \sqrt 2 &         0 \\
 25 \sqrt 2 &         -2 & 9 \sqrt 6 \\ 
          0 &  9 \sqrt 6 &       -43 
\end{array}
\right ) 
+ \frac{1}{15} \, A_{ld} 
\left ( 
\begin{array}{ccc} 
         10 & 25 \sqrt 2 &         0 \\
 25 \sqrt 2 &         12 & 9 \sqrt 6 \\ 
          0 &  9 \sqrt 6 &       -22 
\end{array}
\right ) 
\nonumber \\ 
&& \hspace*{17mm} 
+  b 
\left ( 
\begin{array}{ccc} 
   -128/1125 &      \sqrt 2/4 &   -3 \sqrt 3/5 \\
   \sqrt 2/4 &      -146/1125 & 81 \sqrt 6/100 \\ 
-3 \sqrt 3/5 & 81 \sqrt 6/100 &       274/1125  
\end{array}
\right ) .   
\label{eq:MM-P1-3/2-so-TLM}  
\end{eqnarray}
There is a standard procedure to find the eigenvalues,~$\lambda_i$, 
$i = 1,\, 2,\, 3$, of the traceless symmetric $(3 \times 3)$ matrix. 
Denoting such a matrix as 
\begin{equation} 
A = 
\left ( 
\begin{array}{ccc} 
A_{11} & A_{12} & A_{13} \\ 
A_{12} & A_{22} & A_{23} \\ 
A_{13} & A_{23} & A_{33} \\ 
\end{array}
\right ) , 
\label{eq:matrix-3-times-3}
\end{equation}
its characteristic equation 
\begin{equation} 
\left ( \lambda - \lambda_1 \right )
\left ( \lambda - \lambda_2 \right )
\left ( \lambda - \lambda_3 \right ) = 
\lambda^3 - I_1 \lambda^2 - 3 I_2 \lambda - 2 I_3 = 0 , 
\label{eq:CE-3-times-3-matrix}
\end{equation}
is expressed in terms of three rotational invariants: 
\begin{equation} 
I_1 = \Tr (A) = A_{11} + A_{22} + A_{33} = 
\lambda_1 + \lambda_2 + \lambda_3 = 0 , 
\label{eq:I1-def} 
\end{equation} 
\begin{equation} 
I_2 = \frac{1}{3} \left [ 
A_{12}^2 + A_{13}^2 + A_{23}^2 - A_{11} A_{22} - A_{11} A_{33} - A_{22} A_{33}    
\right ]  = 
- \frac{1}{3} \left ( 
\lambda_1 \lambda_2 + \lambda_1 \lambda_3 + \lambda_2 \lambda_3 
\right ) ,  
\label{eq:I2-def} 
\end{equation} 
\begin{equation} 
I_3 = \frac{1}{2} \det (A) = 
\frac{1}{2} \left ( A_{11} A_{22} A_{33} + 2 A_{12} A_{13} A_{23} - 
A_{11} A_{23}^2 - A_{22} A_{13}^2 - A_{33} A_{12}^2 
\right ) = 
\frac{1}{2} \, \lambda_1 \lambda_2 \lambda_3 .  
\label{eq:I3-def}  
\end{equation}

Since real eigenvalues are of our interest, two non-trivial 
invariants,~$I_2$ and~$I_3$, must satisfy the following conditions: 
\begin{equation} 
I_2 > 0, 
\qquad 
I_2^3 > I_3^2 . 
\label{eq:invariant-conditions}
\end{equation}
If these conditions are fulfilled, we can define the angle: 
\begin{equation}
\phi = \frac{1}{3} \, \arccos \sqrt{\frac{I_3^2}{I_2^3}} , 
\label{eq:phi-def}
\end{equation}
through which the sorted eigenvalues ($\lambda_1 > \lambda_2 > \lambda_3$)
can be written as follows: 
\begin{equation}
\lambda_1 = 2 \sqrt{I_2} \cos (\phi) , 
\qquad 
\lambda_2 = - 2 \sqrt{I_2} \cos (\pi/3 + \phi) , 
\qquad 
\lambda_3 = - 2 \sqrt{I_2} \cos (\pi/3 - \phi) .  
\label{eq:lambda-1-2-3}
\end{equation}
After these eigenvalues are obtained, the masses are calculated easily: 
\begin{equation}  
m^{P1}_{11,12,13} = M_0  
+ \frac{1}{4} \left ( \mathcal{K}_{\bar c q} + \mathcal{K}_{\bar c c} \right )  
\left ( 2 + r_{hd} + r_{ld} \right ) 
+ B - \frac{4}{3} \left ( A_t + A_{ld} \right ) 
- \frac{3622}{1125} \, b + \lambda_{3,2,1} . 
\label{eq:mass-P1-11-12-13-app}  
\end{equation}

\subsection[The states with $J^P = 3/2^+$ and triquark spin $S_t = 1/2$]
           {\boldmath The states with $J^P = 3/2^+$ and triquark spin $S_t = 1/2$} 
\label{ssec:J-3/2-St-1/2}

To derive the expressions for the masses of the four states with the spin-parity 
$J^P = 3/2^+$ and triquark spin $S_t = 1/2$ from table~\ref{tab:P-wave-pentaquarks-bad-ld}, 
one needs to find the eigenvalues of the following non-diagonal symmetric $(4 \times 4)$ matrix: 
\begin{eqnarray}  
&& 
M^{P1}_{J = 3/2} = M_0 + 
\frac{1}{2} \, (\mathcal{K}_{c q})_{\bar 3} 
\left ( 
\begin{array}{cccc} 
-3 &  0 &  0 &  0 \\
 0 & -3 &  0 &  0 \\ 
 0 &  0 &  1 &  0 \\
 0 &  0 &  0 &  1 \\ 
\end{array}
\right ) 
+ \frac{1}{2} \, (\mathcal{K}_{q^\prime q^{\prime\prime}})_{\bar 3} 
\nonumber \\ 
&& \hspace*{11mm}
+ \frac{1}{2} \left ( \mathcal{K}_{\bar c q} + \mathcal{K}_{\bar c c} \right )  
\left ( 
\begin{array}{cccc} 
       0 &       0 & \sqrt 3 &       0 \\
       0 &       0 &       0 & \sqrt 3 \\ 
 \sqrt 3 &       0 &      -2 &       0 \\
       0 & \sqrt 3 &       0 &      -2 \\ 
\end{array}
\right ) 
+ B  
- \frac{1}{3} \, A_t 
\left ( 
\begin{array}{cccc} 
         1 & -2 \sqrt 5 &          0 &          0 \\
-2 \sqrt 5 &          2 &          0 &          0 \\
         0 &          0 &          1 & -2 \sqrt 5 \\ 
         0 &          0 & -2 \sqrt 5 &          2
\end{array}
\right ) 
\nonumber \\ 
&& \hspace*{11mm}
+ \frac{2}{3} \, A_{ld} 
\left ( 
\begin{array}{cccc} 
       2 & \sqrt 5 &       0 &       0 \\
 \sqrt 5 &      -2 &       0 &       0 \\
       0 &       0 &       2 & \sqrt 5 \\ 
       0 &       0 & \sqrt 5 &      -2
\end{array}
\right ) 
+ \frac{1}{30} \, b 
\left ( 
\begin{array}{cccc} 
       -40 & 3 \sqrt 5 &         0 &         0 \\
 3 \sqrt 5 &       -16 &         0 &         0 \\
         0 &         0 &       -40 & 3 \sqrt 5 \\ 
         0 &         0 & 3 \sqrt 5 &       -16
\end{array}
\right ) . 
\label{eq:MM-P1-3/2-ss-so}  
\end{eqnarray}
Noting that the last three matrices are block-diagonal and their non-trivial 
$(2 \times 2)$ blocks are identical. A diagonalization of these matrices 
is reduced to the diagonalization of the blocks: 
\begin{eqnarray} 
&& 
- \frac{A_t}{3}  
\left ( 
\begin{array}{cc} 
         1 & -2 \sqrt 5 \\
-2 \sqrt 5 &          2 
\end{array}
\right ) 
+ \frac{2}{3} \, A_{ld} 
\left ( 
\begin{array}{cc} 
       2 & \sqrt 5 \\
 \sqrt 5 &      -2 
\end{array}
\right ) 
+ \frac{b}{30}  
\left ( 
\begin{array}{cc} 
       -40 & 3 \sqrt 5 \\
 3 \sqrt 5 &       -16 
\end{array}
\right ) 
\nonumber \\ 
&& \hspace*{33mm}
\Longrightarrow  
- \frac{A_t}{2} - \frac{14}{15} \, b + 
%  \frac{1}{2} \left ( \mathcal{K}_{\bar c q} + \mathcal{K}_{\bar c c} \right ) 
\left ( 
\begin{array}{cc} 
 \mu^{(3/2)} &            0 \\
           0 & -\mu^{(3/2)} 
\end{array}
\right ) , 
\label{eq:eigenvalues-3/2} 
\end{eqnarray}
where the following parameter is introduced: 
\begin{equation} 
\mu^{(3/2)} = \frac{1}{30} \, 
%  \frac{1}{15 \left ( \mathcal{K}_{\bar c q} + \mathcal{K}_{\bar c c} \right )} \, 
\sqrt{\left ( 5 A_t + 40 A_{ld} - 12 b \right )^2 + 5 \left ( 20 A_t + 20 A_{ld} + 3 b \right )^2} . 
\label{eq:mu-3/2-def-app} 
\end{equation}
Coming back to the $(4 \times 4)$ mass matrix~(\ref{eq:MM-P1-3/2-ss-so}), 
it is necessary to point out that the transformation to the diagonal form 
of the sum of the last three matrices in~(\ref{eq:MM-P1-3/2-ss-so}) can 
be done by using the orthogonal matrix: 
\begin{equation} 
a (\alpha) = 
\left ( 
\begin{array}{cccc}
 \cos\alpha &  \sin\alpha &           0 &           0 \\ 
-\sin\alpha &  \cos\alpha &           0 &           0 \\ 
          0 &           0 &  \cos\alpha &  \sin\alpha \\ 
          0 &           0 & -\sin\alpha &  \cos\alpha  
\end{array}
\right ) , 
\label{eq:orthogonal-matrix}
\end{equation} 
with some fixed value of the angle~$\alpha$. Under such a transformation, the 
two matrices in the first row in~(\ref{eq:MM-P1-3/2-ss-so}) remain unchanged. 
Hence, the $(4 \times 4)$ mass matrix~(\ref{eq:MM-P1-3/2-ss-so}) can 
be presented in the form: 
\begin{eqnarray}  
&& 
M^{P1}_{J = 3/2} = M_0 - 
\frac{1}{4} \left ( \mathcal{K}_{\bar c q} + \mathcal{K}_{\bar c c} \right )  
\left ( 2 + r_{hd} - r_{ld} \right ) + 
B - \frac{A_t}{2} - \frac{14}{15} \, b 
+ \frac{1}{2} \left ( \mathcal{K}_{\bar c q} + \mathcal{K}_{\bar c c} \right )  
\nonumber \\  
&& \hspace*{7mm} \times 
\left ( 
\begin{array}{cccc} 
 1 - r_{hd} + \tilde\mu^{(3/2)} &                              0 & 
                        \sqrt 3 &                              0 \\
                              0 & 1 - r_{hd} - \tilde\mu^{(3/2)} & 
                              0 &                        \sqrt 3 \\ 
                        \sqrt 3 &                              0 & 
 r_{hd} - 1 + \tilde\mu^{(3/2)} &                              0 \\
                              0 &                        \sqrt 3 & 
                              0 & r_{hd} - 1 - \tilde\mu^{(3/2)}  
\end{array}
\right ) , 
\qquad  
\label{eq:MM-P1-3/2-ss-so-rotated}  
\end{eqnarray}
where $\tilde\mu^{(3/2)} = 2 \mu^{(3/2)}/( \mathcal{K}_{\bar c q} + \mathcal{K}_{\bar c c})$. 
Applying again the orthogonal transformations, the last matrix can be diagonalized, 
and we get the following set of masses:   
\begin{equation}  
m^{P1}_7 = M_0 + B - \frac{A_t}{2} - \frac{14}{15} \, b - 
\frac{1}{4} \left ( \mathcal{K}_{\bar c q} + \mathcal{K}_{\bar c c} \right )  
\left ( 2 + r_{hd} - r_{ld} + 2 \sqrt{3 + (1 - r_{hd})^2} \right ) - \mu^{(3/2)} , 
\label{eq:mass-P1-7-app} 
\end{equation}
\begin{equation}  
m^{P1}_8 = M_0 + B - \frac{A_t}{2} - \frac{14}{15} \, b - 
\frac{1}{4} \left ( \mathcal{K}_{\bar c q} + \mathcal{K}_{\bar c c} \right )  
\left ( 2 + r_{hd} - r_{ld} + 2 \sqrt{3 + (1 - r_{hd})^2} \right ) + \mu^{(3/2)} , 
\label{eq:mass-P1-8-app}   
\end{equation}
\begin{equation}  
m^{P1}_9 = M_0 + B - \frac{A_t}{2} - \frac{14}{15} \, b - 
\frac{1}{4} \left ( \mathcal{K}_{\bar c q} + \mathcal{K}_{\bar c c} \right )  
\left ( 2 + r_{hd} - r_{ld} - 2 \sqrt{3 + (1 - r_{hd})^2} \right ) - \mu^{(3/2)} , 
\label{eq:mass-P1-9-app}   
\end{equation}
\begin{equation}  
m^{P1}_{10} = M_0 + B - \frac{A_t}{2} - \frac{14}{15} \, b - 
\frac{1}{4} \left ( \mathcal{K}_{\bar c q} + \mathcal{K}_{\bar c c} \right )  
\left ( 2 + r_{hd} - r_{ld} - 2 \sqrt{3 + (1 - r_{hd})^2} \right ) + \mu^{(3/2)} .  
\label{eq:mass-P1-10-app}   
\end{equation}

\subsection[The states with $J^P = 1/2^+$ and triquark spin $S_t = 1/2$]
           {\boldmath The states with $J^P = 1/2^+$ and triquark spin $S_t = 1/2$} 
\label{ssec:J-1/2-St-1/2}

For the first four states with the spin-parity $J^P = 1/2^+$ and triquark spin $S_t = 1/2$ 
from table~\ref{tab:P-wave-pentaquarks-bad-ld}, the mass matrix is again the non-diagonal 
symmetric $(4 \times 4)$ matrix: 
\begin{eqnarray}  
&& 
M^{P1}_{J = 1/2} = M_0 + 
\frac{1}{2} \, (\mathcal{K}_{c q})_{\bar 3} 
\left ( 
\begin{array}{cccc} 
-3 &  0 &  0 &  0 \\
 0 & -3 &  0 &  0 \\ 
 0 &  0 &  1 &  0 \\
 0 &  0 &  0 &  1 \\ 
\end{array}
\right ) 
+ \frac{1}{2} \, (\mathcal{K}_{q^\prime q^{\prime\prime}})_{\bar 3} 
\nonumber \\ 
&& \hspace*{11mm}
+ \frac{1}{2} \left ( \mathcal{K}_{\bar c q} + \mathcal{K}_{\bar c c} \right )  
\left ( 
\begin{array}{cccc} 
       0 &       0 & \sqrt 3 &       0 \\
       0 &       0 &       0 & \sqrt 3 \\ 
 \sqrt 3 &       0 &      -2 &       0 \\
       0 & \sqrt 3 &       0 &      -2 \\ 
\end{array}
\right ) 
+ B  
+ \frac{1}{3} \, A_t 
\left ( 
\begin{array}{cccc} 
         2 & 2 \sqrt 2 &         0 &         0 \\
 2 \sqrt 2 &        -5 &         0 &         0 \\
         0 &         0 &         2 & 2 \sqrt 2 \\ 
         0 &         0 & 2 \sqrt 2 &        -5
\end{array}
\right ) 
\nonumber \\ 
&& \hspace*{11mm}
+ \frac{2}{3} \, A_{ld} 
\left ( 
\begin{array}{cccc} 
      -4 & \sqrt 2 &       0 &       0 \\
 \sqrt 2 &      -5 &       0 &       0 \\
       0 &       0 &      -4 & \sqrt 2 \\ 
       0 &       0 & \sqrt 2 &      -5
\end{array}
\right ) 
+ \frac{1}{30} \, b 
\left ( 
\begin{array}{cccc} 
        -16 & 21 \sqrt 2 &          0 &          0 \\
 21 \sqrt 2 &        -46 &          0 &          0 \\
          0 &          0 &        -16 & 21 \sqrt 2 \\ 
          0 &          0 & 21 \sqrt 2 &        -46
\end{array}
\right ) . 
\label{eq:MM-P1-1/2-ss-so}  
\end{eqnarray}
The rotational matrix~(\ref{eq:orthogonal-matrix}) with some fixed value 
of the angle~$\alpha$ transforms the sum of three matrices in the second 
line to the diagonal form. As in the case of $J^P = 3/2^+$ pentaquarks, 
the diagonalization procedure can be applied to the $(2 \times 2)$ blocks: 
\begin{eqnarray} 
&& 
\frac{A_t}{3}  
\left ( 
\begin{array}{cc} 
         2 & 2 \sqrt 2 \\
 2 \sqrt 2 &        -5 
\end{array}
\right ) 
+ \frac{2}{3} \, A_{ld} 
\left ( 
\begin{array}{cc} 
      -4 & \sqrt 2 \\
 \sqrt 2 &      -5 
\end{array}
\right ) 
+ \frac{b}{30}  
\left ( 
\begin{array}{cc} 
        -16 & 21 \sqrt 2 \\
 21 \sqrt 2 &        -46 
\end{array}
\right ) 
\nonumber \\ 
&& \hspace*{33mm}
\Longrightarrow 
- \frac{A_t}{2} - 3 A_{ld} - \frac{31}{30} \, b + 
%  \frac{1}{2} \left ( \mathcal{K}_{\bar c q} + \mathcal{K}_{\bar c c} \right ) 
\left ( 
\begin{array}{cc} 
 \mu^{(1/2)} &            0 \\
           0 & -\mu^{(1/2)} 
\end{array}
\right ) , 
\label{eq:eigenvalues-1/2} 
\end{eqnarray}
where the following parameter is introduced: 
\begin{equation} 
\mu^{(1/2)} = \frac{1}{30} \, 
%  \frac{1}{15 \left ( \mathcal{K}_{\bar c q} + \mathcal{K}_{\bar c c} \right )} \, 
\sqrt{25 \left ( 7 A_t + 2 A_{ld} + 3 b \right )^2 + 2 \left ( 20 A_t + 20 A_{ld} + 21 b \right )^2} . 
\label{eq:mu-1/2-def-app} 
\end{equation}
After this transformation is done, the $(4 \times 4)$ mass 
matrix~(\ref{eq:MM-P1-1/2-ss-so}) takes the form: 
\begin{eqnarray}  
&& 
M^{P1}_{J = 1/2} = M_0 - 
\frac{1}{4} \left ( \mathcal{K}_{\bar c q} + \mathcal{K}_{\bar c c} \right )  
\left ( 2 + r_{hd} - r_{ld} \right ) + 
B - \frac{A_t}{2} - 3 A_{ld} - \frac{31}{30} \, b 
+ \frac{1}{2} \left ( \mathcal{K}_{\bar c q} + \mathcal{K}_{\bar c c} \right )  
\nonumber \\  
&& \hspace*{11mm}
\times \left ( 
\begin{array}{cccc} 
 1 - r_{hd} + \tilde\mu^{(1/2)} &                              0 &
                        \sqrt 3 &                              0 \\
                              0 & 1 - r_{hd} - \tilde\mu^{(1/2)} & 
                              0 &                        \sqrt 3 \\ 
                        \sqrt 3 &                              0 & 
 r_{hd} - 1 + \tilde\mu^{(1/2)} &                              0 \\
                              0 &                        \sqrt 3 & 
                              0 & r_{hd} - 1 - \tilde\mu^{(1/2)}  
\end{array}
\right ) , 
\label{eq:MM-P1-1/2-ss-so-rotated} 
\end{eqnarray}
where $\tilde\mu^{(1/2)} = 2 \mu^{(1/2)}/( \mathcal{K}_{\bar c q} + \mathcal{K}_{\bar c c})$. 
The last matrix can be diagonalized by using the orthogonal transformations 
and we get the following set of masses:   
\begin{eqnarray} 
&& 
m^{P1}_1 = M_0 + B - \frac{A_t}{2} - 3 A_{ld} - \frac{31}{30} \, b 
\nonumber \\ 
&& \hspace*{11mm}
- \frac{1}{4} \left ( \mathcal{K}_{\bar c q} + \mathcal{K}_{\bar c c} \right )  
\left ( 2 + r_{hd} - r_{ld} + 2 \sqrt{3 + (1 - r_{hd})^2} \right ) - \mu^{(1/2)} , 
\label{eq:mass-P1-1-app} \\  
&& 
m^{P1}_2 = M_0 + B - \frac{A_t}{2} - 3 A_{ld} - \frac{31}{30} \, b 
\nonumber \\ 
&& \hspace*{11mm}
- \frac{1}{4} \left ( \mathcal{K}_{\bar c q} + \mathcal{K}_{\bar c c} \right )  
\left ( 2 + r_{hd} - r_{ld} + 2 \sqrt{3 + (1 - r_{hd})^2} \right ) + \mu^{(1/2)} , 
\label{eq:mass-P1-2-app} \\  
&& 
m^{P1}_3 = M_0 + B - \frac{A_t}{2} - 3 A_{ld} - \frac{31}{30} \, b 
\nonumber \\ 
&& \hspace*{11mm}
- \frac{1}{4} \left ( \mathcal{K}_{\bar c q} + \mathcal{K}_{\bar c c} \right )  
\left ( 2 + r_{hd} - r_{ld} - 2 \sqrt{3 + (1 - r_{hd})^2} \right ) - \mu^{(1/2)} , 
\label{eq:mass-P1-3-app} \\  
&& 
m^{P1}_4 = M_0 + B - \frac{A_t}{2} - 3 A_{ld} - \frac{31}{30} \, b 
\nonumber \\ 
&& \hspace*{11mm}
- \frac{1}{4} \left ( \mathcal{K}_{\bar c q} + \mathcal{K}_{\bar c c} \right )  
\left ( 2 + r_{hd} - r_{ld} - 2 \sqrt{3 + (1 - r_{hd})^2} \right ) + \mu^{(1/2)} .  
\label{eq:mass-P1-4-app}   
\end{eqnarray}

\section{\boldmath $\chi^2$-analysis of orbitally-excited $\Omega_c^*$-baryons}
\label{sec:chi2-Omega-c}

Existing experimental data on the masses of orbitally-excited $\Omega_c^*$-baryons 
were obtained by the LHCb~\cite{Aaij:2017nav} and Belle~\cite{Yelton:2017qxg} 
collaborations and collected in table~\ref{tab:Omega-c-*-data}. 
\begin{table}[tb] 
\begin{center}
\begin{tabular}{|cccc|} \hline 
     Baryon       & $J^P$~\cite{Karliner:2017kfm} &          LHCb            &          Belle           \\ \hline 
$\Omega_c (3000)$ &            $1/2^-$            & $3000.4 \pm 0.2 \pm 0.1$ & $3000.7 \pm 1.0 \pm 0.2$ \\ 
$\Omega_c (3050)$ &            $1/2^-$            & $3050.2 \pm 0.1 \pm 0.1$ & $3050.2 \pm 0.4 \pm 0.2$ \\ 
$\Omega_c (3066)$ &            $3/2^-$            & $3065.5 \pm 0.1 \pm 0.3$ & $3064.9 \pm 0.6 \pm 0.2$ \\ 
$\Omega_c (3090)$ &            $3/2^-$            & $3090.2 \pm 0.3 \pm 0.5$ & $3089.3 \pm 1.2 \pm 0.2$ \\ 
$\Omega_c (3119)$ &            $5/2^-$            & $3119.0 \pm 0.3 \pm 0.9$ &         $\cdots$         \\ \hline
\end{tabular} 
\end{center}
\caption{
Experimental data from the LHCb~\cite{Aaij:2017nav} and Belle~\cite{Yelton:2017qxg} 
collaborations. (Masses are given in MeV.) All the measurements have the uncertainty 
$^{+0.3}_{-0.5}$ MeV coming from the mass of the ground-state $\Xi_c^+$-baryon, which 
is common in both analysis. The spin-parity,~$J^P$, assignment assumes that all the 
states are orbitally-excited baryons as suggested in~\cite{Karliner:2017kfm}.   
}
\label{tab:Omega-c-*-data}

\end{table}

The theoretical expressions for the orbitally-exciting $\Omega_c^*$-baryon 
masses follow from the effective Hamiltonian~\cite{Karliner:2017kfm}\footnote{
The coefficients of the spin-orbit,~$a_1$ and~$a_2$, and spin-spin,~$c$, interactions 
in the spin-dependent part of the Hamiltonian,~$V_{\rm SD}$, differ by a factor~2 
from the ones defined in~\cite{Karliner:2017kfm,Ali:2017wsf}.}:
\begin{eqnarray}
H_{\rm eff} & = & m_c + m_{\{ss\}} + 
2 \kappa_{ss} \left ( {\bf S}_{s_1} \cdot {\bf S}_{s_2} \right ) + 
\frac{B_{\mathcal Q}}{2}\, {\bf L}^2 + V_{\rm SD}, 
\label{Hamiltonian-OC} \\
V_{\rm SD} & = & 
2 a_1 \left ( {\bf L} \cdot {\bf S}_{\{ ss \}} \right ) + 
2 a_2 \left ( {\bf L} \cdot {\bf S}_c \right ) + b\, \frac{S_{12}}{4} + 
2 c \left ( {\bf S}_{\{ ss \}} \cdot {\bf S}_c \right ) . 
\nonumber
\end{eqnarray}
In eq.~(\ref{Hamiltonian-OC}),~$m_c$ and~$m_{\{ss\}}$ are the masses of the $c$-quark 
and the spin-1 $\{ss\}$-diquark, respectively, $\kappa_{ss}$ is the spin-spin coupling 
of the quarks in the diquark, ${\bf L}$ is the orbital angular momentum of the 
$\Omega_c^*$-baryon, and $B_{\mathcal Q}$ is the orbital coupling. The coefficients~$a_1$ 
and~$a_2$ are the strengths of the spin-orbit terms involving the spin of the diquark 
${\bf S}_{\{ss\}}$ and the charm-quark spin ${\bf S}_c$, respectively, $c$~is the strength 
of the spin-spin interaction between the diquark and the charm quark, and $S_{12}/4$ 
represents the tensor interaction, defined by 
\begin{equation}
\frac{S_{12}}{4} = Q ({\bf S}_{\{ss\}}, {\bf S}_c) = 
3 \left ( {\bf S}_{\{ss\}} \cdot {\bf n} \right ) \left ( {\bf S}_c \cdot {\bf n} \right ) - 
\left ( {\bf S}_{\{ss\}} \cdot {\bf S}_c \right ) ,  
\label{tensor}
\end{equation}
where 
% ${\bf S}_{\{ ss \}}$ and ${\bf S}_c$ are the spins of the diquark and charm quark, respectively, and 
${\bf n} = {\bf r}/r$ is the unit vector in the direction 
from charged quark to the doubly-strange diquark. 

One should remember that several required internal parameters are fixed: charm-quark 
spin $S_c = 1/2$, diquark spin $S_{\{ss\}} = 1$, and $\Omega_c^*$-baryon orbital 
angular momentum $L = 1$. So, in the $L - S$ coupling scheme, two possible values 
of the total spin $S = 1/2$ and $S = 3/2$, after their coupling to~$L$, allow to get 
five $P$-wave states: two with the total angular momentum $J = 1/2$, two with $J = 3/2$ 
and the last one with $J = 5/2$ (see table~\ref{tab:Omega-c-*-data}). In this scheme 
the spin-dependent part,~$V_{\rm SD}$, is represented by the block-diagonal matrix, 
each block of which corresponds to the states with a fixed value of~$J$~\cite{Karliner:2017kfm}: 
\begin{equation} 
\Delta\mathcal{M}_{1/2} = \frac{1}{3} \left ( 
\begin{array}{cc} 
      2 \left ( a_2 - 4 a_1 \right ) & 2 \sqrt 2 \left ( a_2 - a_1 \right ) \\ 
2 \sqrt 2 \left ( a_2 - a_1 \right ) &     - 5 \left ( 2 a_1 + a_2 \right )
\end{array}
\right ) 
+ \frac{b}{\sqrt 2} \left (  
\begin{array}{rr} 
 0 &         1 \\ 
 1 & - \sqrt 2 
\end{array}
\right ) 
+ c \left (  
\begin{array}{rr} 
-2 & 0 \\ 
 0 & 1 
\end{array}
\right ) , 
\label{eq:Delta-M-12}
\end{equation}
\begin{equation} 
\Delta\mathcal{M}_{3/2} = \frac{1}{3} \left ( 
\begin{array}{cc} 
                         4 a_1 - a_2 & 2 \sqrt 5 \left ( a_2 - a_1 \right ) \\ 
2 \sqrt 5 \left ( a_2 - a_1 \right ) &     - 2 \left ( 2 a_1 + a_2 \right )
\end{array}
\right ) 
+ \frac{b}{10} \left (  
\begin{array}{rr} 
         0 & - \sqrt 5 \\ 
 - \sqrt 5 &         8 
\end{array}
\right ) 
+ c \left (  
\begin{array}{rr} 
-2 & 0 \\ 
 0 & 1 
\end{array}
\right ) , 
\label{eq:Delta-M-32}
\end{equation}
\begin{equation} 
\Delta\mathcal{M}_{5/2} = 2 a_1 + a_2 - \frac{b}{5} + c .  
\label{eq:Delta-M-52}
\end{equation}

After diagonalizing the matrices and adding the common mass term 
\begin{equation} 
M_0^{(\Omega_c)} = m_c + m_{\{ss\}} + \frac{\kappa_{ss}}{2} + B_{\mathcal Q} ,  
\label{eq:M0-Omega-c-def}
\end{equation}
we get the set of five mass formulae: 
\begin{eqnarray} 
&&  % \hspace*{-11mm}
m^{(1/2)}_1 = M_0^{(\Omega_c)} - \frac{1}{2} \left ( 6 a_1 + a_2 + b + c \right ) 
\nonumber \\ 
&& \hspace*{11mm}
- \frac{1}{6} 
\sqrt{\left ( 2 a_1 + 7 a_2 + 3 b - 9 c \right )^2 + 2 \left ( 4 a_1 - 4 a_2 - 3 b \right )^2} , 
\label{eq:m-12-1} \\ 
&& % \hspace*{-11mm} 
m^{(1/2)}_2 = M_0^{(\Omega_c)} - \frac{1}{2} \left ( 6 a_1 + a_2 + b + c \right ) 
\nonumber \\ 
&& \hspace*{11mm}
+ \frac{1}{6} 
\sqrt{\left ( 2 a_1 + 7 a_2 + 3 b - 9 c \right )^2 + 2 \left ( 4 a_1 - 4 a_2 - 3 b \right )^2} , 
\label{eq:m-12-2} \\ 
% \end{eqnarray}
% 
% \begin{eqnarray} 
&& % \hspace*{-11mm} 
m^{(3/2)}_1 = M_0^{(\Omega_c)} - \frac{1}{10} \left ( 5 a_2 - 4 b + 5 c \right ) 
\nonumber \\ 
&& \hspace*{11mm}
- \frac{1}{30} 
\sqrt{\left ( 40 a_1 + 5 a_2 - 12 b - 45 c \right )^2 + 5 \left ( 20 a_1 - 20 a_2 + 3 b \right )^2} , 
\label{eq:m-32-1} \\ 
&& % \hspace*{-11mm} 
m^{(3/2)}_2 = M_0^{(\Omega_c)} - \frac{1}{10} \left ( 5 a_2 - 4 b + 5 c \right ) 
\nonumber \\ 
&& \hspace*{11mm}
+ \frac{1}{30} 
\sqrt{\left ( 40 a_1 + 5 a_2 - 12 b - 45 c \right )^2 + 5 \left ( 20 a_1 - 20 a_2 + 3 b \right )^2} ,
\label{eq:m-32-2} \\ 
&& % \hspace*{-11mm} 
m^{(5/2)} = M_0^{(\Omega_c)} + 2 a_1 + a_2 - \frac{b}{5} + c . 
\label{eq:m-52} 
\end{eqnarray}
There are five unknown variables $\{ M_0^{(\Omega_c)},\, a_1,\, a_2,\, b,\, c \}$ in these 
five equations. If there are five experimentally measured states which can be assigned with 
the $P$-wave $\Omega_c$-baryons, all five variables, in general, can be determined but their 
values do not always satisfy physical requirements.  

After the assignment is done as specified in table~\ref{tab:Omega-c-*-data}, it is possible 
to perform a $\chi^2$-analysis of the data. We start from the LHCb data alone as all five states 
are present there and then add the Belle measurements which allow to find~$\chi^2_{\rm min}$  
for four degrees of freedom. The best-fit values and~$1\sigma$ uncertainties of free parameters 
are presented in table~\ref{tab:Omega-c*-chi2-analysis}. There are no free degrees of freedom 
for the fit based on the LHCb data alone (the first row in table~\ref{tab:Omega-c*-chi2-analysis})
but combining them with the Belle data results $\chi^2_{\rm min}/{\rm ndf} = 0.87/4$.  
\begin{table}[tb] 
\begin{center}
\begin{tabular}{|cccccc|} \hline 
         & $M_0^{(\Omega_c)}$ &      $a_1$       &      $a_2$       &        $b$       &       $c$       \\ \hline 
    LHCb & $3079.89 \pm 0.40$ & $13.47 \pm 0.14$ & $12.86 \pm 0.38$ & $13.48 \pm 0.54$ & $2.00 \pm 0.22$ \\ 
Combined & $3079.80 \pm 0.39$ & $13.45 \pm 0.13$ & $12.94 \pm 0.36$ & $13.30 \pm 0.48$ & $2.01 \pm 0.20$ \\ \hline 
\end{tabular} 
\end{center}  
\caption{
The $\chi^2$-analysis of the orbitally-excited $\Omega_c^*$-baryon masses
based on the measurements by the LHCb and Belle collaborations. This analysis 
is performed based on the LHCb data alone and on the combined data set from 
both the LHCb and Belle collaborations presented in table~\ref{tab:Omega-c-*-data}.
All values of parameters are given in~MeV. The combine fit results 
$\chi^2_{\rm min}/{\rm ndf} = 0.87/4$. 
} 
\label{tab:Omega-c*-chi2-analysis}
\end{table}

The correlation matrices in both cases are as follows: 
\begin{equation}
R_{\rm LHCb} = \left ( 
\begin{array}{rrrrr} 
1 & 0.71 & 0.43 & -0.32 &  0.37 \\ 
  &    1 & 0.21 & -0.31 &  0.08 \\ 
  &      &    1 & -0.73 &  0.53 \\ 
  &      &      &     1 & -0.18 \\ 
  &      &      &       &     1  
\end{array}
\right ) , 
\quad 
R_{\rm comb} = \left ( 
\begin{array}{rrrrr} 
1 & 0.77 & 0.53 & -0.44 &  0.45 \\ 
  &    1 & 0.33 & -0.38 &  0.18 \\ 
  &      &    1 & -0.75 &  0.62 \\ 
  &      &      &     1 & -0.32 \\ 
  &      &      &       &     1  
\end{array}
\right ) .  
\end{equation}

The common mass~$M_0^{(\Omega_c)}$ is determined quite precisely. 
With the values of the input parameters from subsec.~\ref{ssec:input-parameters}: 
$m_c = (1710 \pm 10)$~MeV, $m_{\{ss\}} = (1099 \pm 15)$~MeV, and 
$\mathcal{K}_{ss} = (23 \pm 2)$~MeV, the strength of the orbital 
interaction can be estimated as: 
\begin{equation} 
B_{\mathcal Q} = (259 \pm 18)~\mathrm{MeV} . 
\label{eq:B-Omega-c-extimate}
\end{equation}


\begin{thebibliography}{99}

%\cite{Aaij:2019vzc}
\bibitem{Aaij:2019vzc} 
  R.~Aaij {\it et al.} [LHCb Collaboration],
  {\it Observation of a narrow pentaquark state, $P_c(4312)^+$, and of two-peak structure of the $P_c(4450)^+$},
  {\it Phys.\ Rev.\ Lett.}\  {\bf 122} (2019) 222001 
%   doi:10.1103/PhysRevLett.122.222001
  [arXiv:1904.03947 [hep-ex]].
  %%CITATION = doi:10.1103/PhysRevLett.122.222001;%%

%\cite{Aaij:2015tga}
\bibitem{Aaij:2015tga} 
  R.~Aaij {\it et al.} [LHCb Collaboration],
  {\it Observation of $J/\psi p$ Resonances Consistent with Pentaquark States in $\Lambda_b^0 \to J/\psi K^- p$ Decays},
  {\it Phys.\ Rev.\ Lett.}\  {\bf 115} (2015) 072001 
%   doi:10.1103/PhysRevLett.115.072001
  [arXiv:1507.03414 [hep-ex]].
  %%CITATION = doi:10.1103/PhysRevLett.115.072001;%%

%%%%%%%%%%%%%%%%%%%%%%%%%%%%%%%%%%%%%%%%%%%%%%%%%%%%%%%%%%%%%%%%%
%\cite{Tanabashi:2018oca}
\bibitem{Tanabashi:2018oca} 
  M.~Tanabashi {\it et al.} [Particle Data Group],
  {\it Review of Particle Physics},
  {\it Phys.\ Rev.}\ D {\bf 98} (2018) 030001.
%   doi:10.1103/PhysRevD.98.030001
  %%CITATION = doi:10.1103/PhysRevD.98.030001;%%

%\cite{Ali:2017jda}
\bibitem{Ali:2017jda} 
  A.~Ali, J.\,S.~Lange and S.~Stone,
  {\it Exotics: Heavy Pentaquarks and Tetraquarks},
  {\it Prog.\ Part.\ Nucl.\ Phys.}\  {\bf 97} (2017) 123 
%   doi:10.1016/j.ppnp.2017.08.003
  [arXiv:1706.00610 [hep-ph]].
  %%CITATION = doi:10.1016/j.ppnp.2017.08.003;%%


%\cite{Esposito:2016noz}
\bibitem{Esposito:2016noz} 
  A.~Esposito, A.~Pilloni and A.\,D.~Polosa,
  {\it Multiquark Resonances},
  {\it Phys.\ Rept.}\  {\bf 668} (2016) 1 
%   doi:10.1016/j.physrep.2016.11.002
  [arXiv:1611.07920 [hep-ph]].
  %%CITATION = doi:10.1016/j.physrep.2016.11.002;%%


%\cite{Chen:2016qju}
\bibitem{Chen:2016qju} 
  H.~X.~Chen, W.~Chen, X.~Liu and S.~L.~Zhu,
  {\it The hidden-charm pentaquark and tetraquark states},
  {\it Phys.\ Rept.}\  {\bf 639} (2016) 1 
%   doi:10.1016/j.physrep.2016.05.004
  [arXiv:1601.02092 [hep-ph]].
  %%CITATION = doi:10.1016/j.physrep.2016.05.004;%%


%\cite{Guo:2017jvc}
\bibitem{Guo:2017jvc} 
  F.\,K.~Guo, C.~Hanhart, U.\,G.~Meissner, Q.~Wang, Q.~Zhao and B.\,S.~Zou,
  {\it Hadronic molecules},
  {\it Rev.\ Mod.\ Phys.}\  {\bf 90} (2018) 015004 
%   doi:10.1103/RevModPhys.90.015004
  [arXiv:1705.00141 [hep-ph]].
  %%CITATION = doi:10.1103/RevModPhys.90.015004;%%

%\cite{Olsen:2017bmm}
\bibitem{Olsen:2017bmm} 
  S.\,L.~Olsen, T.~Skwarnicki and D.~Zieminska,
  {\it Nonstandard heavy mesons and baryons: Experimental evidence},
  {\it Rev.\ Mod.\ Phys.}\  {\bf 90} (2018) 015003 
%   doi:10.1103/RevModPhys.90.015003
  [arXiv:1708.04012 [hep-ph]].
  %%CITATION = doi:10.1103/RevModPhys.90.015003;%%

%\cite{Lebed:2016hpi}
\bibitem{Lebed:2016hpi} 
  R.\,F.~Lebed, R.\,E.~Mitchell and E.\,S.~Swanson,
  {\it Heavy-Quark QCD Exotica},
  {\it Prog.\ Part.\ Nucl.\ Phys.}\  {\bf 93} (2017) 143 
%   doi:10.1016/j.ppnp.2016.11.003
  [arXiv:1610.04528 [hep-ph]].
  %%CITATION = doi:10.1016/j.ppnp.2016.11.003;%% 

%\cite{Hosaka:2016pey}
\bibitem{Hosaka:2016pey} 
  A.~Hosaka, T.~Iijima, K.~Miyabayashi, Y.~Sakai and S.~Yasui,
  {\it Exotic hadrons with heavy flavors: $X$, $Y$, $Z$, and related states},
  {\it PTEP} {\bf 2016} (2016) 062C01 
%   doi:10.1093/ptep/ptw045
  [arXiv:1603.09229 [hep-ph]].
  %%CITATION = doi:10.1093/ptep/ptw045;%% 

%\cite{Karliner:2017qhf}
\bibitem{Karliner:2017qhf} 
  M.~Karliner, J.\,L.~Rosner and T.~Skwarnicki,
  {\it Multiquark States},
  {\it Ann.\ Rev.\ Nucl.\ Part.\ Sci.}\  {\bf 68} (2018) 17 
%   doi:10.1146/annurev-nucl-101917-020902
  [arXiv:1711.10626 [hep-ph]].
  %%CITATION = doi:10.1146/annurev-nucl-101917-020902;%% 

%\cite{Yuan:2018inv}
\bibitem{Yuan:2018inv} 
  C.\,Z.~Yuan,
  {\it The $XYZ$ states revisited},
  {\it Int.\ J.\ Mod.\ Phys.}\ A {\bf 33} (2018) 1830018 
%   doi:10.1142/S0217751X18300181
  [arXiv:1808.01570 [hep-ex]].
  %%CITATION = doi:10.1142/S0217751X18300181;%% 

%\cite{Liu:2019zoy}
\bibitem{Liu:2019zoy} 
  Y.\,R.~Liu, H.\,X.~Chen, W.~Chen, X.~Liu and S.\,L.~Zhu,
  {\it Pentaquark and Tetraquark states},
  {\it Prog.\ Part.\ Nucl.\ Phys.}\  {\bf 107} (2019) 237 
%   doi:10.1016/j.ppnp.2019.04.003
  [arXiv:1903.11976 [hep-ph]].
  %%CITATION = doi:10.1016/j.ppnp.2019.04.003;%%

%\cite{Brambilla:2019esw}
\bibitem{Brambilla:2019esw} 
  N.~Brambilla, S.~Eidelman, C.~Hanhart, A.~Nefediev, C.\,P.~Shen, C.\,E.~Thomas, A.~Vairo and C.\,Z.~Yuan,
  {\it The $XYZ$ states: experimental and theoretical status and perspectives},
  arXiv:1907.07583 [hep-ex].
  %%CITATION = ARXIV:1907.07583;%% 

%\cite{Ali:2019roi}
\bibitem{Ali:2019roi}
  A.~Ali, L.~Maiani, and A.\,D.~Polosa,
  {\it Multiquark Hadrons}.   
  Cambridge University Press, Cambridge U.K. (2019). 
%   doi:10.1017/9781316761465
  %%CITATION = doi:10.1017/9781316761465;%%
  

%%%%%%%%%%%%%%%%%%%%%%%%%%%%%%%%%%%%%%%%%%%%%%%%%%%%%%%%%%%%%%%%%%%5


%\cite{Chen:2019bip}
\bibitem{Chen:2019bip} 
  H.\,X.~Chen, W.~Chen and S.\,L.~Zhu,
  {\it Possible interpretations of the $P_c(4312)$, $P_c(4440)$, and $P_c(4457)$},
  {\it Phys.\ Rev.}\ D {\bf 100} (2019) 051501 
%   doi:10.1103/PhysRevD.100.051501
  [arXiv:1903.11001 [hep-ph]].
  %%CITATION = doi:10.1103/PhysRevD.100.051501;%% 

%\cite{Chen:2019asm}
\bibitem{Chen:2019asm} 
  R.~Chen, X.~Liu, Z.\,F.~Sun and S.\,L.~Zhu,
  {\it Strong LHCb evidence for supporting the existence of hidden-charm molecular pentaquarks},
  {\it Phys.\ Rev}.\ D {\bf 100} (2019) 011502
%   doi:10.1103/PhysRevD.100.011502
  [arXiv:1903.11013 [hep-ph]].
  %%CITATION = doi:10.1103/PhysRevD.100.011502;%%
 
%\cite{Guo:2019fdo}
\bibitem{Guo:2019fdo} 
  F.\,K.~Guo, H.\,J.~Jing, U.\,G.~Meissner and S.~Sakai,
  {\it Isospin breaking decays as a diagnosis of the hadronic molecular structure of the $P_c(4457)$},
  {\it Phys.\ Rev.}\ D {\bf 99} (2019) 091501 
%   doi:10.1103/PhysRevD.99.091501 
  [arXiv:1903.11503 [hep-ph]].
  %%CITATION = ARXIV:1903.11503;%% 
 
 %\cite{Liu:2019tjn}
\bibitem{Liu:2019tjn} 
  M.\,Z.~Liu, Y.\,W.~Pan, F.\,Z.~Peng, M.~Sanchez Sanchez, L.\,S.~Geng, A.~Hosaka and M.\,P.~Valderrama,
  {\it Emergence of a complete heavy-quark spin symmetry multiplet: seven molecular pentaquarks in light of the latest LHCb analysis},
  {\it Phys.\ Rev.\ Lett.}\  {\bf 122} (2019) 242001 
%   doi:10.1103/PhysRevLett.122.242001
  [arXiv:1903.11560 [hep-ph]].
  %%CITATION = doi:10.1103/PhysRevLett.122.242001;%%
  
  %\cite{He:2019ify}
\bibitem{He:2019ify} 
  J.~He,
  {\it Study of $P_c(4457)$, $P_c(4440)$, and $P_c(4312)$ in a quasipotential Bethe-Salpeter equation approach},
  {\it Eur.\ Phys.\ J.}\ C {\bf 79} (2019) 393 
%   doi:10.1140/epjc/s10052-019-6906-1
  [arXiv:1903.11872 [hep-ph]].
  %%CITATION = doi:10.1140/epjc/s10052-019-6906-1;%%

%\cite{Xiao:2019aya}
\bibitem{Xiao:2019aya} 
  C.\,W.~Xiao, J.~Nieves and E.~Oset,
  {\it Heavy quark spin symmetric molecular states from $\bar D^{(*)} \Sigma_c^{(*)}$ and other coupled channels in the light of the recent LHCb pentaquarks},
  {\it Phys.\ Rev.}\ D {\bf 100} (2019) 014021
%   doi:10.1103/PhysRevD.100.014021
  [arXiv:1904.01296 [hep-ph]].
  %%CITATION = doi:10.1103/PhysRevD.100.014021;%%

%\cite{Huang:2019jlf}
\bibitem{Huang:2019jlf} 
  H.~Huang, J.~He and J.~Ping,
  {\it Looking for the hidden-charm pentaquark resonances in $J/\psi p$ scattering},
  arXiv:1904.00221 [hep-ph].
  %%CITATION = ARXIV:1904.00221;%%
 
 %\cite{Shimizu:2019ptd}
\bibitem{Shimizu:2019ptd} 
  Y.~Shimizu, Y.~Yamaguchi and M.~Harada,
  {\it Heavy quark spin multiplet structure of $P_c(4312)$, $P_c(4440)$, and $P_c(4457)$},
  arXiv:1904.00587 [hep-ph].
  %%CITATION = ARXIV:1904.00587;%%

 %\cite{Xiao:2019mvs}
\bibitem{Xiao:2019mvs} 
  C.\,J.~Xiao, Y.~Huang, Y.\,B.~Dong, L.\,S.~Geng and D.\,Y.~Chen,
  {\it Partial decay widths of $P_c(4312)$, $P_c(4440)$, and $P_c(4457)$ into $J/\psi p$ in a molecular scenario},
  {\it Phys.\ Rev.}\ D {\bf 100} (2019) 014022 
%   doi:10.1103/PhysRevD.100.014022
  [arXiv:1904.00872 [hep-ph]].
  %%CITATION = doi:10.1103/PhysRevD.100.014022;%%

%\cite{Zhang:2019xtu}
\bibitem{Zhang:2019xtu} 
  J.\,R.~Zhang,
  {\it Exploring a $\Sigma_c \bar D$ state: with focus on $P_c (4312)^+$},
  arXiv:1904.10711 [hep-ph].
  %%CITATION = ARXIV:1904.10711;%%

%\cite{Skwarnicki:2019}
\bibitem{Skwarnicki:2019}
  Tomasz Skwarnicki, 
  {\it Recent results on exotic hadrons from LHCb},
  talk at {\it the Workshop on Exotic Hadrons: Theory and Experiment at Lepton and Hadron Colliders},
  T.\,D.~Lee Institute, Jiao-Tong University, Shanghai, China, June 25--27, 2019.
  https://indico.leeinst.sjtu.edu.cn/event/57/ 
  
%\cite{Fernandez-Ramirez:2019koa}
\bibitem{Fernandez-Ramirez:2019koa} 
  C.~Fernandez-Ramirez {\it et al.} [JPAC Collaboration],
  {\it Interpretation of the LHCb $P_c (4312)$ Signal},
  {\it Phys.\ Rev.\ Lett.}\ {\bf 123} (2019) 092001 
%   doi:10.1103/PhysRevLett.123.092001
  [arXiv:1904.10021 [hep-ph]].
  %%CITATION = doi:10.1103/PhysRevLett.123.092001;%%

%\cite{Weng:2019ynv}
\bibitem{Weng:2019ynv} 
  X.\,Z.~Weng, X.\,L.~Chen, W.\,Z.~Deng and S.\,L.~Zhu,
  {\it Hidden-charm pentaquarks and $P_c$ states},
  {\it Phys.\ Rev.}\ D {\bf 100} (2019) 016014 
%   doi:10.1103/PhysRevD.100.016014
  [arXiv:1904.09891 [hep-ph]].
  %%CITATION = doi:10.1103/PhysRevD.100.016014;%%
  
%\cite{Ali:2019lzf}
\bibitem{Ali:2019lzf} 
  A.~Ali {\it et al.} [GlueX Collaboration],
  {\it First measurement of near-threshold $J/\psi$ exclusive photoproduction off the proton},
  {\it Phys.\ Rev.\ Lett.}\ {\bf 123} (2019) 072001 
%   doi:10.1103/PhysRevLett.123.072001
  [arXiv:1905.10811 [nucl-ex]].
  %%CITATION = doi:10.1103/PhysRevLett.123.072001;%%
  
 %\cite{Blin:2016dlf}
\bibitem{Blin:2016dlf} 
  A.\,N.~Hiller Blin, C.~Fernandez-Ramirez, A.~Jackura, V.~Mathieu, V.\,I.~Mokeev, A.~Pilloni and A.\,P.~Szczepaniak,
  {\it Studying the P$_c$(4450) resonance in $J/\psi$ photoproduction off protons},
  {\it Phys.\ Rev.}\ D {\bf 94} (2016) 034002 
%   doi:10.1103/PhysRevD.94.034002
  [arXiv:1606.08912 [hep-ph]].
  %%CITATION = doi:10.1103/PhysRevD.94.034002;%%
  
%\cite{Cao:2019kst}
\bibitem{Cao:2019kst}
  X.~Cao and J.\,p.~Dai,
  {\it Confronting pentaquark photoproduction with new LHCb observations},
  {\it Phys.\ Rev.}\ D {\bf 100} (2019) 054033 
%   doi:10.1103/PhysRevD.100.054033
  [arXiv:1904.06015 [hep-ph]].
  %%CITATION = ARXIV:1904.06015;%%

%\cite{Wang:2019krd}
\bibitem{Wang:2019krd} 
  X.\,Y.~Wang, X.\,R.~Chen and J.~He,
  {\it Possibility to study pentaquark states $P_c (4312)$, $P_c (4440)$ and $P_c (4457)$ in $\gamma p \rightarrow J/\psi p$ reaction},
  {\it Phys.\ Rev.}\ D {\bf 99} (2019) 114007 
%   doi:10.1103/PhysRevD.99.114007
  [arXiv:1904.11706 [hep-ph]].
  %%CITATION = doi:10.1103/PhysRevD.99.114007;%% 
  
%\cite{Voloshin:2019wxx}
\bibitem{Voloshin:2019wxx} 
  M.\,B.~Voloshin,
  {\it Hidden-charm pentaquark formation in antiproton-deuterium collisions},
  {\it Phys.\ Rev.}\ D {\bf 99} (2019) 093003 
%   doi:10.1103/PhysRevD.99.093003
  [arXiv:1903.04422 [hep-ph]].
  %%CITATION = doi:10.1103/PhysRevD.99.093003;%%

%\cite{Ali:2019npk}
\bibitem{Ali:2019npk} 
  A.~Ali and A.\,Ya.~Parkhomenko,
  {\it Interpretation of the Narrow $J/\psi p$ Peaks in $\Lambda_b \to J/\psi p K^-$ Decay in the Compact Diquark Model},
  {\it Phys.\ Lett.}\ B {\bf 793} (2019) 365 
%   doi:10.1016/j.physletb.2019.05.002
  [arXiv:1904.00446 [hep-ph]].
  %%CITATION = doi:10.1016/j.physletb.2019.05.002;%%
  
  %\cite{Maiani:2004vq}
\bibitem{Maiani:2004vq} 
  L.~Maiani, F.~Piccinini, A.\,D.~Polosa and V.~Riquer,
  {\it Diquark-antidiquarks with hidden or open charm and the nature of $X(3872)$},
  {\it Phys.\ Rev.}\ D {\bf 71} (2005) 014028 
%   doi:10.1103/PhysRevD.71.014028
  [hep-ph/0412098].
  %%CITATION = doi:10.1103/PhysRevD.71.014028;%%
  
  %\cite{Lipkin:1987sk}
\bibitem{Lipkin:1987sk} 
  H.\,J.~Lipkin,
  {\it New Possibilities for Exotic Hadrons: Anticharm Strange Baryons},
  {\it Phys.\ Lett.}\ B {\bf 195} (1987) 484.
%   doi:10.1016/0370-2693(87)90055-4
  %%CITATION = doi:10.1016/0370-2693(87)90055-4;%% 

%\cite{Jaffe:2003sg}
\bibitem{Jaffe:2003sg} 
  R.\,L.~Jaffe and F.~Wilczek,
  {\it Diquarks and exotic spectroscopy},
  {\it Phys.\ Rev.\ Lett.}\ {\bf 91} (2003) 232003 
%   doi:10.1103/PhysRevLett.91.232003
  [hep-ph/0307341].
  %%CITATION = doi:10.1103/PhysRevLett.91.232003;%% 
    
%\cite{Manohar:1992nd}
\bibitem{Manohar:1992nd} 
  A.\,V.~Manohar and M.\,B.~Wise,
  {\it Exotic $Q Q \bar q \bar q$ states in QCD},
  {\it Nucl.\ Phys.}\ B {\bf 399} (1993) 17 
%   doi:10.1016/0550-3213(93)90614-U
  [hep-ph/9212236].
  %%CITATION = doi:10.1016/0550-3213(93)90614-U;%%
  
%\cite{Esposito:2013fma}
\bibitem{Esposito:2013fma} 
  A.~Esposito, M.~Papinutto, A.~Pilloni, A.\,D.~Polosa and N.~Tantalo,
  {\it Doubly charm tetraquarks in $B_c$ and $\Xi_{bc}$ decays},
  {\it Phys.\ Rev.}\ D {\bf 88} (2013) 054029 
%   doi:10.1103/PhysRevD.88.054029
  [arXiv:1307.2873 [hep-ph]].
  %%CITATION = doi:10.1103/PhysRevD.88.054029;%%

%\cite{Luo:2017eub}
\bibitem{Luo:2017eub} 
  S.\,Q.~Luo, K.~Chen, X.~Liu, Y.\,R.~Liu and S.\,L.~Zhu,
  {\it Exotic tetraquark states with the $qq\bar{Q}\bar{Q}$ configuration},
  {\it Eur.\ Phys.\ J.}\ C {\bf 77} (2017) 709 
%   doi:10.1140/epjc/s10052-017-5297-4
  [arXiv:1707.01180 [hep-ph]].
  %%CITATION = doi:10.1140/epjc/s10052-017-5297-4;%%

%\cite{Karliner:2017qjm}
\bibitem{Karliner:2017qjm} 
  M.~Karliner and J.\,L.~Rosner,
  {\it Discovery of doubly-charm $\Xi_{cc}$ baryon implies a stable ($b b \bar u \bar d$) tetraquark},
  {\it Phys.\ Rev.\ Lett.}\  {\bf 119} (2017) 202001 
%   doi:10.1103/PhysRevLett.119.202001
  [arXiv:1707.07666 [hep-ph]].
  %%CITATION = doi:10.1103/PhysRevLett.119.202001;%%

%\cite{Eichten:2017ffp}
\bibitem{Eichten:2017ffp} 
  E.\,J.~Eichten and C.~Quigg,
  {\it Heavy-quark symmetry implies stable heavy tetraquark mesons $Q_i Q_j \bar q_k \bar q_l$},
  {\it Phys.\ Rev.\ Lett.}\  {\bf 119} (2017) 202002 
%   doi:10.1103/PhysRevLett.119.202002
  [arXiv:1707.09575 [hep-ph]].
  %%CITATION = doi:10.1103/PhysRevLett.119.202002;%%

%\cite{Francis:2016hui}
\bibitem{Francis:2016hui} 
  A.~Francis, R.\,J.~Hudspith, R.~Lewis and K.~Maltman,
  {\it Lattice Prediction for Deeply Bound Doubly Heavy Tetraquarks},
  {\it Phys.\ Rev.\ Lett.}\  {\bf 118} (2017) 142001 
%   doi:10.1103/PhysRevLett.118.142001
  [arXiv:1607.05214 [hep-lat]].
  %%CITATION = doi:10.1103/PhysRevLett.118.142001;%%


%\cite{Bicudo:2017szl}
\bibitem{Bicudo:2017szl} 
  P.~Bicudo, M.~Cardoso, A.~Peters, M.~Pflaumer and M.~Wagner,
  {\it $u d \bar b \bar b$ tetraquark resonances with lattice QCD potentials and the Born-Oppenheimer approximation},
  {\it Phys.\ Rev.}\ D {\bf 96} (2017) 054510 
%   doi:10.1103/PhysRevD.96.054510
  [arXiv:1704.02383 [hep-lat]].
  %%CITATION = doi:10.1103/PhysRevD.96.054510;%%

%\cite{Junnarkar:2017sey}
\bibitem{Junnarkar:2017sey} 
  P.~Junnarkar, M.~Padmanath and N.~Mathur,
  {\it Heavy light tetraquarks from Lattice QCD},
  {\it EPJ Web Conf.}\ {\bf 175} (2018) 05014 
%   doi:10.1051/epjconf/201817505014
  [arXiv:1712.08400 [hep-lat]].
  %%CITATION = doi:10.1051/epjconf/201817505014;%%
  
%\cite{Mehen:2017nrh}
\bibitem{Mehen:2017nrh} 
  T.~Mehen,
  {\it Implications of Heavy Quark-Diquark Symmetry for Excited Doubly Heavy Baryons and Tetraquarks},
  {\it Phys.\ Rev.}\ D {\bf 96} (2017) 094028 
%   doi:10.1103/PhysRevD.96.094028
  [arXiv:1708.05020 [hep-ph]].
  %%CITATION = doi:10.1103/PhysRevD.96.094028;%%

%\cite{Czarnecki:2017vco}
\bibitem{Czarnecki:2017vco} 
  A.~Czarnecki, B.~Leng and M.\,B.~Voloshin,
  {\it Stability of tetrons},
  {\it Phys.\ Lett.}\ B {\bf 778} (2018) 233 
%   doi:10.1016/j.physletb.2018.01.034
  [arXiv:1708.04594 [hep-ph]].
  %%CITATION = doi:10.1016/j.physletb.2018.01.034;%%

%\cite{Maiani:2019cwl}
\bibitem{Maiani:2019cwl} 
  L.~Maiani, A.\,D.~Polosa and V.~Riquer,
  {\it The Hydrogen Bond of QCD},
  {\it Phys.\ Rev.}\ D {\bf 100} (2019) 014002 
%   doi:10.1103/PhysRevD.100.014002
  [arXiv:1903.10253 [hep-ph]].
  %%CITATION = doi:10.1103/PhysRevD.100.014002;%% 

%\cite{Maiani:2019lpu}
\bibitem{Maiani:2019lpu} 
  L.~Maiani, A.\,D.~Polosa and V.~Riquer,
  {\it The Hydrogen Bond of QCD in Doubly Heavy Baryons and Tetraquarks},
  arXiv:1908.03244 [hep-ph].

%\cite{Lebed:2015tna}
\bibitem{Lebed:2015tna} 
  R.\,F.~Lebed,
  {\it The Pentaquark Candidates in the Dynamical Diquark Picture},
  {\it Phys.\ Lett.}\ B {\bf 749} (2015) 454 
%   doi:10.1016/j.physletb.2015.08.032
  [arXiv:1507.05867 [hep-ph]].
  %%CITATION = doi:10.1016/j.physletb.2015.08.032;%% 
 
%\cite{Ali:2016dkf}
\bibitem{Ali:2016dkf} 
  A.~Ali, I.~Ahmed, M.\,J.~Aslam and A.~Rehman,
  {\it Heavy quark symmetry and weak decays of the $b$-baryons in pentaquarks with a $c \bar c$ component},
  {\it Phys.\ Rev.}\ D {\bf 94} (2016) 054001 
%   doi:10.1103/PhysRevD.94.054001
  [arXiv:1607.00987 [hep-ph]].
  %%CITATION = doi:10.1103/PhysRevD.94.054001;%%
  
%\cite{Ali:2017ebb}
\bibitem{Ali:2017ebb} 
  A.~Ali, I.~Ahmed, M.\,J.~Aslam and A.~Rehman,
  {\it Mass spectrum of spin-1/2 pentaquarks with a $c \bar c$ component and their anticipated discovery modes in $b$-baryon decays},
  arXiv:1704.05419 [hep-ph].
  
  %\cite{Li:2015gta}
\bibitem{Li:2015gta} 
  G.\,N.~Li, X.\,G.~He and M.~He,
  {\it Some Predictions of Diquark Model for Hidden Charm Pentaquark Discovered at the LHCb},
  {\it JHEP} {\bf 1512} (2015) 128 
%   doi:10.1007/JHEP12(2015)128
  [arXiv:1507.08252 [hep-ph]].
  %%CITATION = doi:10.1007/JHEP12(2015)128;%% 

 %\cite{Giron:2019bcs}
\bibitem{Giron:2019bcs} 
  J.\,F.~Giron, R.\,F.~Lebed and C.\,T.~Peterson,
  {\it The Dynamical Diquark Model: First Numerical Results},
  {\it JHEP} {\bf 1905} (2019) 061 
%   doi:10.1007/JHEP05(2019)061
  [arXiv:1903.04551 [hep-ph]].
  %%CITATION = doi:10.1007/JHEP05(2019)061;%%

  
%\cite{Eides:2019tgv}
\bibitem{Eides:2019tgv} 
  M.\,I.~Eides, V.\,Y.~Petrov and M.\,V.~Polyakov,
  {\it New LHCb pentaquarks as hadrocharmonium states},
  arXiv:1904.11616 [hep-ph].
  %%CITATION = ARXIV:1904.11616;%%
  
  
%%%%%%%%%%%%%%%%%%%%%%%%%%%%%%%%%%%%%%%%%%%%%%%%%%%%%

%%%%%%%%%%%%%%%%%%%%%%%%%%%%%%%%%%%%
%\cite{Maiani:2014aja}
\bibitem{Maiani:2014aja} 
  L.~Maiani, F.~Piccinini, A.\,D.~Polosa and V.~Riquer,
  {\it The Z(4430) and a New Paradigm for Spin Interactions in Tetraquarks},
  {\it Phys.\ Rev.}\ D {\bf 89} (2014) 114010 
%   doi:10.1103/PhysRevD.89.114010
  [arXiv:1405.1551 [hep-ph]].
  %%CITATION = doi:10.1103/PhysRevD.89.114010;%%

%\cite{Ali:2017wsf}
\bibitem{Ali:2017wsf} 
  A.~Ali, L.~Maiani, A.\,V.~Borisov, I.~Ahmed, M.\,J.~Aslam, A.\,Ya.~Parkhomenko,
  A.\,D.~Polosa and A.~Rehman,
  {\it A New Look at the $Y$ Tetraquarks and $\Omega_c$ Baryons in the Diquark Model},
  {\it Eur.\ Phys.\ J.}\ C {\bf 78} (2018) 29 
%   doi:10.1140/epjc/s10052-017-5501-6
  [arXiv:1708.04650 [hep-ph]].
  %%CITATION = ARXIV:1708.04650;%%

\bibitem{Landau:1977} 
  L.\,D.~Landau and E.\,M.~Lifshitz, 
  {\it Quantum Mechanics (Nonrelativistic Theory)},  % 3rd edition. 
  Pergamon Press, Oxford (1977).

 %\cite{Karliner:2017kfm}
\bibitem{Karliner:2017kfm} 
  M.~Karliner and J.\,L.~Rosner,
  {\it Very narrow excited $\Omega_c$ baryons},
  {\it Phys.\ Rev.}\ D {\bf 95} (2017) 114012 
%   doi:10.1103/PhysRevD.95.114012
  [arXiv:1703.07774 [hep-ph]].
  %%CITATION = doi:10.1103/PhysRevD.95.114012;%%

%\cite{Edmonds:1957} 
\bibitem{Edmonds:1957} 
  A.\,R.~Edmonds, 
  {\it Angular Momentum in Quantum Mechanics},  
  Princeton University Press, Princeton, New Jersey (1957). 
  
%\cite{Karliner:2014gca}
\bibitem{Karliner:2014gca} 
  M.~Karliner and J.\,L.~Rosner,
  {\it Baryons with two heavy quarks: Masses, production, decays, and detection},
  {\it Phys.\ Rev.}\ D {\bf 90} (2014) 094007 
%   doi:10.1103/PhysRevD.90.094007
  [arXiv:1408.5877 [hep-ph]].
  %%CITATION = doi:10.1103/PhysRevD.90.094007;%%
   

%\cite{Karliner:2018bms}
\bibitem{Karliner:2018bms} 
  M.~Karliner and J.\,L.~Rosner,
  {\it Scaling of $P$-wave excitation energies in heavy-quark systems},
  {\it Phys.\ Rev.}\ D {\bf 98} (2018) 074026 
%   doi:10.1103/PhysRevD.98.074026
  [arXiv:1808.07869 [hep-ph]].
  %%CITATION = doi:10.1103/PhysRevD.98.074026;%% 

%\cite{Aaij:2017nav}
\bibitem{Aaij:2017nav} 
  R.~Aaij {\it et al.} [LHCb Collaboration],
  {\it Observation of five new narrow $\Omega_c^0$ states decaying to $\Xi_c^+ K^-$},
  {\it Phys.\ Rev.\ Lett.}\  {\bf 118} (2017) 182001 
%   doi:10.1103/PhysRevLett.118.182001
  [arXiv:1703.04639 [hep-ex]].
  %%CITATION = doi:10.1103/PhysRevLett.118.182001;%%

%\cite{Yelton:2017qxg}
\bibitem{Yelton:2017qxg}
  J.~Yelton {\it et al.} [Belle Collaboration],
  {\it Observation of Excited $\Omega_c$ Charm Baryons in $e^+ e^-$ Collisions},
  {\it Phys.\ Rev.}\ D {\bf 97} (2018) 051102   
%   doi:10.1103/PhysRevD.97.051102
  [arXiv:1711.07927 [hep-ex]].
  %%CITATION = doi:10.1103/PhysRevD.97.051102;%%

%\cite{Stancu:2019qga}
\bibitem{Stancu:2019qga} 
  F.~Stancu,
  {\it Spectrum of the $uudc \bar c$ hidden charm pentaquark with an $SU(4)$ flavor-spin hyperfine interaction},
  arXiv:1902.07101 [hep-ph].
  %%CITATION = ARXIV:1902.07101;%%

%\cite{Maiani:2017kyi}
\bibitem{Maiani:2017kyi} 
  L.~Maiani, A.\,D.~Polosa and V.~Riquer,
  {\it A Theory of $X$ and $Z$ Multiquark Resonances},
  {\it Phys.\ Lett.}\ B {\bf 778} (2018) 247 
%   doi:10.1016/j.physletb.2018.01.039
  [arXiv:1712.05296 [hep-ph]].
  %%CITATION = doi:10.1016/j.physletb.2018.01.039;%%

%\cite{Brodsky:2014xia}
\bibitem{Brodsky:2014xia} 
  S.\,J.~Brodsky, D.\,S.~Hwang and R.\,F.~Lebed,
  {\it Dynamical Picture for the Formation and Decay of the Exotic $XYZ$ Mesons},
  {\it Phys.\ Rev.\ Lett.}\  {\bf 113} (2014) 112001 
%   doi:10.1103/PhysRevLett.113.112001
  [arXiv:1406.7281 [hep-ph]].
  %%CITATION = doi:10.1103/PhysRevLett.113.112001;%%

%\cite{Lebed:2017min}
\bibitem{Lebed:2017min} 
  R.\,F.~Lebed,
  {\it Spectroscopy of Exotic Hadrons Formed from Dynamical Diquarks},
  {\it Phys.\ Rev.}\ D {\bf 96} (2017) 116003 
%   doi:10.1103/PhysRevD.96.116003
  [arXiv:1709.06097 [hep-ph]].
  %%CITATION = doi:10.1103/PhysRevD.96.116003;%%

%\cite{Esposito:2018cwh}
\bibitem{Esposito:2018cwh}
  A.~Esposito and A.\,D.~Polosa,
  {\it A $bb\bar b\bar b$ di-bottomonium at the LHC?},
  {\it Eur.\ Phys.\ J.}\ C {\bf 78} (2018) 782 
%   doi:10.1140/epjc/s10052-018-6269-z
  [arXiv:1807.06040 [hep-ph]].
  %%CITATION = doi:10.1140/epjc/s10052-018-6269-z;%%

%\cite{An:2018cln}
\bibitem{An:2018cln} 
  H.~An and M.\,B.~Wise,
  {\it The Direct Coupling of Light Quarks to Heavy Di-quarks},
  {\it Phys.\ Lett.}\ B {\bf 788} (2019) 131  
%   doi:10.1016/j.physletb.2018.11.004
  [arXiv:1809.02139 [hep-ph]].
  %%CITATION = doi:10.1016/j.physletb.2018.11.004;%%

\end{thebibliography}
\end{document}